\documentclass[openany]{nature}
\usepackage[T1]{fontenc}
\usepackage[left]{lineno}
\usepackage{amsthm,amsmath,amssymb}
\usepackage{mathrsfs}
\usepackage{overpic}
\usepackage{multirow}
\usepackage[T1]{fontenc}
\usepackage{graphicx}
\usepackage{pdflscape}
\usepackage{hyperref}
\usepackage{threeparttable}
\usepackage{threeparttablex}
\usepackage{xcolor}
\usepackage{ulem}
\usepackage{cancel}
\usepackage[figuresleft]{rotating}
\usepackage{caption}
\usepackage{subfigure}
\usepackage{tablefootnote}
\usepackage{hyperref}
\usepackage{graphicx}
\usepackage{longtable}
\usepackage{supertabular}
\usepackage{float} 
\usepackage{authblk}
\usepackage{multibib}
\usepackage{subfiles}
\usepackage{enumerate}
\usepackage{bm}

\hypersetup{
colorlinks=true,
linkcolor=red,
filecolor=blue,
urlcolor=cyan,
citecolor=green}

\makeatletter
 
\def\emph#1 {\textit{ #1 } }

\let\saved@includegraphics\includegraphics
\AtBeginDocument{\let\includegraphics\saved@includegraphics}
\renewenvironment*{figure}{\@float{figure}}{\end@float}
\makeatother

\def\@fnsymbol#1{\ensuremath{\ifcase#1\or \dagger\or \ddagger\or
 \mathsection\or \mathparagraph\or \|\or **\or \dagger\dagger
 \or \ddagger\ddagger \else\@ctrerr\fi}}
\makeatother

\newcommand{\apj}{Astrophys. J.}

\newcommand{\pasp}{Publ. Astron. Soc. Pac.}
\newcommand{\apjs}{Astrophys. J. Supp.}

\newcommand{\mnras}{Mon. Not. R. Astron. Soc.}
\newcommand{\apjl}{Astrophys. J. Let.}
\newcommand{\aap}{Astron. Astrophys.}
\newcommand{\aj}{Astron. J.}
\newcommand{\nat}{Nature}

\newcommand{\prl}{Phys. Rev. Lett.}
\newcommand{\prd}{Phys. Rev. D}

\newcommand{\cjaa}{Chinese Journal of Astronomy and Astrophysics}

\def\be{\begin{eqnarray}}
\def\ee{\end{eqnarray}}

\makeatletter

\let\oldequation\equation
\let\oldendequation\endequation
\renewenvironment{equation}{\linenomathNonumbers\oldequation}{\oldendequation\endlinenomath}

\let\oldalign\align
\let\oldendalign\endalign
\renewenvironment{align}{\linenomathNonumbers\oldalign}{\oldendalign\endlinenomath}

\let\oldgather\gather
\let\oldendgather\endgather
\renewenvironment{gather}{\linenomathNonumbers\oldgather}{\oldendgather\endlinenomath}

\definecolor{dkblue}{RGB}{54, 86, 169}

\title{Magnetar emergence in a peculiar gamma-ray burst from a compact star merger}

\author{H. Sun$^{1}$\thanks{These authors contributed equally to this work}, C.-W. Wang$^{2,3*}$, J. Yang$^{4,5*}$, B.-B. Zhang$^{4,5,6}$\thanks{E-mail: bbzhang@nju.edu.cn}, S.-L. Xiong$^{2}$\thanks{E-mail: xiongsl@ihep.ac.cn}, Y.-H. I. Yin$^{4}$, Y. Liu$^{1}$, Y. Li$^{6}$, W.-C. Xue$^{2,3}$, Z.-Y. Yan$^{4}$, C. Zhang$^{1,3}$, W.-J. Tan$^{2,3}$, H.-W. Pan$^{1}$, J.-C. Liu$^{2,3}$, H.-Q. Cheng$^{1}$, Y.-Q. Zhang$^{2,3}$, J.-W. Hu$^{1}$, C. Zheng$^{2,3}$, Z.-H. An$^{2}$, C. Cai$^{7}$, Z.-M. Cai$^{8}$, L. Hu$^{6}$, C. Jin$^{1,3}$, D.-Y. Li$^{1}$, X.-Q. Li$^{2}$, H.-Y. Liu$^{1}$, M. Liu$^{1,3}$, W.-X. Peng$^{2}$, L.-M. Song$^{2,3}$, S.-L. Sun$^{9}$, X.-J. Sun$^{9}$, X.-L. Wang$^{2}$, X.-Y. Wen$^{2}$, S. Xiao$^{10}$, S.-X. Yi$^{2}$, F. Zhang$^{2}$, W.-D. Zhang$^{1}$, X.-F. Zhang$^{8}$, Y.-H. Zhang$^{8}$, D.-H. Zhao$^{1}$, S.-J. Zheng$^{2}$, Z.-X. Ling$^{1,3}$\thanks{E-mail: lingzhixing@nao.cas.cn}, S.-N. Zhang$^{2,3}$, W. Yuan$^{1,3}$, B. Zhang$^{11,12}$\thanks{E-mail: bing.zhang@unlv.edu}}

\begin{document}

\maketitle

\begin{affiliations}
 \item National Astronomical Observatories, Chinese Academy of Sciences, Beijing 100101, China.
 \item Key Laboratory of Particle Astrophysics, Institute of High Energy Physics, Chinese Academy of Sciences, Beijing 100049, China.
 \item University of Chinese Academy of Sciences, Chinese Academy of Sciences, Beijing 100049, China.
 \item School of Astronomy and Space Science, Nanjing University, Nanjing 210023, China.
 \item Key Laboratory of Modern Astronomy and Astrophysics (Nanjing University), Ministry of Education, Nanjing 210022, China.
 \item Purple Mountain Observatory, Chinese Academy of Sciences, Nanjing 210023, China.
 \item College of Physics and Hebei Key Laboratory of Photophysics Research and Application, Hebei Normal University, Shijiazhuang, Hebei 050024, China.
 \item Innovation Academy for Microsatellites, Chinese Academy of Sciences, Shanghai, 201304, China.
 \item Shanghai Institute of Technical Physics, Chinese Academy of Sciences, Shanghai, 200083, China.
 \item Guizhou Provincial Key Laboratory of Radio Astronomy and Data Processing, Guizhou Normal University, Guiyang 550001, China. 
 \item Nevada Center for Astrophysics, University of Nevada Las Vegas, NV 89154, USA.
 \item Department of Physics and Astronomy, University of Nevada Las Vegas, NV 89154, USA.
\end{affiliations}

\clearpage

\begin{abstract}
The central engine that powers gamma-ray bursts (GRBs), the most powerful explosions in the universe, is still not identified. Besides hyper-accreting black holes, rapidly spinning and highly magnetized neutron stars, known as millisecond magnetars, have been suggested to power both long and short GRBs\cite{Usov1992Nat, Dai1998PhR, Zhang2001ApJ, Dai2006Sci, GaoFan2006, Metzger2011MNRAS, Zhang2013ApJ, Piro2019MNRAS}. The presence of a magnetar engine following compact star mergers is of particular interest as it would provide essential constraints on the poorly understood equation of state for neutron stars\cite{Gao2016PRD, Margalit2019ApJ}. Indirect indications of a magnetar engine in these merger sources have been observed in the form of plateau features present in the X-ray afterglow light curves of some short GRBs\cite{Rowlinson2013MNRAS, Lu2015ApJ}. Additionally, some X-ray transients lacking gamma-ray bursts have been identified as potential magnetar candidates originating from compact star mergers\cite{Zhang2013ApJ, Xue2019Nat, Sun2019ApJ}. Nevertheless, smoking gun evidence is still lacking for a magnetar engine in short GRBs, and associated theoretical challenges have been raised\cite{Ciolfi2020MNRAS}. 
Here we present a comprehensive analysis of the broad-band prompt emission data of a peculiar, very bright GRB 230307A. Despite its apparently long duration, the prompt emission and host galaxy properties are consistent with a compact star merger origin, as suggested by its association with a kilonova\cite{Levan2023arXiv}. Intriguingly, an extended X-ray emission component shows up as the $\gamma$-ray emission dies out, signifying the likely emergence of a magnetar central engine. We also identify an achromatic temporal break in the high-energy band during the prompt emission phase, which was never observed in previous bursts and reveals a narrow jet with half opening angle of $\sim 3.4^\circ (R_{\rm GRB}/10^{15}~{\rm cm})^{-1/2}$, where $R_{\rm GRB}$ is the GRB prompt emission radius.
\end{abstract}

\begin{figure}
\centering
\includegraphics[width=0.96\textwidth]{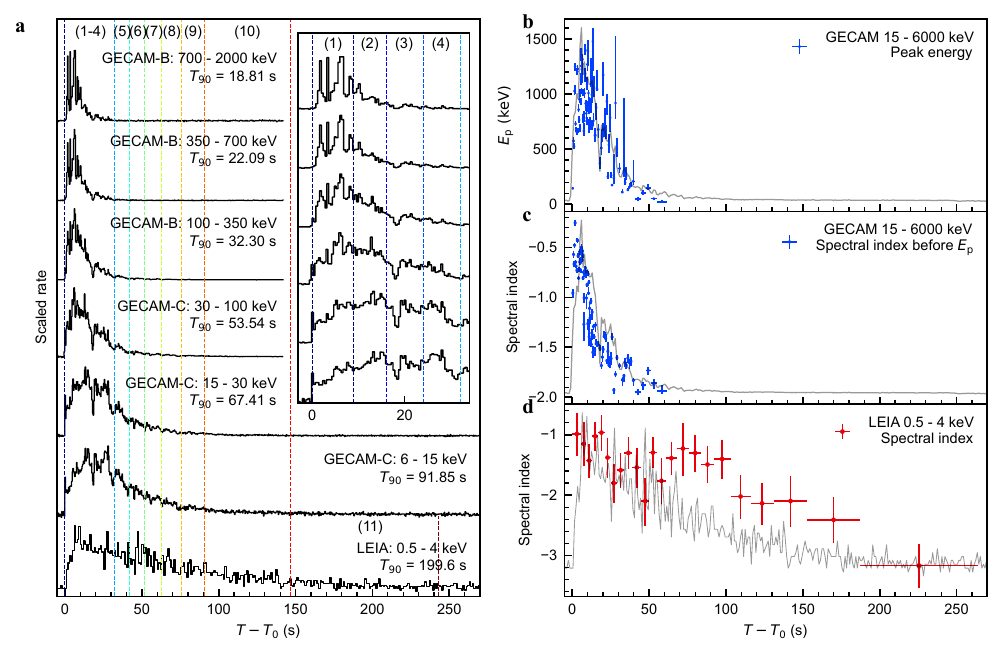}
\caption{\noindent\textbf{The temporal and spectral behaviors of GRB 230307A.} \textbf{a}, The multi-wavelength net light curves of GRB 230307A. The spectral energy distribution time intervals are demarcated by vertical dashed lines, accompanied by numerical labels at the top. Insets within the panel provide detailed profiles of prompt light curves in GECAM energy bands. \textbf{b} and \textbf{c}, The evolution of peak energy $E_{\rm p}$ and spectral power law index derived from GECAM spectral fittings in 15--6000 $\rm keV$. \textbf{d}, The evolution of spectral power law index derived from LEIA spectral fittings in 0.5--4 $\rm keV$. The light curves (gray lines) are plotted in the background as a reference. All error bars represent the 1$\sigma$ confidence level.}
\label{fig:lc}
\end{figure}

At 15:44:06.650 UT on 7 March 2023 (denoted as $T_0$) Gravitational wave high-energy Electromagnetic Counterpart All-sky Monitor (GECAM)\cite{Li2021RDTM, ZhangDL2023arXiv} was triggered by the extremely bright GRB 230307A\cite{Xiong2023GCN}. The burst was also reported by the {\it Fermi} Gamma-ray Burst Monitor (GBM)\cite{Fermi2023GCN33405}. Utilizing the unsaturated GECAM data (Methods), we determined the burst's duration ($T_{90}$) to be $41.52 \pm 0.03$ s in the 10--1000 keV energy range (Table \ref{tab:obs_prob}, see Fig. \ref{fig:lc}a for the energy-band-dependent light curves). The peak flux and total fluence in the same energy range were found to be $4.26_{-0.07}^{+0.08} \times 10^{-4}$ $\rm erg \ cm^{-2} \ s^{-1}$ and $(3.10 \pm 0.01) \times 10^{-3}$ $\rm erg \ cm^{-2}$, respectively, making it the second brightest GRB observed, only dwarfed by the brightest-of-all-time GRB 221009A\cite{An2023arXiv}. The pathfinder of the Einstein Probe mission\cite{Yuan2022} named Lobster Eye Imager for Astronomy (LEIA)\cite{Zhang2022ApJL, Ling2023arXiv}, with its large field of view of 340 $\rm deg^2$, caught the prompt emission of this burst in the soft X-ray band (0.5--4 keV) exactly at its trigger time\cite{Liu2023GCN} (Methods), revealing a significantly longer duration of $199.6_{-2.2}^{+5.1}$ s and a peak flux of $3.6_{-0.5}^{+0.6} \times 10^{-7}$ $\rm erg \ cm^{-2} \ s^{-1}$ (Fig. \ref{fig:lc}a and Table \ref{tab:obs_prob}). Subsequent follow-up observations\cite{Levan2023arXiv} indicate that the burst is most likely associated with a nearby galaxy at a redshift of $z=0.065$. Despite its long duration, the association of a kilonova signature\cite{Levan2023arXiv, YHYang2023prep} implies that this burst originates from a binary compact star merger. 

The broad-band (0.5--6000 $\rm keV$, Methods) prompt emission data we have collected from GECAM and LEIA also provide supportive evidence for a compact star merger origin. The burst's placement on various correlation diagrams is consistent with the so-called type I GRBs\cite{Zhang2009ApJ}, i.e., those with a compact star merger origin (Methods and Extended Data Fig. \ref{fig:classification}). First, its relatively small minimum variability timescale is more consistent with type I GRBs. Second, it deviates from the Amati relation of type II GRBs (massive star core collapse origin) but firmly falls into the 1$\sigma$ scattering region of type I GRBs (also see Ref.\cite{YHYang2023prep}). Third, it is a significant outlier of the anti-correlation between the spectral lags and peak luminosities of type II GRBs but is mixed in with other type I GRBs. Besides, the optical host galaxy data\cite{Levan2023arXiv} adds yet one more support: the location of the burst has a significant offset from the host galaxy, which is at odds with type II GRBs but is fully consistent with type I GRBs. This is the second strong instance for long-duration type I GRBs after GRB 211211A\cite{Yang2022Natur, Rastinejad2022Natur, Troja2022Natur,Mei2022Natur, Dichiara2023ApJ}.

\begin{figure}
\centering
\begin{tabular}{cc}
\begin{overpic}[width=0.35\textwidth]{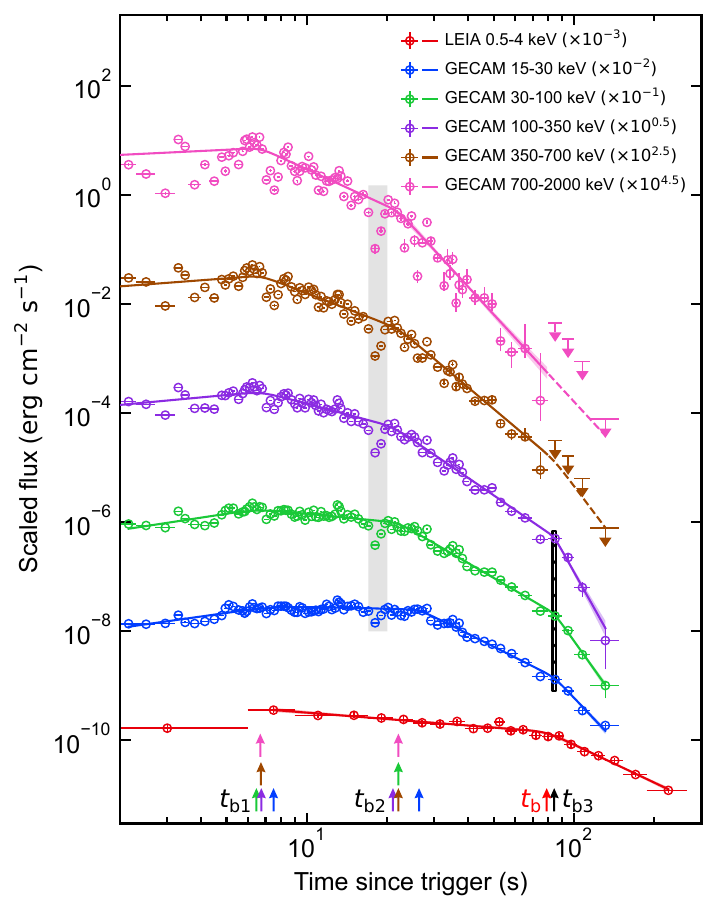}\put(0, 97){\bf a}\end{overpic} &
\begin{overpic}[width=0.36\textwidth]{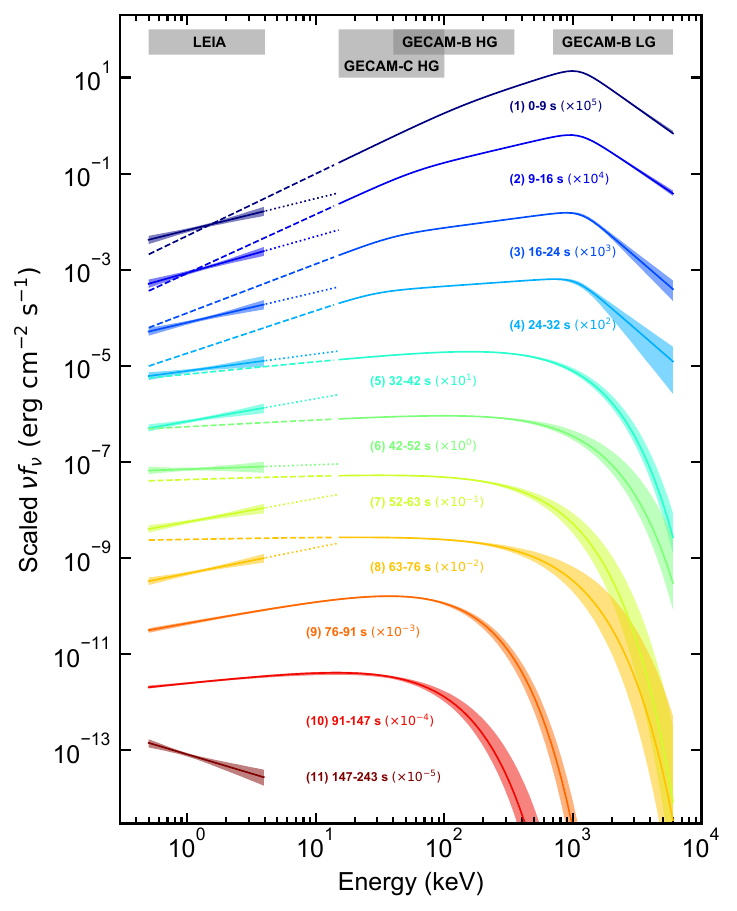}\put(0, 97){\bf b}\end{overpic} \\
\multicolumn{2}{c}{\begin{overpic}[width=0.60\textwidth]{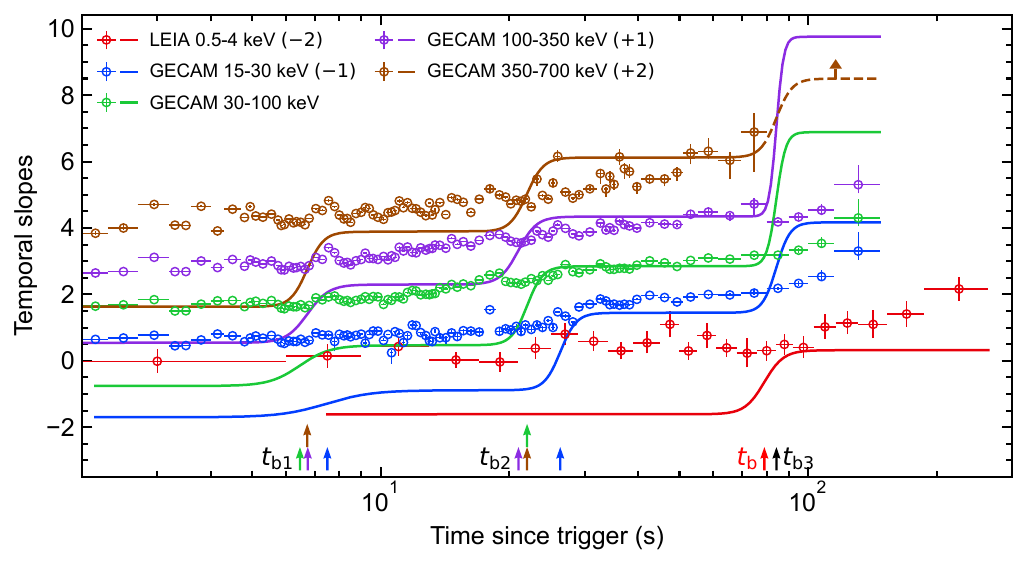}\put(0, 52){\bf c}\end{overpic}} \\
\end{tabular}
\caption{\noindent\textbf{The flux light curves and spectral energy distributions (SEDs) of GRB 230307A.} \textbf{a}, The multi-wavelength flux light curves. The fluxes (data points) are derived from time-resolved spectral fittings, and the lines represent the best fits of smoothly broken power law (SBPL) functions to the multi-wavelength flux light curves. The downward arrows represent the 3$\sigma$ upper limits of fluxes. The grey shaded area marks the dip phase (17--20 s since trigger time). The black hatched area represents the achromatic break in GECAM energy bands. \textbf{b}, The evolution of SEDs. The SEDs are derived from the spectral fittings at different time intervals as indicated by the labels. The dotted and dashed lines represent the natural extrapolations of the best-fit models of LEIA and GECAM independently spectral fittings, respectively. \textbf{c}, The temporal slopes. The data points ($\hat{\beta}+2$) represent the temporal slopes predicted by the curvature effect, and the lines ($\hat{\alpha}$) depict the temporal slopes corresponding to the SBPL fits to the multi-wavelength flux light curves. In \textbf{a} and \textbf{c}, all the break times derived from the SBPL fits to the multi-wavelength flux light curves are labeled by the upward arrows. All error bars on data points represent their 1$\sigma$ confidence level. All shaded areas around the best-fit lines represent their 1$\sigma$ confidence bands.}
\label{fig:sed}
\end{figure}

With the broad-band coverage jointly provided by GECAM-B/GECAM-C and LEIA throughout the prompt emission phase, one can perform a detailed temporal and spectral analysis of the data of GRB 230307A (Methods and Fig. \ref{fig:lc}). The light curves in the energy range of GECAM-B and GECAM-C exhibit synchronized pulses with matching peak and dip features (Fig. \ref{fig:lc}a). The time-resolved spectrum in the 15--6000 keV displays significant evolution (Fig. \ref{fig:lc}b and \ref{fig:lc}c), aligning with the ``intensity tracking'' pattern (i.e., the peak energy tracks the intensity evolution\cite{Golenetskii1983Nat}). When plotting the energy flux light curves in the logarithmic-logarithmic space (Fig. \ref{fig:sed}a), a break is identified around 21--26 s post-trigger in all five GECAM bands (15--30 keV, 30--100 keV, 100--350 keV, 350--700 keV, and 700--2000 keV), 
after which the light curves decay with indices ($\hat{\alpha}$) of $2.45_{-0.02}^{+0.02}$, $2.85_{-0.01}^{+0.01}$, $3.34_{-0.02}^{+0.02}$, $4.12_{-0.06}^{+0.06}$, and $5.3_{-0.3}^{+0.3}$ in the convention of $F_{\rm\nu} \propto t^{-\hat{\alpha}}\nu^{-\hat{\beta}}$ for these bands respectively (Methods and Extended Data Table \ref{tab:lc_sbpl}).
These measured slopes are fully consistent with the theoretical prediction of the relationship between temporal decay slope ($\hat{\alpha}$) and spectral slope ($\hat{\beta}$) solely due to the high-latitude emission after a sudden cessation of prompt emission, namely, $\hat{\alpha} = \hat{\beta} + 2 $ (the so-called curvature effect\cite{Kumar2000ApJ}). This suggests that the prompt high-energy emission abruptly ceased or significantly reduced its emission amplitude at about 21--26 s post-trigger (Methods and Fig. \ref{fig:sed}c). The variability observed in the light curves after the ceasation of the central engine is naturally expected within the framework of the internal-collision-induced magnetic reconnection and turbulence (ICMART) model, because there are mini-jet emission at high latitudes\cite{Zhang2011ApJ,Zhang2014ApJ}.
There is an additional achromatic break around 84 s, after which the three light curves (15--30 keV, 30--100 keV, and 100--350 keV) decay with much steeper slopes (Methods and Fig. \ref{fig:sed}a and \ref{fig:sed}c). This is consistent with the edge effect of a narrow jet that powers the prompt $\gamma$-ray emission, with a half opening angle of $\sim 3.4^\circ (R_{\rm GRB}/10^{15}~{\rm cm})^{-1/2}$, where $R_{\rm GRB}$ is the unknown GRB prompt emission radius from the central engine (Methods). The jet opening angle is generally consistent with that derived from multi-band modeling of the afterglows\cite{YHYang2023prep}, where the jet break is observed at around a few days. This brings the collimation-corrected jet energy to $\sim5.6\times 10^{49}$ erg, typical for type I GRBs\cite{wang2018}. 

In contrast to the hard X-rays and gamma-rays, the soft X-ray emission in the 0.5--4 keV LEIA-band exhibits a different behavior. The emission sustains for a much longer duration of $>250$ s in the form of a plateau followed by a decline. Its spectrum shows much less significant evolution within the first 100 s (Fig. \ref{fig:lc}d and Extended Data Table \ref{tab:leia_spec_fit}) compared to the high-energy GECAM spectrum. Notably, its spectral shape from the beginning up to $\sim$76 s deviates strongly from the extrapolation to lower energies of the spectral energy distributions derived from the GECAM data (Methods and Fig. \ref{fig:sed}b). These deviations cannot easily be ascribed to a simple spectral break at low energies as sometimes seen in GRBs\cite{Oganesyan2017ApJ}, but rather hint at a different radiation process dominating the LEIA band. As the high-energy emission suddenly ceases at around 21–26 s (with the decay slope controlled by the curvature effect), the late decay slope in the LEIA band is shallower than the curvature effect prediction, suggesting an intrinsic temporal evolution from the central engine (Fig. \ref{fig:sed}c). These facts suggest that the LEIA-band soft X-ray emission comes from a distinct emission component from the GRB, which emerges already from the onset of the burst (Fig. \ref{fig:sed}a). After 76 seconds, as the $\gamma$-ray emission dies out, the joint LEIA and GECAM spectra can be well fitted using a single cutoff power law model (Fig. \ref{fig:sed}b). This indicates that the entire spectra after 76s is dominated by the new emergent component observed by LEIA.

\begin{figure}
\centering
\begin{tabular}{cc}
\begin{overpic}[width=0.48\textwidth]{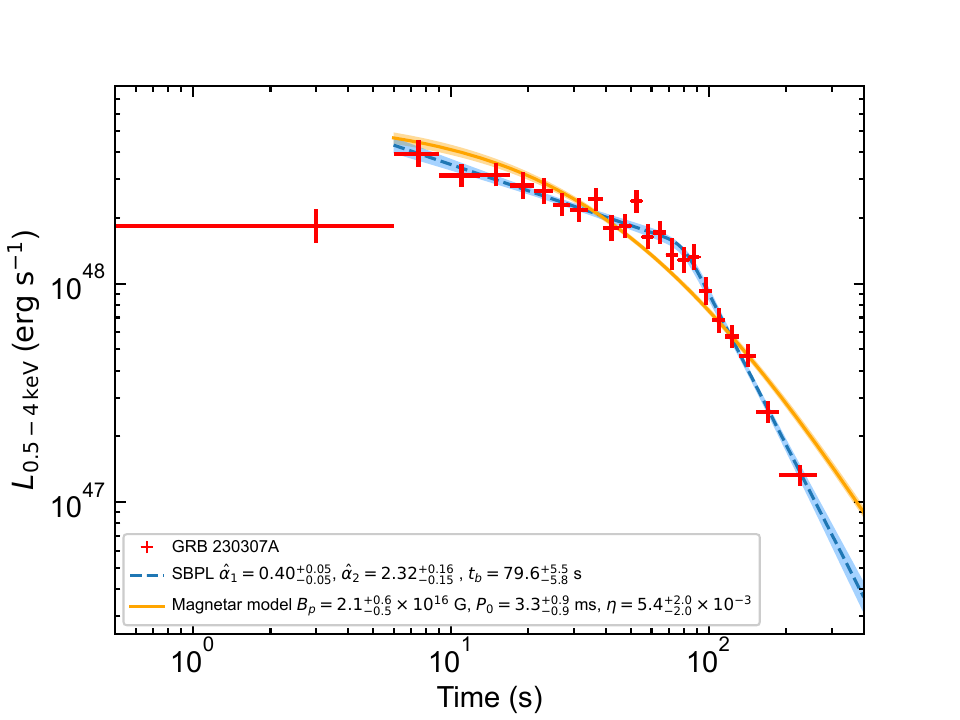}\put(5, 65){\bf a}\end{overpic} &
\begin{overpic}[width=0.48\textwidth]{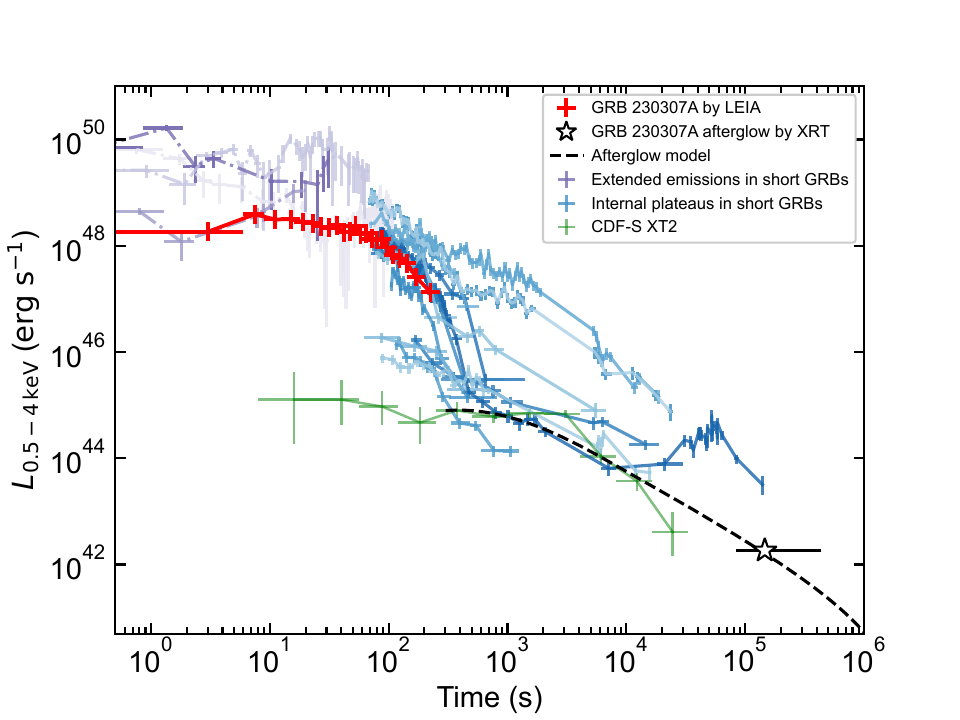}\put(0, 65){\bf b}\end{overpic} \\
\end{tabular}
\caption{\noindent\textbf{X-ray luminosity light curve of GRB 230307A.} \textbf{a}, The unabsorbed X-ray luminosity light curve in the energy range of $0.5 - 4$ $\rm keV$, 
excluding the first data point,
is fitted using both the smoothly broken power law model and the magnetar dipole radiation model, as described in Methods. The shaded area represents the 1$\sigma$ confidence bands. \textbf{b}, GRB 230307A compared with the X-ray afterglows of the internal plateau sample (blue solid line) and the extended emissions (purple dash-dotted line) in short GRBs\cite{Lu2015ApJ} (blue and purple population) and CDF-S XT2, all corrected to the LEIA energy band. The Swift/XRT-detected X-ray afterglow of GRB 230307A is represented as an unfilled black star. The extrapolation of the afterglow model\cite{YHYang2023prep} is plotted as a black dashed curve, which is significantly below the LEIA data points, indicating that the LEIA emission may not be the early afterglow emission (Methods).}
\label{fig:magnetar}
\end{figure}

A smoothly broken power law fit to the LEIA light curve gives decay slopes of {$0.40_{-0.05}^{+0.05}$} and $2.32_{-0.15}^{+0.16}$ before and after the break time at $79.6_{-5.8}^{+5.5}$ s (Methods and Fig. \ref{fig:magnetar}a). This pattern is generally consistent with the magnetic dipole spin-down law of a newborn, rapidly spinning magnetar. The best fit of the luminosity light curve with the magnetar model\cite{Zhang2001ApJ, Lu2015ApJ} yields a dipole magnetic field of $2.1^{+0.6}_{-0.5} \times 10^{16}$ G, an initial spin period of $3.3^{+0.9}_{-0.9}$ ms, and a radiation efficiency of $5.4^{+2.0}_{-2.0}\times 10^{-3}$ (Methods and Fig. \ref{fig:magnetar}a).  An X-ray plateau may also be interpreted within a black hole engine with a long-term accretion disk\cite{Lu2023}. However, such a long-lived disk would be eventually quickly evaporated so that the jet engine would cease abruptly. The fact that the final decay slope of the LEIA light curve is shallower than the curvature effect prediction rules out such a possibility and reinforces the magnetar interpretation. The considerable discrepancy between the extrapolation of the afterglow model\cite{YHYang2023prep} and the observed data suggests that the soft X-ray component is unlikely to be the X-ray afterglow (Fig. \ref{fig:magnetar}b). Further modeling, by introducing a low-energy spectral break due to synchrotron self-absorption at low energies\cite{Shen2009MNRAS,Troja2022Natur}, indicates that the entire spectral energy distribution of the prompt emission can be interpreted by a combination of the prompt emission, which drops quickly at low energies, and a new magnetar component (Methods).

Indirect evidence of a magnetar engine in compact star mergers has been collected before in the form of extended emissions or internal plateaus in short GRBs\cite{Rowlinson2013MNRAS, Lu2015ApJ} and some short-GRB-less X-ray transients such as CDF-S XT2\cite{Xue2019Nat}. Fig. \ref{fig:magnetar}b shows the comparison of the X-ray luminosity light curves between GRB 230307A and other magnetar candidates. It shows that GRB 230307A is consistent with the other sources but displays the full light curve right from the trigger directly in the X-ray band, thanks to the prompt detection by the wide-field X-ray camera of LEIA. It reveals the details of the emergence of the magnetar emission component and lends further support to the magnetar interpretation of other events. GRB 230307A marks the first simultaneous observation of a compact object merger in both X-ray and gamma-ray energies. In the near future, the synergy between GRB monitors and wide-field soft X-ray telescopes (such as the Einstein Probe) may detect more cases and will generally provide more observational information to diagnose the physics of GRBs during the prompt emission stage.

The identification of a magnetar engine from a merger event suggests that the neutron star (NS) equation of state is relatively stiff\cite{Gao2016PRD,Margalit2019ApJ}. The association with a regular kilonova\cite{Levan2023arXiv, YHYang2023prep} suggests that the energy injection into the ejecta from the magnetar engine is moderate (Methods). The magnetar engine also challenges modelers who currently fail to generate a relativistic jet from new-born magnetars\cite{Ciolfi2020MNRAS}. One possibility is that a highly magnetized jet is launched seconds after the birth of the magnetar when the proto-neutron star cools down and the wind becomes clean enough\cite{Metzger2011MNRAS}. In any case, the concrete progenitor of GRB 230307A remains an enigma. With a magnetar engine the progenitor can only be a binary neutron star merger or a (near Chandrasekhar limit) white dwarf -- NS merger\cite{Yang2022Natur}. For the former possibility, one must explain why this burst is particularly long. A ``tip-of-iceberg'' test\cite{Lu2014MNRAS} suggests that it is hard to make this burst to be a bright short GRB when one arbitrarily raises the background flux or moves the source to higher redshift (Methods and Table \ref{tab:obs_prob}). For the latter scenario, the fact that the light curves and spectral evolution between GRBs 230307A and 211211A\cite{Yang2022Natur} do not fully resemble each other would suggest that the mechanism must be able to produce diverse light curves. Unfortunately, GRB 230307A was detected prior to the fourth operation run (O4) of LIGO-Virgo-KAGRA. Future multi-messenger observations of similar events hold the promise of eventually unveiling the identity of the progenitor of these peculiar systems\cite{Yin2023arXiv}.

\clearpage

\captionsetup[table]{name={\bf Table}}
\captionsetup[figure]{name={\bf Fig.}}

\begin{table*}
\centering
\renewcommand\arraystretch{0.8}
\begin{threeparttable}
\caption{\textbf{Observational properties.} All errors represent the 1$\sigma$ uncertainties.}
\label{tab:obs_prob}
\begin{tabular}{lc}
\hline
Observed Properties & GRB 230307A \\
\hline
\textbf{Gamma-Ray [10--1000 keV]:} & \\
Duration ($\rm s$) & $41.52 \pm 0.03$ \\
Effective amplitude & $1.23 \pm 0.07$ \\
Minimum variability timescale ($\rm ms$) & $9.35$ \\
Rest-frame spectral lag\tnote{*} ($\rm ms$) & $1.6_{-1.2}^{+1.4}$ \\
Spectral index $\alpha_1$ & $-0.92_{-0.03}^{+0.05}$ \\
Spectral index $\alpha_2$ & $-1.274_{-0.008}^{+0.005}$ \\
Spectral index $\beta$ & $-3.85_{-0.09}^{+0.03}$ \\
Break energy $E_{\rm b}$ ($\rm keV$) & $24_{-2}^{+3}$ \\
Peak energy $E_{\rm p}$ ($\rm keV$) & $1052_{-8}^{+16}$ \\
Peak flux ($\rm erg~cm^{-2}~s^{-1}$) & $4.26_{-0.07}^{+0.08} \times 10^{-4}$ \\
Total fluence ($\rm erg~cm^{-2}$) & $(3.10\pm0.01) \times 10^{-3}$ \\
Peak luminosity ($\rm erg~s^{-1}$) & $4.64_{-0.08}^{+0.09} \times 10^{51}$ \\
Isotropic energy ($\rm erg$) & $(3.18\pm0.01) \times 10^{52}$ \\
\hline
\textbf{Soft X-Ray [0.5--4 keV]:} & \\
Duration ($\rm s$) & $199.6_{-2.2}^{+5.1}$ \\
Spectral index $\alpha$ & $-1.70_{-0.06}^{+0.06}$ \\
Peak flux ($\rm erg~cm^{-2}~s^{-1}$) & $3.6_{-0.5}^{+0.6} \times 10^{-7}$\\
Total fluence ($\rm erg~cm^{-2}$) & $2.24_{-0.06}^{+0.07} \times 10^{-5}$ \\
Peak luminosity ($\rm erg~s^{-1}$) & $3.9_{-0.5}^{+0.6} \times 10^{48}$ \\ 
Isotropic energy ($\rm erg$) & $2.44_{-0.06}^{+0.07} \times 10^{50}$\\
\hline
\textbf{Host Galaxy:} & \\
Redshift & $0.065$ \\
Half-light radius ($\rm kpc$) & $4.0$ \\
Offset ($\rm kpc$) & $36.60$ \\
Normalized offset & $9.2$ \\
Probability of chance coincidence & $0.11$ \\
\hline
\textbf{Associations:} & \\
Kilonova & Yes \\
Supernova & No \\
\hline
\end{tabular}
\begin{tablenotes}
\item [*] The rest-frame spectral lag is measured between rest-frame energy bands 100–150 and 200–250 keV.
\end{tablenotes}
\end{threeparttable}
\end{table*}

\clearpage

\clearpage

\clearpage

\clearpage

\section*{Methods}

\subsection{Multi-mission observations of GRB 230307A \\}

GRB 230307A triggered in real-time the Gravitational wave high-energy Electromagnetic Counterpart All-sky Monitor (GECAM) and the {\it Fermi} Gamma-ray Burst Monitor (GBM) almost simultaneously\footnote{GECAM trigger time is 2023-03-07T15:44:06.650 UTC, which is 21 ms earlier than that of {\it Fermi}/GBM.}. The extreme brightness of GRB 230307A was first reported by GECAM-B with its real-time alert data downlinked by the Beidou satellite navigation system\cite{Xiong2023GCN}, which was subsequently confirmed by {\it Fermi}/GBM\cite{Fermi2023GCN33407, Fermi2023GCN33414} and Konus-Wind\cite{Konus2023GCN}. Preliminary location of this burst measured by GECAM is consistent with that of GBM within error\cite{Xiong2023GCN}. Refined localization in gamma-ray band was provided by the Inter-Planetary Network triangulation\cite{Kozyrev2023GCN} and later improved by follow-up observations of the X-ray Telescope (XRT) aboard the Neil Gehrels Swift Observatory\cite{Evans2023GCN,Burrows2023GCN}. The Lobster Eye Imager for Astronomy (LEIA) detected the prompt emission of this burst in the 0.5--4 keV soft X-ray band\cite{Liu2023GCN}. The burst was also followed up by the Gemini-South\cite{O'Connor2023GCN, Gillanders2023GCN} and the James Webb Space Telescope (JWST) at several epochs\cite{Levan2023GCN33569, Levan2023GCN33580, Levan2023GCN33747}, unveiling a fading afterglow with possible signatures of kilonova in the optical and infrared bands and a candidate host galaxy at a redshift of $z = 0.065$\cite{Levan2023arXiv} (Table \ref{tab:obs_prob}).

\subsection{Data reduction \\}

\noindent\textbf{LEIA.} The Lobster Eye Imager for Astronomy\cite{Zhang2022ApJL, Ling2023arXiv}, a pathfinder of the Einstein Probe mission of the Chinese Academy of Sciences, is a wide-field ($18.6^\circ \times 18.6^\circ$) X-ray focusing telescope built from novel technology of lobster eye micro-pore optics. The instrument operates in the 0.5--4 keV soft X-ray band, with an energy resolution of 130 eV (at 1.25 keV) and a time resolution of 50 ms. LEIA is onboard the Space Advanced Technology demonstration satellite (SATech-01), which was launched on 2022 July 27 and is operating in a Sun-synchronous orbit with an altitude of 500 $\rm km$ and an inclination of $97.4^\circ$. Since LEIA operates only in the Earth's shadow to eliminate the effects of the Sun, the observable time out of the radiation belt at high geo-latitude regions is $\sim$1000 s for each orbit. The observation of GRB 230307A was conducted from 15:42:32 UT (94 s earlier than the GECAM trigger time) to 16:00:50 UT on 7 March 2023 with a net exposure of 761\,s. GRB 230307A was detected within the extended field of view (about $0.6^\circ$ outside the nominal field of view) of LEIA\cite{Liu2023GCN} (Extended Data Fig. \ref{fig:leia_image}). 

The X-ray photon events were processed and calibrated using the data reduction software and calibration database (CALDB) designed for the Einstein Probe mission (Liu et al. in prep.). The CALDB is generated based on the results of on-ground and in-orbit calibration campaigns (Ref.\cite{Cheng2023arXiv} and Cheng et al. in prep.). The energy of each event was corrected using the bias and gain stored in CALDB. Bad/flaring pixels were also flagged. Single-, double-, triple-, and quadruple-events without anomalous flags were selected to form the cleaned event file. The image in the 0.5--4 keV range was extracted from the cleaned events (Extended Data Fig. \ref{fig:leia_image}a). The position of each photon was projected into celestial coordinates (J2000). The light curve and the spectrum of the source and background in a given time interval were extracted from the regions indicated in Extended Data Fig. \ref{fig:leia_image}a. Since the peak count rate is only 2 ct/frame over the extended source region, the pile-up effect is negligible in the LEIA data.

LEIA is the very first of its kind with a considerably large field of view ever flown. To fully characterize its instrumental performance before launch, a series of tests and calibrations have been performed for the mirror assembly and the detectors separately, and for the complete module of the LEIA instrument\cite{Zhang2022ApJL, Ling2023arXiv, 10.1117/12.2690354, Cheng2023arXiv}. The in-flight calibration, which took over 8 months, was performed after launch and aimed to achieve accurate source localization and flux derivation (Cheng et al. in prep). The systematic uncertainty of source positioning is $1-2$ arcmin and that of the effective area (of the central focal spot) is about 10 percent, both at the 90 percent confidence level. However, as LEIA detected the cruciform arms of the source image (point spread function, PSF) of GRB 230307A instead of its focal spot, in order to verify the effective area (of the photon extraction region on the arm, i.e. the red-dotted box in Extended Data Fig. \ref{fig:leia_image}a) employed in the spectral fitting, which is derived from simulations, we performed observations of the Crab nebula (the standard calibration source) for 5 orbits on March 14, 2023 with a total exposure time of 4398 seconds. The Crab nebula was placed at the same direction of GRB 230307A relative to the same detector in these observations, and the same analysis procedure was conducted to the data. The fitted photon index and estimated flux of the Crab are consistent with those derived from previous in-flight calibrations (also using the Crab nebular) at a level of less than 10 percent. This value is consistent with the aforementioned systematic error.

\noindent\textbf{GECAM.} GECAM is a dedicated all-sky gamma-ray monitor constellation funded by the Chinese Academy of Sciences. The original GECAM mission\cite{Li2021RDTM} is composed of two microsatellites (GECAM-A and GECAM-B) launched in December 2020. GECAM-C\cite{ZhangDL2023arXiv} is the 3rd GECAM spacecraft, also onboard the SATech-01 satellite as LEIA. Each GECAM spacecraft has an all-sky field of view unblocked by the Earth, capable of triggering bursts in real-time\cite{Zhao2021arxiv} and distributing trigger alerts promptly with the Global Short Message Communication of Beidou satellite navigation system\cite{BDS2023IEEE}. As the main instrument of GECAM, most gamma-ray detectors (GRDs) operate in two readout channels: high gain (HG) and low gain (LG), which are independent in terms of data processing, transmission, and dead-time\cite{An2021RDTM}. Comprehensive ground and cross calibrations have been conducted on the GRDs of both GECAM-B and GECAM-C\cite{Zheng2022NIMA, ZhangDL2023arXiv, Zheng2023arXiv, ZhangYQ2023arXiv}. Both GECAM-B and GECAM-C have charged particle detectors (CPDs) to monitor the space environment, which is designed to be sensitive to charged particles and much less sensitive to photons\cite{Xu2022RDTM}. However, for some extremely bright GRBs, CPDs can record some photons.

GRB 230307A was detected by GECAM-B and GECAM-C, while GECAM-A was offline. GECAM-B was triggered by this burst and automatically distributed a trigger alert to General Coordinates Network (GCN) Notice about 1 minute post-trigger\footnote{GECAM real-time alert for GRB 230307A: \url{https://gcn.gsfc.nasa.gov/other/160.gecam}}. GECAM-C also made the real-time trigger onboard, while the trigger alert was disabled due to the high latitude region setting\cite{ZhangDL2023arXiv}. With the automatic pipeline processing the GECAM-B real-time alert data\cite{Huang2023arxiv}, we promptly noticed and reported that this burst features an extreme brightness\cite{Xiong2023GCN}, which initiated the follow-up observations.

Both GECAM-B and GECAM-C were working in inertial pointing mode during the course of GRB 230307A. Among all 25 GRDs of GECAM-B, GRD04 maintains a constant minimum 
incident angle of $12.7^\circ$ throughout the duration of the burst. GRD01, GRD05 also exhibit small incident angles of $32.4^\circ$, $32.7^\circ$, respectively. Thus these three detectors are selected for subsequent analysis. For GECAM-B GRDs, the HG channel operates from $\sim$ 40 to 350 keV, while the LG channel from $\sim$ 700 keV to 6 MeV. Among all 12 GRDs of GECAM-C, GRD01 exhibits the most optimal incident angle of $10.1^\circ$ throughout the burst and is selected in the subsequent analysis. For GECAM-C/GRD01, the HG channel operates from $\sim$ 6 to 350 keV. Since the detector response for 6--15 keV is affected by the electronics and is subject to further verification\cite{ZhangYQ2023arXiv}, we only use $>$15 keV for spectral analysis in this work. 

During the detection of GRB 230307A, the background of GECAM-B was found to be very stable and could be well fitted by a first-order polynomial. However, GECAM-C experienced background evolution due to the motion of the satellite in orbit, which can also be seen from the light curve of CPDs. The background estimation methodology for GECAM-C/GRD01 involves fitting an empirical function with a combination of first and second-order exponential polynomials (i.e., exp($c_0+c_1t$)+exp($c_2+c_3t+c_4t^2$)) (Extended Data Fig. \ref{fig:gecam_reduction}a). Though the CPD is much less sensitive to photons, a clear profile of GRB 230307A can be seen on the residual of CPD background fitting owing to the extreme photon flux during the peak region of this GRB. The reasonable residuals at other time intervals rather than the peak time period of GRB 230307A indicate that the background of GECAM-C is well modelled and subtracted. More directly, this background estimation of GECAM-C is verified by the comparison with GECAM-B data (Extended Data Fig. \ref{fig:gecam_reduction}b).

We note that GRB 230307A was so bright that the {\it Fermi}/GBM observation suffered from data saturation\cite{Fermi2023GCN33551}. GECAM has dedicated designs to minimize data saturation for bright bursts\cite{Li2021RDTM, liu2021arxiv}. For GECAM, the engineering count rate records the number of events processed onboard, while the event count rate records the number of events received on ground. In the case of data saturation, these two count rates would differ significantly. As shown in Extended Data Fig. \ref{fig:gecam_reduction}c, the negligible discrepancy between these two count rates, due to the limited digital accuracy on the count numbering, confirms no count loss in event data and indicates the absence of saturation for both GECAM-B and GECAM-C.

\subsection{Temporal analysis \\}

\noindent\textbf{Duration.} The light curves are obtained through the process of photon counts binning (Fig. \ref{fig:lc}). The light curve for LEIA is obtained in 0.5--4 $\rm keV$ with a bin size of 1 s. The light curves for GECAM-B are generated using the bin size of 0.4 s in the energy ranges of 100--350 $\rm keV$, 350--700 $\rm keV$, and 700--2000 $\rm keV$, by combining the data from three selected detectors (namely, GRD04, GRD01, and GRD05). The GECAM-C light curves are derived by binning the photon counts from GRD01 with a bin size of 0.4 s in the energy ranges of 6--15 $\rm keV$, 15--30 $\rm keV$, and 30--100 $\rm keV$. The burst duration, denoted as $T_{\rm 90}$, is determined by calculating the time interval between the epochs when the total accumulated net photon counts reach the $5\%$ and $95\%$ levels. The durations obtained from the multi-wavelength light curve are annotated in Fig. \ref{fig:lc}. It is found that the duration of the burst significantly increases towards the lower energy range. We also calculate the duration within 10--1000 $\rm keV$ based on the data from GECAM-B and list it in Table \ref{tab:obs_prob}.

\noindent\textbf{Amplitude parameter.} The amplitude parameter\cite{Lu2014MNRAS} is a metric used to classify GRBs and is defined as $f=F_{\rm p}/F_{\rm b}$, denoting the ratio between the peak flux $F_{\rm p}$ and the background flux $F_{\rm b}$ at the same time epoch. A long-duration type II GRB may be disguised as a short-duration type I GRB due to the tip-of-iceberg effect. To distinguish intrinsically short-duration type I GRBs from ostensible short-duration type II GRBs, an effective amplitude parameter, $f_{\rm eff}=\epsilon f=F_{\rm p}^{\prime}/F_{\rm b}$, can be defined for long-duration type II GRBs by quantifying the tip-of-iceberg effect, where $F_{\rm p}^{\prime}$ is the peak flux of a pseudo GRB whose amplitude is lower by a factor $\epsilon$ from an original long-duration type II GRB so that its duration is just shorter than 2 s. Generally speaking, the $f_{\rm eff}$ values of long-duration type II GRBs are systematically smaller than the $f$ values of short-duration type I GRBs, thereby facilitating their distinction from one another. Utilizing the procedure presented in Ref.\cite{Lu2014MNRAS}, the effective amplitude of GRB 230307A is determined to be $f_{\rm eff}=1.23\pm0.07$ within the energy range of 10--1000 keV (Table \ref{tab:obs_prob}). Such small $f_{\rm eff}$ value aligns with the characteristics typically exhibited by long-duration GRBs.

\noindent\textbf{Variability.} The minimum variability timescale (MVT) is defined as the shortest timescale of significant variation that exceeds statistical noise in the GRB temporal profile\cite{Golkhou2015ApJ, Camisasca2023A&A}. It serves as an indicator of both the central engine's activity characteristics\cite{MacLachlan2013MNRAS} and the geometric dimensions of the emitting region. The median values of the minimum variability timescale in the rest frame (i.e., MVT$/(1+z)$) for type I and type II GRBs are found to be 10 ms and 45 ms, respectively. To determine the MVT, we utilize the Bayesian block algorithm\cite{Scargle2013ApJ} on the entire light curve within the 10--1000 keV energy range to identify the shortest block that satisfies the criterion of encompassing the rising phase of a pulse. We find that the MVT of GRB 230307A is about 9.35 ms (Table \ref{tab:obs_prob}), which is more consistent with type I GRBs rather than type II GRBs in the distribution of the MVTs\cite{Golkhou2015ApJ} (Extended Data Fig. \ref{fig:classification}a). We also utilize the continuous wavelet transform method to derive MVT and obtain a consistent outcome in accordance with the Bayesian block algorithm.

\noindent\textbf{Spectral lag.} Spectral lag refers to the time delay between the soft-band and hard-band background-subtracted light curves. It may be attributed to the curvature effect in the relativistic outflow. Upon reaching the observer, on-axis photons are boosted to higher energies, while off-axis photons receive a smaller boost and must travel a longer distance. Type II GRBs usually exhibit considerable spectral lags, while type I GRBs tend to have tiny lags\cite{Yi2006MNRAS, Bernardini2015MNRAS}, indicating the difference in their emission region sizes. Measurement of the spectral lag can be achieved by determining the time delay corresponding to the maximum value of the cross-correlation function\cite{Norris2000ApJ, Ukwatta2012MNRAS, Zhang2012ApJ}. Following the treatment in Ref.\cite{Ukwatta2012MNRAS}, we use the background-subtracted light curves of GRB 230307A to measure the rest-frame lag between rest-frame energy bands 100--150 and 200--250 keV to be $1.6_{-1.2}^{+1.4}$ ms (Table \ref{tab:obs_prob}). We note that GRB 230307A manifests a very tiny spectral lag, an indicator that points towards being a type I GRB (Extended Data Fig. \ref{fig:classification}c).

\subsection{Spectral analysis\\}

\noindent\textbf{LEIA spectral fitting.} We perform a detailed spectral analysis using the LEIA data. The energy channels in the range of 0.5--4 keV are utilized and re-binned to ensure that each energy bin contains at least ten counts. We employ three kinds of time segmentation approaches to extract LEIA spectra:

\begin{itemize}
 \item LEIA S-I: the time interval from 0 to 250 s is divided into three slices (0--50 s, 50--150 s, 150--250 s) to investigate the possible evolution of X-ray absorption;
 \item LEIA S-II: 23 time slices with sufficient time resolution are obtained by accumulating 100 photon counts for each individual slice;
 \item LEIA S-III: 12 time slices with sufficient level of confidence are obtained by accumulating 200 photon counts for each individual slice. This partitioning is designed to align with the temporal divisions of the GECAM S-III spectra, enabling subsequent comparisons and joint spectral fitting.
\end{itemize}

For each of the above time slices, we generate the source spectrum and background spectrum, and the corresponding detector redistribution matrix and the ancillary response. 

First, the spectra of LEIA S-I are individually fitted with the \textit{XSPEC}\cite{Arnaud1996ASPC} using the {\it tbabs*ztbabs*powerlaw} model, where the first and second components are responsible for the Galactic absorption and intrinsic absorption ($N_{\rm H}$), and the third one is a power law function in the observer's frame. We incorporate the absorption model\cite{Wilms2000ApJ} for the cross sections and abundances. We employ CSTAT\cite{Cash1979ApJ} as the statistical metric to evaluate the likelihood of LEIA spectral fitting where both the source and background spectra are Poisson-distributed data. The column density of the Galactic absorption in the direction of the burst is fixed at $1.26 \times 10^{21}~{\rm cm^{-2}}$\cite{Willingale2013} and the redshift is fixed at 0.065. It is found that all three spectra of S-I can be reproduced reasonably well by the absorption modified power law model (Extended Data Fig. \ref{fig:leia_image}b). The best-fit values and confidence contours of the column density and photon index are shown in Extended Data Fig. \ref{fig:leia_image}c. The fitted column densities are generally consistent within their uncertainties, indicating no significant variations of the absorption feature within the observation of LEIA. We perform a simultaneous fitting of all S-III spectra to make full use of all spectra with sufficient level of confidence. A time-averaged absorption of $N_{\rm H}=3.92 \times 10^{21}~{\rm cm^{-2}}$, is obtained and thus fixed in all subsequent spectral analysis. This value is larger than that derived from the X-ray afterglow\cite{Mereghetti2023ApJ} at $\sim$ 4.5 days. The early soft X-ray detection of this bright burst provides an opportunity to observe the absorption variation. Such variation in the absorption density has only been reported for a few GRBs\cite{campana2021}. Given that the burst is located at large galactocentric distances, the decrease in $N_{\rm H}$ over time may be associated with the dynamically evolving circumburst medium during the early afterglow phase.

We then proceed with the fitting of the LEIA S-II spectra. The obtained results for the photon index ($\Gamma_{\rm ph}$), normalization, and the corresponding fitting statistic are presented in Extended Data Table \ref{tab:leia_spec_fit}. By employing a redshift of $z = 0.065$, we further calculated the unabsorbed flux (Fig. \ref{fig:sed}a) and determined the luminosity for each S-II spectrum (Fig. \ref{fig:magnetar}a and \ref{fig:magnetar}b). Additionally, we cross-validated our findings by analyzing spectra with higher photon statistics (S-III), and found that the alternate spectra yielded consistent results in our analysis. We also perform a spectral fit to the time-averaged spectrum during the whole observation interval of S-II, and the derived power-law index ($\alpha = -\Gamma_{\rm ph}$) is shown in Table \ref{tab:obs_prob}. Finally, the LEIA S-III spectra are employed for spectral energy distribution analysis. 

\noindent\textbf{GECAM spectral fitting}. We conduct a thorough time-resolved and time-integrated spectral analysis using the data from the GRD04, GRD01 and GRD05 of GECAM-B, as well as the GRD01 of GECAM-C. Each GRD detector has two independent readout channels, namely, high gain (HG) and low gain (LG). We utilize both the high and low gain data of the GECAM-B detectors with effective energy ranges of 40--350 and 700--6000 keV, respectively. Additionally, we only use the high gain data of the GECAM-C detector with an effective energy range of 15--100 keV but ignored the channels within 35--42 keV around the Iodine K-edge at 38.9 keV. We employ three distinct time segmentation methods to extract GECAM spectra: 

\begin{itemize}
 \item GECAM S-I: the time interval of 1.6--43.2 s, in which 5\% and 95\% of the total net photons accumulate, is utilized as a single time slice for conducting time-integrated spectral analysis;
 \item GECAM S-II: the time interval of 0--147 s is divided into 99 time slices with sufficient photon counts and approximately equal signal-to-noise levels to study the spectral evolution;
 \item GECAM S-III: the time interval of 0--147 s is partitioned to match the time division of LEIA S-III for studying the spectral energy distributions. However, to ensure that there is at least five net photons per detector channel on average, we merge two bins, namely 91--113 s and 113--147 s, into a single bin. Consequently, the time interval of 0--147 s is divided into ten time slices.
\end{itemize}

For each of the above time slices, we generate a source spectrum, a background spectrum, and a response matrix for each gain mode of each detector. Then we perform spectral fitting by utilizing the Python package, {\it MySpecFit}, in accordance with the methodology outlined in Refs.\cite{Yang2022Natur, Yang2023ApJL}. The {\it MySpecFit} package facilitates Bayesian parameter estimation by wrapping the widely-used Fortran nested sampling implementation {\it Multinest}\cite{Feroz2008MNRAS, Feroz2009MNRAS, Buchner2014A&A, Feroz2019OJAp}. PGSTAT\cite{Arnaud1996ASPC} is utilized for GECAM spectral fitting, which is appropriate for Poisson data in the source spectrum with Gaussian background in the background spectrum. To fit the observed spectra, seven empirical models ranging from simple to complex are adopted. Here, we establish the following conventions for the model parameters: $N(E)$ is the photon spectrum, $A$ is the normalization parameter in units of ${\rm photons~cm^{-2}~s^{-1}~keV^{-1}}$, $E_{\rm p}$ is defined as the peak energy of $\nu f_{\nu}$ spectrum, $\beta$ is the high-energy photon index above the peak energy, $E_{\rm b}$ and $E_{\rm c}$ are defined as the low-energy break and high-energy cutoff of photon spectrum respectively, $\alpha_1$ and $\alpha_2$ represent the spectral indices of the two low-energy power law segments smoothly connected at the break energy, and $\alpha$ represents the spectral index at low energies if the break energy does not exist. Then all the spectral models are listed as follows:
\begin{itemize}
\item The power law (PL):
\begin{equation}
N(E)=A\Big(\frac{E}{100\,{\rm keV}}\Big)^{\alpha}
\end{equation}
\item The cutoff power law (CPL):
\begin{equation}
N(E)=A\Big(\frac{E}{100\,{\rm keV}}\Big)^{\alpha}{\rm exp}\Big(-\frac{E}{E_{\rm c}}\Big),
\end{equation}
where the peak energy $E_{\rm p}$ is related to the cutoff energy $E_{\rm c}$ through $E_{\rm p}=(2+\alpha)E_{\rm c}$.
\item The Band function (BAND\cite{Band1993ApJ}):
\begin{equation}
N(E)=\left\{
\begin{array}{l}
A(\frac{E}{100\,{\rm keV}})^{\alpha}{\rm exp}(-\frac{E}{E_{\rm c}}),\,E<(\alpha-\beta)E_{\rm c}, \\
A\big[\frac{(\alpha-\beta)E_{\rm c}}{100\,{\rm keV}}\big]^{\alpha-\beta}{\rm exp}(\beta-\alpha)(\frac{E}{100\,{\rm keV}})^{\beta}, E\geq(\alpha-\beta)E_{\rm c}, \\
\end{array}\right.
\end{equation}
where the peak energy $E_{\rm p}$ is related to the cutoff energy $E_{\rm c}$ through $E_{\rm p}=(2+\alpha)E_{\rm c}$.
\item The smoothly broken power law (SBPL\cite{Ravasio2018AA}):
\begin{equation}
N(E)=AE_{\rm j}^{\alpha}\Big[\Big(\frac{E}{E_{\rm j}}\Big)^{-\alpha n}+\Big(\frac{E}{E_{\rm j}}\Big)^{-\beta n}\big]^{-\frac{1}{n}},
\end{equation}
where $E_{\rm j}=E_{\rm p}(-\frac{\alpha+2}{\beta+2})^{\frac{1}{(\beta-\alpha)n}}$, and the smoothness parameter is fixed to $n=2.69$.
\item The Band function with high-energy exponential cutoff (CBAND\cite{Zheng2012ApJ}):
\begin{equation}
N(E)=\left\{
\begin{array}{l}
AE^{\alpha_1}{\rm exp}(-\frac{E}{E_1}), ~E \leq E_{\rm b}, \\
AE_{\rm b}^{\alpha_1-\alpha_2}{\rm exp}(\alpha_2-\alpha_1)E^{\alpha_2}{\rm exp}(-\frac{E}{E_2}), ~E > E_{\rm b}, \\
\end{array}\right.
\end{equation}
where the break energy is expressed as $E_{\rm b}=\frac{E_1E_2}{E_2-E_1}(\alpha_1-\alpha_2)$, and the peak energy is defined as $E_{\rm p}=(2+\alpha_2)E_2$.
\item The smoothly broken power law with high-energy exponential cutoff (CSBPL):
\begin{equation}
N(E)=AE_{\rm b}^{\alpha_1}\Big[\Big(\frac{E}{E_{\rm b}}\Big)^{-\alpha_1 n}+\Big(\frac{E}{E_{\rm b}}\Big)^{-\alpha_2 n}\Big]^{-\frac{1}{n}}{\rm exp}\Big(-\frac{E}{E_{\rm c}}\Big),
\end{equation}
where the peak energy $E_{\rm p}$ is related to the cutoff energy $E_{\rm c}$ through $E_{\rm p}=(2+\alpha_2)E_{\rm c}$, and the smoothness parameter is fixed to $n=5.38$.
\item The double smoothly broken power law (2SBPL\cite{Ravasio2018AA}):
\begin{equation}
N(E)=AE_{\rm b}^{\alpha_1}\Big[\Big[\Big(\frac{E}{E_{\rm b}}\Big)^{-\alpha_1 n_1}+\Big(\frac{E}{E_{\rm b}}\Big)^{-\alpha_2 n_1}\Big]^{\frac{n_2}{n_1}}+\Big(\frac{E}{E_{\rm j}}\Big)^{-\beta n_2}\Big[\Big(\frac{E_{\rm j}}{E_{\rm b}}\Big)^{-\alpha_1 n_1}+\Big(\frac{E_{\rm j}}{E_{\rm b}}\Big)^{-\alpha_2 n_1}\Big]^{\frac{n_2}{n_1}}\Big]^{-\frac{1}{n_2}},
\end{equation}
where $E_{\rm j}=E_{\rm p}(-\frac{\alpha_2+2}{\beta+2})^{\frac{1}{(\beta-\alpha_2)n_2}}$, and the smoothness parameters are fixed to $n_1=5.38$ and $n_2=2.69$.
\end{itemize}

We then fit them to the spectral data in each time slice. We perform a model comparison based on the Bayesian Information Criterion\cite{Schwarz1978AnSta} (BIC) with the following procedure: 
\begin{enumerate}
    \item A BIC value is calculated for each of the above models;
    \item The minimal value, BIC$_{\text{min}}$, of the above seven BIC values is obtained;
    \item Within the range of [BIC$_{\text{min}}$, BIC$_{\text{min}}$ + 2], we check how many BIC values fall in that range.
    \item If only one model's BIC falls in the range [BIC$_{\text{min}}$, BIC$_{\text{min}}$ + 2], we select such a model as the best model.
    \item If two or more models' BIC values fall in the range [BIC$_{\text{min}}$, BIC$_{\text{min}}$ + 2], we consider those models can equally explain the data\cite{jeffreys1998, kass1995, Mukherjee1998ApJ}. In this case, if those models include the CPL model, we simply select it as the best-fit model; otherwise, we select the one corresponding to BIC$_{\text{min}}$. 
\end{enumerate}
Priority is given to CPL in step 5, as it allows us to study the spectral properties in the most consistent manner possible. Additionally, this consideration takes into account the fact that most GRBs' fine time-resolved spectra can indeed be described by a CPL model.

Extended Data Table \ref{tab:gecam_spec_fit} and \ref{tab:sed} list the spectral fitting results of the best model and corresponding fitting statistics for GECAM spectra. It should be noted that we use cutoff energy as a substitute for peak energy when the 1$\sigma$ lower limit of $\alpha_2$ (or $\alpha$) nears or falls below $-2$. The GECAM S-I spectrum can be best fitted by the 2SBPL model with spectral indices $\alpha_1 = -0.92_{-0.03}^{+0.05}$, $\alpha_2 = -1.274_{-0.008}^{+0.005}$, and $\beta = -3.85_{-0.09}^{+0.03}$, as well as the break energy $E_{\rm b} = 24_{-2}^{+3}$ keV and peak energy $E_{\rm p} = 1052_{-8}^{+16}$ keV. Fig. \ref{fig:lc}b and \ref{fig:lc}c illustrate the significant ``intensity tracking'' spectral evolution in terms of $E_{\rm p}$ and $\alpha_2$ (or $\alpha$) derived from GECAM S-II spectra, respectively. During the time interval before approximately 30 s, we notice that many time-resolved GECAM spectra necessitated the use of a model incorporating a low-energy break (i.e., CBAND, CSBPL and 2SBPL) for fitting. The break energy $E_{\rm b}$ exhibits a decreasing trend from approximately 200 to 30 keV, with the spectral indices (i.e., $\alpha_1$ and $\alpha_2$) on either side of the break energy distributed in the 1$\sigma$ interval of [-0.9, -0.4] and [-1.6, -1.2], respectively. It should be noted that the fitting statistics are acceptable for all time-resolved GECAM S-II spectra, whereas for the time-integrated GECAM S-I spectrum and some GECAM S-III spectra, the relatively large statistics should not be over-interpreted but can be attributed to spectral evolution\cite{Zhang2018NatAs}.

\subsection{Classification \\}
\noindent\textbf{Variability timescale versus duration}. The temporal variability timescale of GRBs may conceal imprints of central engine activity and energy dissipation processes. On average, the minimum variability timescale of short-duration type I GRBs is significantly shorter than that of long-duration type II GRBs, providing a new clue to distinguish the nature of GRB progenitors and central engines\cite{Golkhou2015ApJ, Camisasca2023A&A}. We collected previous research samples, redrawn the ${\rm MVT}-T_{\rm 90}$ diagram, and overlaid GRB 230307A and GRB 211211A on it (Extended Data Fig. \ref{fig:classification}a). It is noteworthy that these two GRBs are outliers as they fall outside the 3$\sigma$ confidence contours of both the type I and type II populations. While in terms of the MVT distribution, we find that their MVTs are more consistent with that of type I GRBs.

\noindent\textbf{Peak energy versus isotropic energy}. The $E_{\rm p,z}-E_{\rm\gamma,iso}$ diagram serves as a unique classification scheme in the study of GRB energy characteristics, as different physical origins of GRBs typically follow distinct tracks\cite{Amati2002A&A, Zhang2009ApJ, Minaev2020MNRAS}. We first replot the $E_{\rm p,z}-E_{\rm\gamma,iso}$ diagram (Extended Data Fig. \ref{fig:classification}b) based on previous samples of type I and type II GRBs with known redshifts\cite{Minaev2020MNRAS}. Here, $E_{\rm p,z}=E_{\rm p}(1 + z)$ represents the rest-frame peak energy, while $E_{\rm\gamma,iso}$ denotes the isotropic energy. The relations between $E_{\rm p,z}$ and $E_{\rm\gamma,iso}$ can be modeled using a linear relationship, ${\rm log}E_{\rm p,z}=b + k{\rm log}E_{\rm\gamma,iso}$, for both GRB samples. The fitting process is implemented using the Python module \emph{emcee}\cite{Foreman2013PASP}, and the likelihood is determined using the orthogonal-distance-regression (ODR) method\cite{Lelli2019MNRAS}. The ODR method is appropriate in this case as the data satisfy two criteria: (1) a Gaussian intrinsic scatter $\sigma_{\rm int}$ along the perpendicular direction; (2) independent errors $\sigma_{x_i}$ and $\sigma_{y_i}$ on both the $x$ and $y$ axes. The log-likelihood function can be expressed as
\begin{equation}
 {\rm ln}\mathcal{L}=-\frac{1}{2}\sum_i \Big[{\rm ln}(2\pi\sigma_i^2)+\frac{\Delta_i^2}{\sigma_i^2}\Big],
\end{equation}
with the perpendicular distance
\begin{equation}
 \Delta_i^2=\frac{(y_i-kx_i-b)^2}{k^2+1},
\end{equation}
and the total perpendicular uncertainties
\begin{equation}
 \sigma_i^2=\frac{k^2\sigma_{x_i}^2+\sigma_{y_i}^2}{k^2+1} + \sigma_{\rm int}^2,
\end{equation}
where the subscript $i$ runs over all data points. The best-fitting parameters with 1$\sigma$ uncertainties are $k_{\rm I}=0.36_{-0.05}^{+0.04}$, $b_{\rm I}=-15.61_{-2.14}^{+2.51}$ and ${\rm log}\sigma_{\rm int,I}=-1.29_{-0.13}^{+0.13}$ for type I GRBs, and $k_{\rm II}=0.39_{-0.02}^{+0.02}$, $b_{\rm II}=-17.82_{-0.98}^{+0.93}$ and ${\rm log}\sigma_{\rm int,II}=-1.42_{-0.05}^{+0.05}$ for type II GRBs. The best-fit correlations and corresponding 1$\sigma$ intrinsic scattering regions are presented in Extended Data Fig. \ref{fig:classification}b. The peak energy of GRB 230307A is constrained to be $E_{\rm p}=1052_{-8}^{+16}$ keV by the spectral fitting to GECAM S-I. Given a redshift of 0.065, the isotropic energy of GRB 230307A can be calculated as $E_{\rm\gamma,iso}=(3.18\pm0.01)\times10^{52}$ erg based on the spectral fits to GECAM S-II. We overplot both long-duration GRB 230307A and GRB 211211A on the $E_{\rm p,z}-E_{\rm\gamma,iso}$ diagram. As can be seen from Extended Data Fig. \ref{fig:classification}b, GRB 211211A resides in an intermediate area between the tracks of type I and type II GRBs. The perpendicular distance, $\Delta$, is approximately $1.5\sigma_{\rm int,I}$ away from the type I track and $1.6\sigma_{\rm int,II}$ away from the type II track. On the other hand, GRB 230307A is firmly located within the 1$\sigma$ intrinsic scattering region of the type I GRB track. Specifically, the perpendicular distance is approximately $0.8\sigma_{\rm int,I}$ away from the type I track and $2.3\sigma_{\rm int,II}$ away from the type II track. In addition, GRB 230307A poses higher total energy and harder spectra, likely indicating a more intense merger event. We also consider the scenario where the host galaxy has a redshift of 3.87\cite{Levan2023GCN33747} and notice that in this case, the isotropic energy of GRB 230307A is $\sim 10^{56}$ erg, which is an order of magnitude larger than that of the brightest-of-all-time GRB 221009A\cite{Yang2023ApJL, An2023arXiv}. Such extremely high energy is very unlikely consistent with the known sample of GRBs.

\noindent\textbf{Peak luminosity versus spectral lag}. An anti-correlation exists between the spectral lag and peak luminosity in the sample of type II GRB with a positive spectral lag\cite{Norris2000ApJ, Ukwatta2012MNRAS}. Such anti-correlation can serve as a physically ambiguous indicator, suggesting that GRBs with short spectral lags have higher peak luminosities. In general, type I GRBs deviate from the anti-correlation of type II GRBs, with type I GRBs tending to exhibit smaller spectral lags than type II GRBs at the same peak luminosity. The significant differences between the two in this regard make the peak luminosity versus spectral lag correlation a useful classification scheme. On the basis of the previous samples of type I and type II GRBs with known redshift\cite{Ukwatta2010ApJ, Ukwatta2012MNRAS, Xiao2022ApJ}, we replot the $L_{\rm\gamma,iso}-\tau_{\rm z}$ diagram, where $L_{\rm\gamma,iso}$ is isotropic peak luminosity and $\tau_{\rm z}=\tau/(1+z)$ is the rest-frame spectral lag (Extended Data Fig. \ref{fig:classification}c). The anti-correlation between $L_{\rm\gamma,iso}$ and $\tau_{\rm z}$ for type II GRBs can be fit with the linear model ${\rm log}L_{\rm\gamma,iso}=b + k{\rm log}\tau_{\rm z}$. The ODR method\cite{Lelli2019MNRAS} gives the best-fitting parameters with 1$\sigma$ uncertainties to be $k=-0.94_{-0.26}^{+0.07}$, $b=54.24_{-0.14}^{+0.49}$ and ${\rm log}\sigma_{\rm int}=-1.61_{-0.17}^{+0.29}$. The best-fit correlation and corresponding 3$\sigma$ intrinsic scattering region are presented in Extended Data Fig. \ref{fig:classification}c. To maintain consistency with the calculation method of the sample, we estimate the rest-frame lag of GRB 230307A between rest-frame energy bands 100--150 and 200--250 keV to be $1.6_{-1.2}^{+1.4}$ ms. Assuming the redshift to be 0.065, the isotropic peak luminosity of GRB 230307A can be calculated as $L_{\rm\gamma,iso}=4.64_{-0.08}^{+0.09} \times 10^{51}$ $\rm erg~s^{-1}$ based on the spectral fits to GECAM S-II. Then we place both GRB 230307A and GRB 211211A on the $L_{\rm\gamma,iso}-\tau_{\rm z}$ diagram. As illustrated in the Extended Data Fig. \ref{fig:classification}c, GRB 230307A and GRB 211211A exhibit significant deviations from the anti-correlation of type II GRBs. Specifically, their perpendicular distances $\Delta$ from the anti-correlation are approximately $8.7\sigma_{\rm int}$ and $7.4\sigma_{\rm int}$, respectively. Notably, despite being long-duration GRBs, they are intermingled with other type I GRBs. We also note that, if the redshift is 3.87\cite{Levan2023GCN33747}, the extremely high peak luminosity of GRB 230307A would make it a significant outlier in the known sample of GRBs.

\subsection{Fit of multi-wavelength flux light curves \\}

On the basis of detailed spectral fitting to the time-resolved spectra of LEIA S-II and GECAM S-II, we calculate the energy flux for each time slice and construct the multi-wavelength flux light curves for six energy bands. These six energy bands, including 0.5--4, 15--30, 30--100, 100--350, 350--700, and 700--2000 keV, are set by referencing the effective energy ranges of the LEIA, GECAM-B, and GECAM-C. It is noteworthy that, during the last four time slices of GECAM S-II, only the high gain data from GECAM-C and GECAM-B exhibit effective spectra that are significantly above the background level, while the low gain data from GECAM-B are already close to the background level and do not provide effective spectral information, which leads to insufficient confidence in determining flux values in the energy bands above 350 keV. Therefore, based on the Bayesian posterior probability distribution generated by {\it Multinest}\cite{Feroz2008MNRAS, Feroz2009MNRAS, Buchner2014A&A, Feroz2019OJAp}, we provide the 3$\sigma$ upper limits of flux for 350--700 and 700--2000 keV in the last four time intervals. The multi-wavelength flux light curves are represented in Fig. \ref{fig:sed}a. All these flux light curves display multi-segment broken power law features. To fit these features, we introduce multi-segment SBPL functions.

In general, smoothly connected functions in a logarithmic-logarithmic scale can be expressed as
\begin{equation}
 F=(F_{\rm l}^{-\omega} + F_{\rm r}^{-\omega})^{-1/\omega},
 \label{eqs:smooth_connect}
\end{equation}
where $F_{\rm l}$ and $F_{\rm r}$ are the functions located on the left and right sides respectively, and $\omega$ describes the smoothness. When both of $F_{\rm l}$ and $F_{\rm r}$ are power law functions, a two-segment SBPL function can be obtained as
\begin{equation}
 F_{12}=(F_{1}^{-\omega_{1}} + F_{2}^{-\omega_1})^{-1/\omega_1},
 \label{eqs:sbpl}
\end{equation}
where $F_{1}=A(t/t_{\rm b1})^{-\alpha_1}$, $F_{2}=A(t/t_{\rm b1})^{-\alpha_2}$. The power law slopes before and after the break time $t_{\rm b1}$ are $\alpha_1$ and $\alpha_2$, respectively, and $A$ is the normalization coefficient at $t_{\rm b1}$. In the case where another break occurs after two-segment SBPL function, a three-segment SBPL function can be expressed as
\begin{equation}
 F_{123}=(F_{12}^{-\omega_2} + F_{3}^{-\omega_2})^{-1/\omega_2},
\end{equation}
where $F_{3}=F_{12}(t_{\rm b2})(t/t_{\rm b2})^{-\alpha_3}$ describes the third power law function. By extension, we can further expand the three-segment SBPL to include a third break and a forth power law function, namely the four-segment SBPL function:
\begin{equation}
 F_{1234}=(F_{123}^{-\omega_3} + F_{4}^{-\omega_3})^{-1/\omega_3},
\end{equation}
where $F_{4}=F_{123}(t_{\rm b3})(t/t_{\rm b3})^{-\alpha_4}$ describes the forth power law function.

The fitting process is implemented by the Python module {\it PyMultinest}\cite{Buchner2014A&A}, a Python interface to the widely-used Fortran nested sampling implementation {\it Multinest}\cite{Feroz2008MNRAS, Feroz2009MNRAS, Buchner2014A&A, Feroz2019OJAp}. We employ $\chi^2$ as the statistical metric to evaluate the likelihood. It should be noted that a dip covering 17--20 s exists in the flux light curves of all energy bands of GECAM. Since the dip is an additional component superimposed on the multi-segment broken power law, the time interval containing the dip is omitted from the fitting procedure and will be examined in a separate analysis, which will be discussed in detail elsewhere\cite{Yi2023arXiv}. For convenience, we refer to the flux light curves of the six energy bands as LF (0.5--4 keV) and GFi (where i ranges from 1 to 5, representing the five energy bands of GECAM from low to high). We note that LF consists of a shallow power law decay followed by a steeper decline, which is significantly different from GFs, where they include an initial power law rise and several gradually steepening power law decay phases. GF1-3 require the four-segment SBPL functions to describe their features, while GF4 and GF5 can be described using three-segment SBPL functions as the late-time features cannot be constrained due to the last four data points being 3$\sigma$ upper limits. Interestingly, in the GECAM energy bands, the first three power law segments and the corresponding two break times exhibit clear spectral evolution features. From low to high energy, the power law decay index gradually increases. Additionally, there appears to be a tendency for the break times to shift towards earlier intervals.
However, we note that the final breaks in GF1-3 appear to be a simultaneous feature. Such an achromatic break is typically attributed to the geometric effects of the emission region.

To test the simultaneity of the final break and determine its break time, we performed a joint fit to GF1-3. The fitting process is described as follows. We simultaneously use three four-segment SBPL functions to fit GF1, GF2, and GF3 independently but allow the three $t_{\rm b3}$ parameters to degenerate into a common parameter. The statistic of the joint fit is the sum of $\chi^2$ values for GF1-3. The fitting process is also implemented by the Python module {\it PyMultinest}\cite{Buchner2014A&A}. The best-fitting parameter values and their 1$\sigma$ uncertainties are presented in Extended Data Table \ref{tab:lc_sbpl}. Extended Data Fig. \ref{fig:lc_sbpl}a exhibits the corresponding corner plot of posterior probability distributions of the parameters for the joint fit, where all the parameters are well constrained, and the common parameter $t_{\rm b3}$ is also well constrained to be $84.3_{-1.2}^{+1.2}$ s. It should be noted that for visual clarity, only the posterior probability distributions of the achromatic break time ($t_{\rm b3}$) and its previous decay indices ($\hat{\alpha}_3$) are displayed. For GF4, if $t_{\rm b3}$ is fixed at 84.3, the flux upper limits of the last four time intervals can constrain the lower limit of $\hat{\alpha}_4$ to be 6.5.

\subsection{Curvature effect \\}

The ``curvature effect'' refers to the phenomenon of photons arriving progressively later at the observer from higher latitudes with respect to the line of sight\cite{Kumar2000ApJ, Dermer2004ApJ}. It has been proposed that this effect plays an important role in shaping the decay phase of the light curve after the sudden cessation of the GRB's emitting shell\cite{Zhang2006ApJ, Liang2006ApJ}. By assuming a power law spectrum with spectral index $\hat{\beta}$ for the GRB, the most straightforward relation of the curvature effect can be given as $\hat{\alpha}=2+\hat{\beta}$\cite{Kumar2000ApJ, Dermer2004ApJ, Uhm2015ApJ}, where $\hat{\alpha}$ and $\hat{\beta}$ are the temporal decay index and spectral index in the convention $F_{\rm\nu} \propto t^{-\hat{\alpha}}\nu^{-\hat{\beta}}$, respectively. If the aforementioned assumptions are released, and the intrinsically curved spectral shape and strong spectral evolution are taken into account, the above relationship is no longer applicable\cite{ZhangBB2007ApJ, ZhangBB2009ApJ}. Nevertheless, for a narrow energy band, the intrinsically curved spectrum can be approximated, on average, by a power law, and the time-dependent $\hat{\alpha}$ and $\hat{\beta}$ still approximately follow the relation $\hat{\alpha}(t)=2+\hat{\beta}(t)$\cite{ZhangBB2009ApJ}. 

The multi-wavelength flux light curves of GRB 230307A exhibit an initial rise and subsequent power law decay phases that gradually become steeper. To test whether the decay phases are dominated by the curvature effect, we compare the time-dependent $\hat{\alpha}(t)$ and $2+\hat{\beta}(t)$ in each narrow energy band. Here, $\hat{\alpha}(t)$ is obtained by assuming the trigger time $T_0$ as the time zero point and numerically calculating $-\Delta{\rm log}F/\Delta{\rm log}t$ of the best-fitting SBPL functions, while $\hat{\beta}(t)$ is the average spectral index $-\Delta{\rm log}F_{\rm\nu}/\Delta{\rm log}E$ calculated in the corresponding narrow energy band based on the spectral fitting results for each time slice of LEIA S-II or GECAM S-II. Fig. \ref{fig:sed}c displays the time-dependent $\hat{\alpha}(t)$ and $2+\hat{\beta}(t)$ for each energy band. We note that the power law decay indices of the segments between the second and third breaks (see $t_{\rm b2}$ and $t_{\rm b3}$ in Extended Data Table \ref{tab:lc_sbpl}) of GECAM multi-wavelength flux light curves are consistent with the prediction of the curvature effect, implying that the jet's emitting shell stops shining at $t_{\rm b2}$ ($\sim$ 23 s) and then high-latitude emission dominates the prompt emission. On the contrary, LEIA flux light curve is in a shallower decay phase, with a decay index much lower than the prediction of the curvature effect. Such a behavior suggests that the soft X-ray emission detected by LEIA is intrinsic to the central engine, not related to the narrow jet but consistent with the emission due to direct dissipation of the magnetar wind.

As stated above, the relation $\hat{\alpha}(t)=2+\hat{\beta}(t)$ is only an approximation for an intrinsically curved spectrum. To more precisely test the curvature effect, we also employ another more sophisticated approach. The curvature effect postulates that the spectrum in the comoving frame is the same for different latitudes with respect to the line of sight. In order to maintain consistency with the observed spectral shape, we consider a time-dependent cutoff power law spectrum in the observer frame with the form 
\begin{equation}
N(E,t)=A(t)\Big(\frac{E}{\rm 100~keV}\Big)^{\alpha}{\rm exp}\Big[-\frac{(2+\alpha)E}{E_{\rm p}(t)}\Big],
\label{eqs:NEt}
\end{equation}
where $\alpha=-(\hat{\beta}+1)$ is the photon spectral index, $E_{\rm p}(t)$ is time-dependent peak energy, and $A(t)$ is the time-dependent normalization parameter in units of ${\rm photons~cm^{-2}~s^{-1}~keV^{-1}}$. The curvature effect predicts $E_{\rm p}(t)\propto t^{-1}$ and $N(E_{\rm p}, t)\propto t^{-1}$\cite{Kumar2000ApJ}. Considering the $t_0$ effect\cite{Zhang2006ApJ, Liang2006ApJ}, one can get\cite{ZhangBB2009ApJ}
\begin{equation}
E_{\rm p}(t)=E_{\rm p, c}\Big(\frac{t-t_0}{t_{\rm c}-t_0}\Big)^{-1},
\label{eqs:Ept}
\end{equation}
and
\begin{equation}
A(t)=A_{\rm c}\Big(\frac{t-t_0}{t_{\rm c}-t_0}\Big)^{\alpha-1},
\label{eqs:At}
\end{equation}
where $t_0$ refers to the time zero point of the pulse being considered for the curvature effect, $t_{\rm c}$ is the epoch when the curvature effect begins, while $E_{\rm p, c}$ and $A_{\rm c}$ denote the peak energy and normalization parameter at $t_{\rm c}$, respectively. On the basis of Eqs. \ref{eqs:NEt}, \ref{eqs:Ept} and \ref{eqs:At}, the model spectrum predicted by the curvature effect can be determined at any time $t$ once the parameter set $\{\alpha,~E_{\rm p,c},~A_{\rm c},~t_0,~t_{\rm c}\}$ is given. In order to compare with the observed spectra and test for consistency with the curvature effect, we fix $t_{\rm c}$ at 23 s based on the results of the aforementioned approximation method and vary the other parameters. Then, we fit the model $N(E,~t,~\mathcal{P})$, where $\mathcal{P}=\{\alpha,~E_{\rm p,c},~A_{\rm c},~t_0\}$, to the GECAM S-II spectra within the time range spanning from $t_{\rm b2}$ to $t_{\rm b3}$. After achieving a successful fit with statistically acceptable reduced statistic PGSTAT/dof $\sim 1$, we present the corner plot of the posterior probability distributions of the free parameters and the comparison between the data and the model in Extended Data Fig. \ref{fig:lc_sbpl}b and \ref{fig:lc_sbpl}c, respectively. We note that the best-fitting parameters of $\alpha$, $E_{\rm p,c}$ and $A_{\rm c}$ agree with the fitting parameters obtained through empirical model fitting in Extended Data Table \ref{tab:gecam_spec_fit}. Additionally, the time zero point is constrained to the trigger time $T_0$, which is the prerequisite for the aforementioned approximation method. This indicates that the entire prompt emission is the broad pulse under consideration for the curvature effect. According to Ref.\cite{Zhang2011ApJ, Zhang2014ApJ}, such a broad pulse corresponds to the ``slow variability'' component associated with the central engine activity, while the ``fast variability'' component with multiple sharp pulses superposed on the slow component can be attributed to the presence of mini-jets resulting from turbulent reconnection within the emission region. We note that despite our efforts to minimize the bin size while performing spectral fitting through data binning, we unavoidably sacrifice temporal information of the ``fast variability'' component. Nonetheless, it does not affect the overall shape of the broad pulse. In summary, we successfully conducted a self-consistency check on the existence of curvature effect between approximately 23--84 s in GECAM data using the two methods mentioned above.

\subsection{Fit of spectral energy distribution \\}

The broad-band prompt emission data enables us to examine the spectral energy distribution (SED) from soft X-rays to gamma-rays. We first perform an independent spectral fitting on the LEIA S-III spectra, by employing the PL model with Galactic absorption ($N_{\rm H}=1.26\times 10^{21}~{\rm cm^{-2}}$) and intrinsic absorption ($N_{\rm H}=3.92\times 10^{21}~{\rm cm^{-2}}$), to derive the theoretical spectra in the LEIA band. We then perform independent spectral fitting on the GECAM S-III spectra,
utilizing the same seven empirical models applied to the GECAM S-II spectra.
The spectral fitting results of the preferred models, along with the corresponding fitting statistics, are presented in Extended Data Table \ref{tab:sed}. The model-predicted SEDs derived from the spectral fittings at different time intervals are displayed in Fig. \ref{fig:sed}b.

As shown in Fig. \ref{fig:sed}b, it is evident that during the first eight time intervals, the extrapolations of the LEIA spectra deviate strongly from the GECAM spectra. Similarly, the extrapolations of the GECAM spectra deviate significantly from the LEIA spectra, regardless of whether the GECAM spectra exhibit breaks at low energies. Despite conducting the spectral fittings by treating the intrinsic absorption as a free parameter, we found that the inconsistency between the LEIA and GECAM spectra still exists. Such inconsistency suggests the presence of a distinct radiation process that dominates the LEIA band. As we can see, during the last two time intervals (after 76 s), the LEIA and GECAM spectra can be jointly fitted well using the single CPL model. This suggests that during this time interval, the high-energy emission component gradually diminishes and that the entire spectrum is dominated by the emergent new component LEIA observed throughout.

In order to better justify the superposition between two emission components, we present a possible model for joint SED fitting by superimposing two spectral models that respectively account for the  high-energy (jet) and low-energy (magnetar) spectral components. Below we describe these two spectral models:
\begin{itemize}
    \item High-energy spectral model: Firstly, we notice that, in Fig. \ref{fig:sed}b, the natural extrapolations of the GECAM spectral models either partially or entirely over-predict the soft X-ray component. This implies that the contribution of the high-energy emission component to the soft X-ray spectrum should not be the simple extrapolation of the GECAM spectral model into the LEIA band, but rather requires a spectral break between 4--15 keV to reduce the contribution from the high-energy spectral component. However, the energy of such a spectral break and the power law index below the break are relatively arbitrary and cannot be constrained well by data with the absence of the data between 4--15 keV. On the other hand, considering that the prompt emission of GRBs is typically interpreted as synchrotron radiation originating from a group of non-thermal electrons, one can introduce a synchrotron self-absorption frequency break at $\nu_{\rm sa}$, which is physically expected. The power law index below $E_{\rm sa}$ (i.e. $h\nu_{\rm sa}$) is predicted to be $N(E)\propto E$ or $\propto E^{3/2}$, depending on if a 2SBPL or CPL spectrum is observed in the high-energy band for our case\cite{Shen2009MNRAS}. Therefore, we employ two physically-motivated new spectral models to describe the high-energy (jet) spectral component, namely PL$_{\rm sa}$-2SBPL and PL$_{\rm sa}$-CPL, where PL$_{\rm sa}$ represents the new hard power law segment below $E_{\rm sa}$, and the smooth connection between PL$_{\rm sa}$ and 2SBPL (or CPL) occurs at $E_{\rm sa}$ in accordance with Eq. \ref{eqs:smooth_connect}.
    \item Low-energy spectral model: Considering that the low energy component dominates the entire spectrum at late times, we utilize the CPL model fitted to the tenth joint SED as the template, with the cutoff energy fixed at 53 keV (obtained from the tenth SED's CPL model fit, where the magnetar component dominates, see Extended Data Fig. \ref{fig:SED} and Extended Data Table \ref{tab:sed}), while allowing the index and normalization parameters to vary freely.
\end{itemize}
During the joint SED fitting, taking into account the calibration uncertainties of less than 10\% for both LEIA and GECAM, we therefore introduce two additional free calibration constants with a prior that allows for 10\% deviation. The fitting results, along with the corresponding fitting statistics, are presented in Extended Data Table \ref{tab:sed}. The comparisons between the observed and model-predicted count spectra, as well as the residuals (defined as $({\rm data}-{\rm model})/{\rm data~error}$) are depicted in Extended Data Fig. \ref{fig:SED}a. The model-predicted SEDs, along with the high- and low-energy spectral components derived from the spectral fittings at different time intervals, are displayed in Extended Data Fig. \ref{fig:SED}b. The results confirm the successful fit of the model and suggest that the time-dependent SEDs can be interpreted as a superposition of two distinct components: the high-energy component, which diminishes quickly both at later times and at lower energies, and the low-energy component, which is observed consistently by LEIA throughout the observation period.

\subsection{Explanation of multi-wavelength flux light curves \\}

We conduct a smoothly broken power law fit to the multi-wavelength flux light curves (Fig. \ref{fig:sed}a) and explain each of the segments using our schematic model depicted in Extended Data Fig. \ref{fig:schematic}. The jet and magnetar-powered emission processes are displayed separately. The emission of the jet-dominated GRB component undergoes a rise followed by a general trend of decline contributed by photons within the jet core of $\theta_{\rm c} = 1/ \Gamma$, where $\Gamma$ is the Lorentz factor. After the emitting shell terminates, the emission is contributed by photons from higher latitude regions with respect to the line of sight of the observer. The best-fit temporal slopes during this phase ($t_{\rm b2}$--$t_{\rm b3}$ in Extended Date Table \ref{tab:lc_sbpl}) are consistent with prediction of the high-latitude curvature effect relation $\hat{\alpha}(t)=2+\hat{\beta}(t)$ (Fig. \ref{fig:sed}c). Following the progressive decrease of the high-latitude emission, an achromatic break occurs, signaling the end of the curvature effect. This enables us to estimate the jet opening angle, for the first time, during the prompt emission phase of a GRB. Finally, the light curve drops with an even steeper slope, possibly contributed by some very weak emission from regions beyond the opening angle of the jet that is likely to have no sharp edges. The detailed energy and Lorentz factor profiles beyond the jet edge can be constrained from the data\cite{Yan24}. In contrast, the soft X-ray emission is powered by a presumably more isotropic magnetar wind. The light curve is dominated by the spin-down law with a plateau followed by a shallow decline.

\subsection{Estimation of jet opening angle \\}
The high-latitude emission effect is observed in the multi-wavelength flux light curves of GECAM, starting from the second break ($t_{\rm b2}$) and continuing until the third break ($t_{\rm b3}$), as depicted in Fig. \ref{fig:sed} (see also Extended Data Table \ref{tab:lc_sbpl}). The third break signifies the moment when photons from the outermost layer of the shell, with a radius $R_{\rm GRB}$, reach the observer. The duration of such tail emission can be described by the following relationship\cite{Zhang2006ApJ}:

\begin{equation}
\Delta t_{\rm b} = t_{\rm b3} - t_{\rm b2} = (1+z) \left(\frac{R_{\rm GRB}}{c}\right) \left(\frac{\theta_j^2}{2}\right),
\label{eqs:TailEmission}
\end{equation}
where $\theta_j$ represents the half opening angle of the jet. By substituting $t_{\rm b3} = 84$ s, $t_{\rm b2} = 23$ s, and assuming a typical radius of $R_{\rm GRB} = 10^{15}$ cm, we can calculate the opening angle of the jet as follows:

\begin{equation}
\theta_{\rm j} = \sqrt{\frac{2c\Delta t_{\rm b}}{(1+z)R_{\rm GRB}}} \approx 3.4^\circ \left(\frac{\Delta t_{\rm b}}{61~\text{s}}\right)^{\frac{1}{2}} \left(\frac{R_{\rm GRB}}{10^{15}~\text{cm}}\right)^{-\frac{1}{2}}.
\label{eqs:JetAngle}
\end{equation}

Such a jet opening angle is generally consistent with that derived from multi-band modeling of the afterglows\cite{YHYang2023prep, Dai2024ApJ}, where the jet break is observed around a few days after the event. However, it is smaller than that reported in another study\cite{Wang2024ApJ}, which used different energy sources for kilonova modeling. The achromatic break observed in our work is due to the edge effect of high-energy photons produced from internal emission, while in the afterglow, it is due to external shocks at a larger radius. However, it is also possible that the jet opening angle measured in the prompt emission could be somewhat different from that measured in the afterglow considering there could be additional angular energy injection at late times from the magnetar wind.


\subsection{Host galaxy \\}

We conducted a search for the host galaxy of GRB 230307A in various public galaxy catalogs and GCN circulars. We estimated the chance coincidence probability $P_{\rm cc}$\cite{Bloom2002AJ} for the most promising candidates, which are listed below.

\begin{enumerate}[(1)]

\item Large Magellanic Cloud (LMC): The LMC is the nearest and brightest galaxy to the Milky Way, located at a distance of 49 kpc. GRB 230307A is 8.15 degrees away from the center of the LMC, corresponding to a physical separation of 7 kpc, and it is situated on the edge of the Magellanic Bridge. The surface density of galaxies, typically used for estimating chance coincidence probabilities, is not applicable to the LMC due to its brightness. Instead, we consider a surface density of galaxies as bright as the LMC, which is $\sigma=1/41252.96$ deg$^{-2}$. Using a half-light radius of $r_{50}=2.2$ deg\cite{Corwin1994AJ}, we estimate the chance coincidence probability $P_{\rm cc}$ to be 0.006. However, its physical properties are not fully consistent with known transients. Firstly, with LMC as the host galaxy, the isotropic energy would be $\sim 9 \times 10^{44}$ erg, much lower than typical GRBs. Secondly, although the isotropic energy is consistent with those of Soft Gamma-ray Repeater (SGR) giant flares, the duration is much longer than the short hard spikes of the giant flares. Furthermore, its large offset with respect to the LMC disk is inconsistent with that of known SGRs. Therefore, we consider this scenario less likely.

\item Galaxy with a redshift of $z=3.87$: According to Ref.\cite{Levan2023GCN33747}, there is a faint galaxy located 0.2 arcsec away from GRB 230307A with a redshift of $z=3.87$. We analyzed the JWST/NIRCam images using the official STScI JWST Calibration Pipeline version 1.9.0 and found the galaxy to be fainter than $28.5$ mag in the JWST/F070W bands and 27.4 mag in the JWST/F277W band. The estimated $P_{\rm cc}$ using the JWST/F277W magnitude is 0.034, while for the JWST/F070W band, it is estimated to be greater than 0.09. Since the F070W band is closer to the $r$ band, with which the galaxy brightness distribution is produced, the latter estimate is considered more reliable. However, the extremely high energy of the GRB and its inconsistency with known GRB transients in Extended Data Fig. \ref{fig:classification}b and \ref{fig:classification}c make this scenario less favored.

\item Galaxy with a redshift of $z=0.065$\cite{Gillanders2023GCN, Levan2023arXiv}: With the half-light radius and $r$ band magnitude from DESI Legacy Survey\cite{Dey2019AJ}\footnote{https://www.legacysurvey.org/dr10/catalogs/}, the chance coincidence probability for this galaxy was estimated to be 0.11 (Table \ref{tab:obs_prob}). Although not statistically significant, the redshift is more consistent with the physical properties of the GRB. Therefore, we consider this galaxy to be the most likely host. The offset of GRB 230307A from the host galaxy is 29.4", which is 36.6 kpc in redshift 0.065. As presented in Extended Data Fig. \ref{fig:classification}d, the offset is consistent with those of type I GRB, and larger than type II GRBs. Specifically, the offset is approximately 1(1.3)$\sigma$ away from the type I track and 2.4(8.9)$\sigma$ away from the type II track in logrithmic (linear) space.

\end{enumerate}

Moreover, we explore other possible host galaxies in DESI Legacy Survey. For objects within 5 arcmin of GRB 230307A, we exclude stars with detected parallax in Gaia, and then calculate the chance coincidence probability with the half-light radius and $r$ band magnitude in the catalog. It turns out that the galaxy with a redshift of $z=0.065$ has the lowest $P_{\rm cc}$, while others have $P_{\rm cc} > 0.2$ or more. 

\subsection{LEIA data compared with other X-ray and optical observations\\}

To investigate whether the LEIA observation is consistent with the early X-ray afterglow, we have incorporated the predicted afterglow model from late-time multi-wavelength observations\cite{YHYang2023prep}, as illustrated in Fig. \ref{fig:magnetar}b. The estimated X-ray flux of the afterglow at approximately a few hundred seconds in the 0.5--4 keV range is $\sim$ $10^{-11}$ $\rm erg~cm^{-2}~s^{-1}$, which is more than two orders of magnitude below the observed flux. We also note that the power law index of the LEIA light curve in the decaying phase is around -2, which differs from the typical value of GRB X-ray afterglows, which is around -1. 

GRB 230307A was also simultaneously detected by TESS\cite{Fausnaugh2023RNAAS} in the optical during the prompt emission and the following afterglow phases. We calculated the X-ray to optical flux density ratios at the epochs when the burst was jointly detected in these two energy bands, as illustrated in Extended Data Fig. \ref{fig:xoptratio}. The results show that the flux density ratios at the early epochs are significantly higher than those at later times. The abnormally large ratios during the first two epochs imply that there is an additional component contributing to the early X-ray data. Additionally, the TESS observation shows that the optical afterglow rises to its peak at approximately 0.05 days, which is much longer than the 200-second duration of the LEIA data.

Therefore, the soft X-ray component detected by LEIA is inconsistent with the X-ray afterglow of the GRB.

\subsection{Magnetar Model\\}
We fit the LEIA (0.5--4 keV) light curve of GRB 230307A in the prompt emission phase with both the smoothly broken power law model (Eq. \ref{eqs:sbpl}) and the magnetar dipole radiation model. 

Considering a millisecond magnetar with rigid rotation, it loses its rotational energy through both magnetic dipole radiation and quadrapole radiation\cite{Shapiro1983,Usov1992Nat,Zhang2001ApJ,Gao2016PRD,Sun2019ApJ}, with 
\begin{equation}
\dot{E}=I\Omega \dot{\Omega}=-\frac{B_{\rm p}^2R^6\Omega^4}{6c^3}-\frac{32GI^2 \epsilon ^2 \Omega^6}{5c^5},
\label{eqs:Edot}
\end{equation}
where $\dot{E}$ is the total spin-down rate, $\Omega=2\pi/P$ is the angular frequency and $\dot{\Omega}$ its time derivative, $I$ is the moment of inertia, $B_{\rm p}$ is the dipolar field strength at the magnetic poles on the NS surface, $R$ is the NS radius, and $\epsilon$ is the ellipticity of the NS. In the following calculations, we use the values $I=3.33 \times 10^{45}$ $\rm g\,cm^{-2}$, $R=1.2 \times 10^6$ cm, based on the Equation of State GM1 as discussed in Ref.\cite{lasky2014}, and $\epsilon = 10^{-4}$. The electromagnetic emission is determined by the dipole spin-down luminosity $L_{\rm sd}$, i.e.,
\begin{equation}
L_{\rm X}(t)=\eta L_{\rm sd} = \frac{\eta B_{\rm p}^2 R^6 \Omega^4(t)}{6c^3},
\label{eqs:X-jet}
\end{equation}
where $\eta$ is the efficiency of converting the dipole spin-down luminosity to the X-ray luminosity. The X-ray luminosity is derived from the observation assuming isotropic emission. In a more realistic situation, the X-ray emission may not be isotropic, and a factor $f_{\rm b}$ (assumed to be 0.1) is introduced to account for this correction between the isotropic and true luminosity $L_{\rm X}$, i.e.,
\begin{equation}
L_{\rm iso}(t) = \frac{L_{\rm X}(t)}{f_{\rm b}} =\frac{\eta}{f_{\rm b}}\frac{B_{\rm p}^2 R^6 \Omega^4(t)}{6c^3}.
\label{eqs:liso}
\end{equation}

We model the unabsorbed luminosity light curves based on Eqs. \ref{eqs:Edot} and \ref{eqs:liso}. The fitting process is performed by using the Markov Chain Monte Carlo code \emph{emcee}\cite{Foreman2013PASP}. We employ $\chi^2$ as the statistical metric to evaluate the likelihood. The prior bounds for the free parameters ($\log(B_{\rm p}/{\rm G})$, $P_0/{\rm ms}$, $\log\eta$) are set to be $(15, 17)$, $(1, 5)$, and $(-3, -2)$, respectively.  
The first data point is excluded, as the soft X-ray data possibly suffer the contamination from the jet emission at the very early time. The fitting results are shown in Fig. \ref{fig:magnetar}a and Extended Data Fig. \ref{fig:magnetar2}. After subtracting the contribution of the jet component based on the SED modeling represented in Extended Data Fig. \ref{fig:SED}b, we calculate the flux of the magnetar component in 0.5--4 keV and fit the light curve using both the smoothly broken power law model and the magnetar dipole radiation model, as shown in Extended Data Fig. \ref{fig:SED}c. The best-fit slopes and magnetar parameters are consistent with those derived in Fig. \ref{fig:magnetar}a within $1\sigma$ uncertainties.

Fast X-ray transients with light curves characteristic of spin-down magnetars have been identified previously in the afterglow of some short GRBs, as well as in a few events without associated GRB such as CDF-S XT2, which are believed to originate from compact star mergers. We compare the X-ray luminosity light curve of GRB 230307A with internal plateaus in the X-ray afterglows of short GRBs with redshifts and in CDF-S XT2. The extended emissions and the afterglow data are retrieved from the BAT/XRT lightcurve repository\cite{Evans2007A&A, Evans2009MNRAS, Evans2010A&A} and corrected to the 0.5--4 $\rm keV$ band assuming an absorbed power law spectrum. Such a correction is also made for the luminosity of CDF-S XT2, using power law indices $\Gamma_1 = 1.45$ and $\Gamma_2 = 2.67$ before and after the break time at 2.3 ks\cite{Xue2019Nat}.

When a magnetar is formed through the merger of binary neutron stars, it is anticipated that there is additional injection of energy into the ejecta that powers the optical/near-IR kilonova. Early modeling assumed an isotropic magnetic wind and a relatively high spin-down injection efficiency, which predicts a kilonova luminosity that is boosted by a factor of ten or even more\cite{Yu2013ApJ}. Recent more detailed modeling suggests that for a Poynting flux dominated wind, the heating efficiency after shock crossing is very low\cite{Ai2022MNRAS}. Furthermore, relativistic magnetohydrodynamic simulations suggest that the magnetar wind is anisotropic due to collimation towards the jet direction. The energy injection into the kilonova ejecta is reduced by a factor of up to 10\cite{Wang2023arXiv}.For this burst, the observed kilonova could be consistent with the central engine of a magnetar if the magnetar wind is Poynting flux dominated without significant dissipation. It is worth noticing that the spectral features predicted from the kilonova model is missing from the late-time data (Extended Data Fig. 4 of Ref.\cite{Gillanders2023arXiv}). This suggests an additional heating source other than radioactive heating, which is consistent with energy injection from a long-lived magnetar.

Although various theories of engine injection have been proposed for both long and short GRBs, such as fallback accretion and refreshed shocks, we have found that these models are not fully consistent with the light curve observed by LEIA. In the case of the BH engine, fallback accretion involves a long-lived disk that would eventually evaporate quickly, causing the jet engine to cease abruptly\cite{Lu2023}. This model predicts long-lasting activities with luminosity decaying as $\sim t^{-40/9}$, which is much steeper than the observed slope from the LEIA data\cite{Kisaka2015ApJ}. For the refreshed shocks scenario, the decay slope would be flatter, similar to that observed in GRB afterglows (around $t^{-1.5}$). This model has been ruled out due to the inconsistency of the LEIA data with the X-ray afterglow.

\section*{Data Availability}

The processed data are presented in the tables and figures of the paper, which are available upon reasonable request. The authors point out that some data used in the paper are publicly available, whether through the UK Swift Science Data Centre website, JWST website, or GCN circulars.

\section*{Code Availability}
Upon reasonable requests, the code (mostly in Python) used to produce the results and figures will be provided.

\bigskip
\bigskip
\bigskip


\bigskip
\bigskip
\bigskip

\begin{addendum}

\item[Acknowledgments] This work is supported by the National Key Research and Development Programs of China (2022YFF0711404, 2021YFA0718500, 2022SKA0130102, 2022SKA0130100). LEIA is a pathfinder of the Einstein Probe mission, which is supported by the Strategic Priority Program on Space Science of CAS (grant Nos. XDA15310000, XDA15052100). The GECAM (Huairou-1) mission is supported by the Strategic Priority Research Program on Space Science (Grant No. XDA15360000, XDA15360102, XDA15360300, XDA15052700) of CAS. We acknowledge the support by the National Natural Science Foundation of China (Grant Nos. 11833003, U2038105, 12121003, 12173055, 11922301, 12041306, 12103089, 12203071, 12103065, 12273042, 12173038, 12173056, 12333007, 12027803 and 13001106), the science research grants from the China Manned Space Project with NO.CMS-CSST-2021-B11, the Natural Science Foundation of Jiangsu Province (Grant No. BK20211000), International Partnership Program of Chinese Academy of Sciences for Grand Challenges (114332KYSB20210018), the Major Science and Technology Project of Qinghai Province (2019-ZJ-A10), the Youth Innovation Promotion Association of the Chinese Academy of Sciences, the Postgraduate Research \& Practice Innovation Program of Jiangsu Province (KYCX23\_0117), the Program for Innovative Talents, Entrepreneur in Jiangsu, and the International Partnership Program of Chinese Academy of Sciences (Grant No.113111KYSB20190020). We thank Y.-H. Yang, E. Troja, Y.-Z. Meng, Rahim Moradi, and Z.-G. Dai for helpful discussions. 

\item[Author Contributions] H.S., S.-L.X. and B.-B.Z. initiated the study. H.S., B.-B.Z., B.Z., J.Y. and S.-L.X. coordinated the scientific investigations of the event. C.-W.W., W.-C.X., J.Y., Y.-H.I.Y., W.-J.T., J.-C.L., Y.-Q.Z., C.Zheng, C.C., S.X., S.-L.X., S.-X.Y. and X.-L.W. processed and analysed the GECAM data. S.-L.X. first noticed the extremely brightness of GRB 230307A from GECAM data. H.S., Y.Liu., H.-W.P. and D.-Y.L. processed and analysed the LEIA data. Y.Liu. first identified GRB 230307A in LEIA data. J.Y., C.-W.W., Y.-H.I.Y., W.-C.X. and Y.-Q. Z. performed the spectral fitting of GECAM data. W.-C.X. and J.-C.L. performed background analysis for GECAM-C. C.Zheng and Y.-Q.Z. performed calibration analysis for GECAM. J.-C.L. performed the data saturation assessment for GECAM. H.S., Y.Liu. and J.Y. performed the spectral fitting of LEIA data. C.Z., Z.-X.L., Y.Liu., H.-Q.C. and D.-H.Z. contributed to the calibration of LEIA data. J.Y. and C.-W. W. contributed the Amati relation and luminosity-lag relation. Y.-H.I.Y., J.Y. and C.-W.W. performed the $T_{90}$ calculation. J.Y. calculated the amplitude parameter. J.Y., W.-C.X., W.-J.T., Y.-H.I.Y. and S.X. calculated the spectral lag. W.-J.T., W.-C.X. and S.X. performed the minimum variability timescale calculation. J.Y. and C.-W.W. fitted the multi-wavelength flux light curves. J.Y. calculated the curvature effect. J.Y., H.S., Y.Liu., C.-W.W. W.-C.X. and Y.-Q.Z. contributed to the SED modeling. J.Y. performed the global fitting to the achromatic break. C.-W.W. and Z.-Y.Y. performed the calculation of jet opening angle. H.S. and J.Y. performed the theoretical modelling with the magnetar dipole radiation model. J.-W.H. contributed to the luminosity correction of the X-ray afterglows. Y. Li, L.H. and J.Y. contributed to the information about the host galaxy. Z.-X.L., C.Z., X.-J.S., S.L.S., X.-F.Z., Y.-H.Z., Z.-M.C. and W.Y. contributed to the development of the LEIA instrument. Y.Liu., H.-Q.C., C.J., W.-D.Z., D.-Y.L., J.-W.H., H.-Y.L., H.S., H.-W.P. and M.L. contributed to the development of LEIA data analysis tools and LEIA operations. Z.-H.A., X.-Q.L., W.-X.P., L.-M.S., X.-Y.W., F.Z., S.-J.Z., C.C., S.X. and S.-L.X. contributed to the development and operation of GECAM. B.Z. led the theoretical investigation of the event. H.S., J.Y., B.-B.Z., B.Z., W.Y., S.-L.X., S.-N.Z., Y. Liu contributed to the interpretation of the observations and the writing of the manuscript with contributions from all authors.

\item[Competing Interests] The authors declare that they have no competing financial interests.

\item[Additional information] Correspondence and requests for materials should be addressed to B.-B.Z. \\ (bbzhang@nju.edu.cn), S.-L.X. (xiongsl@ihep.ac.cn), Z.-X.L. (lingzhixing@nao.cas.cn) and B.Z. \\ (bing.zhang@unlv.edu).

\end{addendum}

\clearpage
\setcounter{figure}{0}
\setcounter{table}{0}

\captionsetup[table]{name={\bf Extended Data Table}}
\captionsetup[figure]{name={\bf Extended Data Fig.}}

\begin{figure}
\centering
\begin{tabular}{cc}
\multicolumn{2}{c}{\begin{overpic}[width=0.75\textwidth]{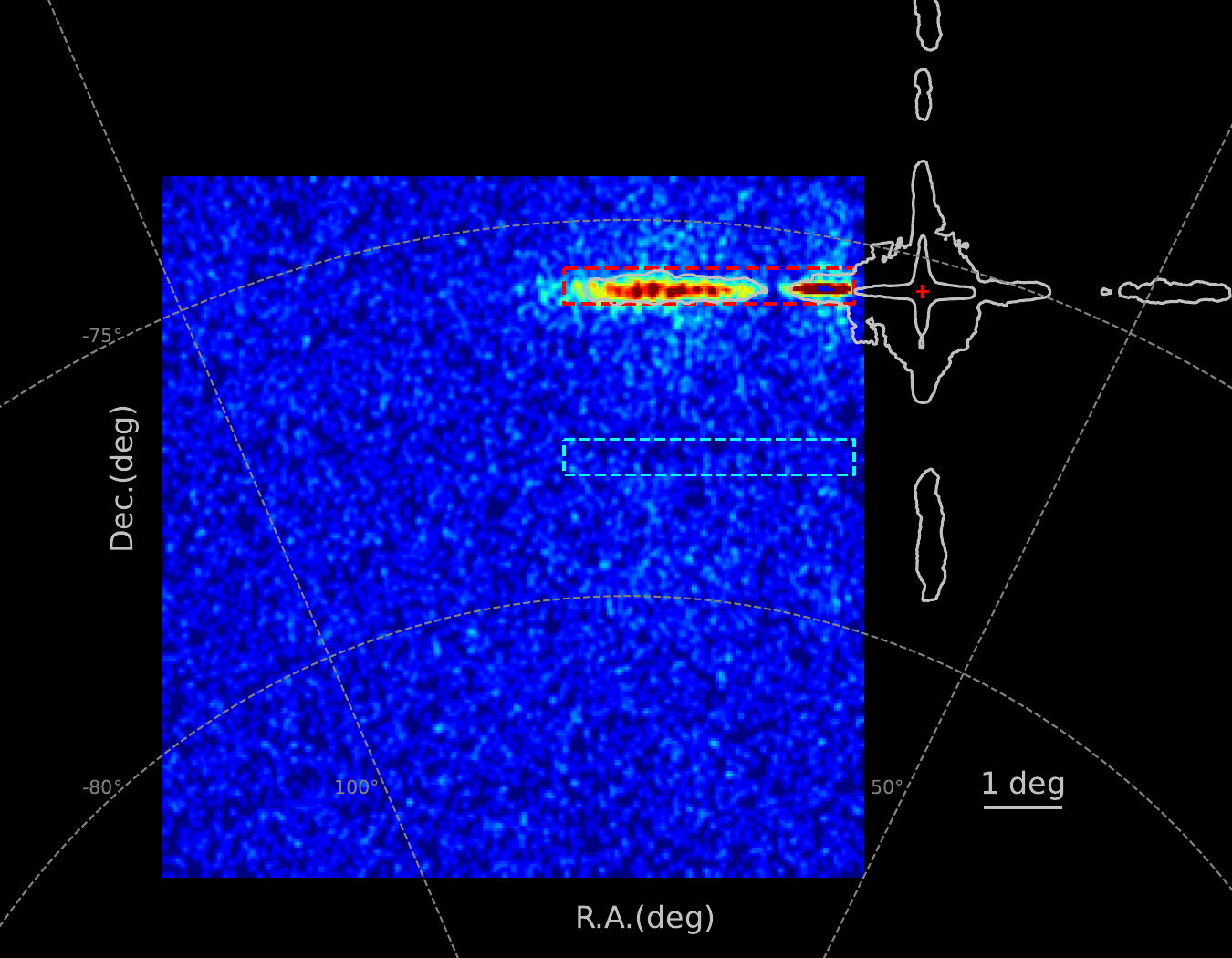}\put(-4, 75){\bf a}\end{overpic}} \\
\begin{overpic}[width=0.40\textwidth]{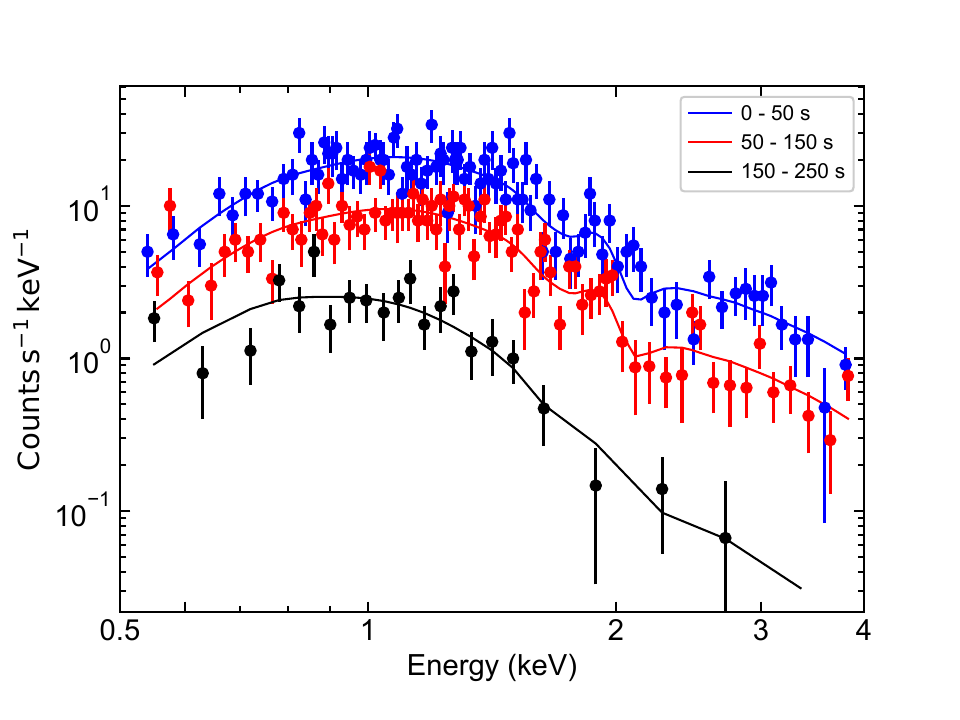}\put(2, 65){\bf b}\end{overpic} &
 \begin{overpic}[width=0.40\textwidth]{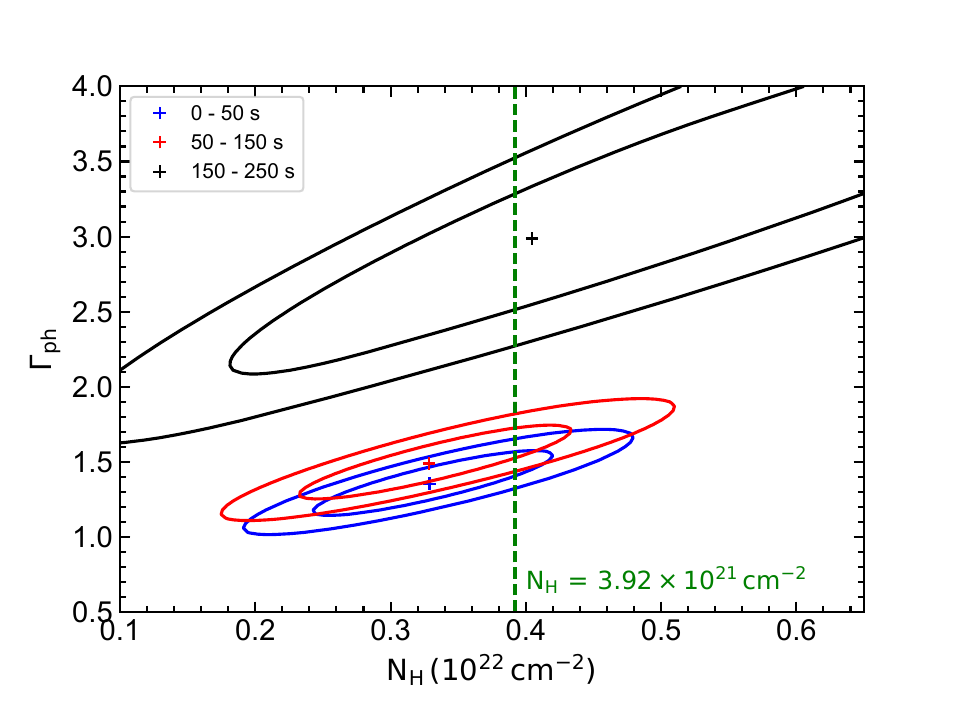}\put(2, 65){\bf c}\end{overpic} \\
\end{tabular}
\caption{\noindent\textbf{LEIA image and spectra.} \textbf{a}, LEIA image. The blue square region represents the quarter of LEIA FoV adjacent to the source. The source with its XRT position denoted by the red cross is located outside of the nominal FoV of LEIA. Therefore only the cruciform arm structure of the point spread function (PSF) was observed, as shown by the brightening regions in the upper right corner. The white solid lines around the source position shows the full PSF profile when the source is detected at this position (the data for generating the PSF profile is taken from on-ground calibration measurements). The regions encircled by red and cyan dash-dotted lines are utilized in the extraction of the source and background photons, respectively. \textbf{b}, LEIA observed and model-predicted spectra in three time segments (LEIA S-I). \textbf{c}, Best-fit values of photon index $\Gamma_{\rm{ph}}$ and hydrogen column density $N_{\rm H}$ and 1$\sigma$, 2$\sigma$ confidence contours.}
\label{fig:leia_image}
\end{figure}

\clearpage

\begin{figure}
\centering
\begin{tabular}{c}
\begin{overpic}[width=0.55\textwidth]{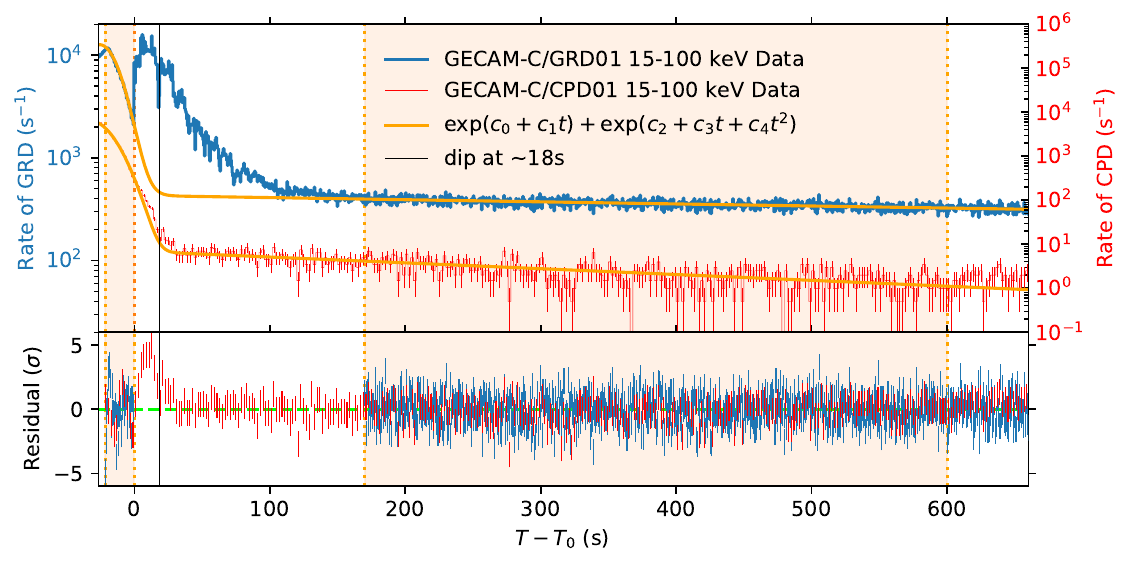}\put(0, 50){\bf a}\end{overpic} \\ 
\begin{overpic}[width=0.50\textwidth]{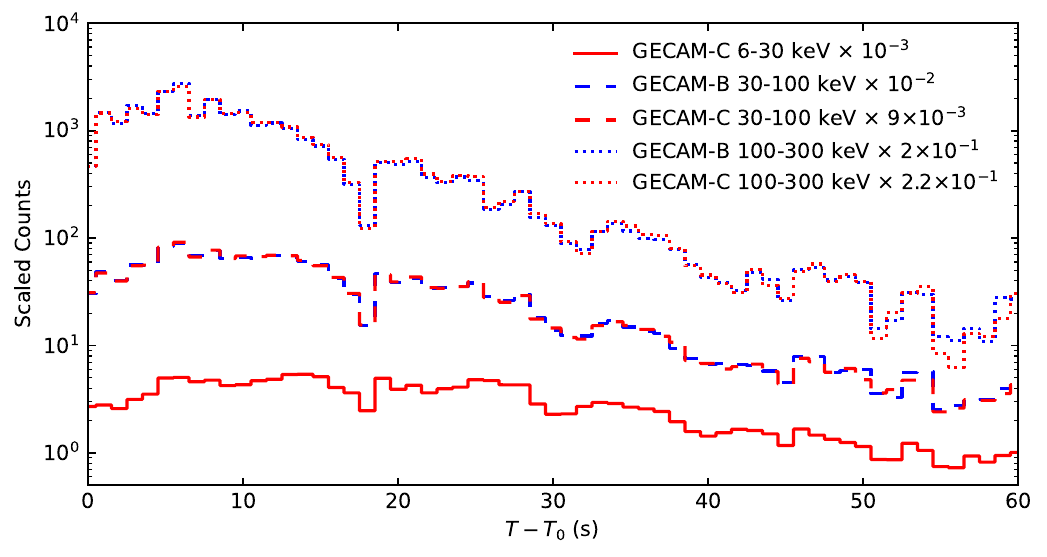}\put(0, 50){\bf b}\end{overpic} \\ 
\begin{overpic}[width=0.50\textwidth]{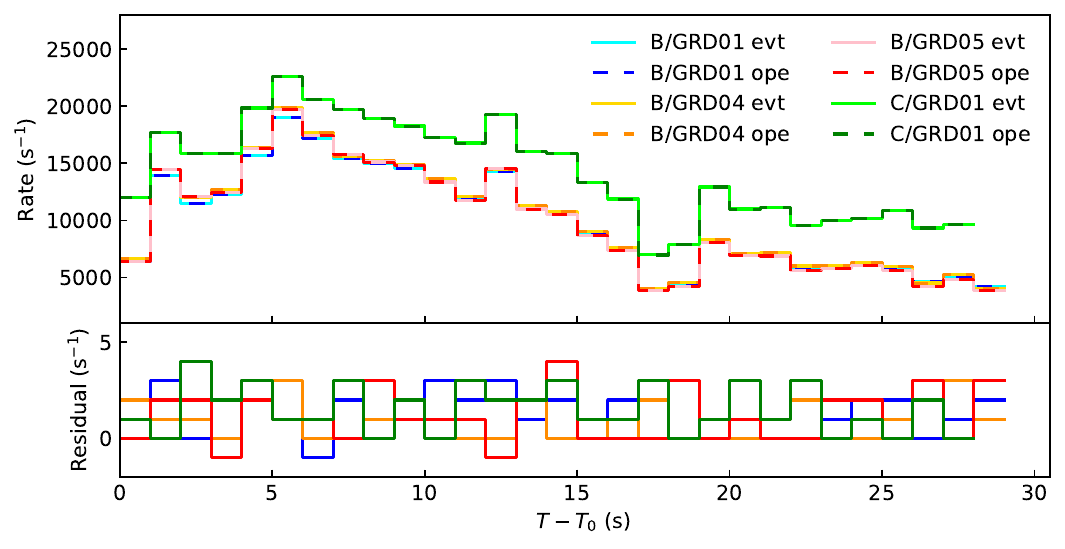}\put(0, 50){\bf c}\end{overpic} \\
\end{tabular}
\caption{\noindent\textbf{GECAM data reduction.} \textbf{a}, The raw light curve of GECAM-C/GRD01 (blue solid line), GECAM-C/CPD01 (red solid line) and the background estimation (orange solid line). The light curve data points within the orange dashed line are used to fit the background model, which is from $T_0-21.6$ s to $T_0-0.1$ s and $T_0+170$ s to $T_0+600$ s, then the background estimation can be obtained by interpolating the fitted model over the burst time interval. The fit starting from $T_0-21.6$ s is due to data saturation before $T_0-21.6$ s caused by the particles in orbit. \textbf{b}, Comparison of the net light curves of GECAM-B and GECAM-C. The consistency between these two instruments validates the background estimation of GECAM-C. \textbf{c}, Comparison of recorded event rate (evt data, solid line) and processed event rate (ope engineering data, dashed line) of GECAM-B and GECAM-C for GRB 230307A. It is shown that the difference between these two event rates is negligible, demonstrating that both GECAM-B and GECAM-C have no data saturation for GRB 230307A.}
\label{fig:gecam_reduction}
\end{figure}

\clearpage

\begin{figure}
\centering
\begin{tabular}{cc}
\begin{overpic}[width=0.45\textwidth]{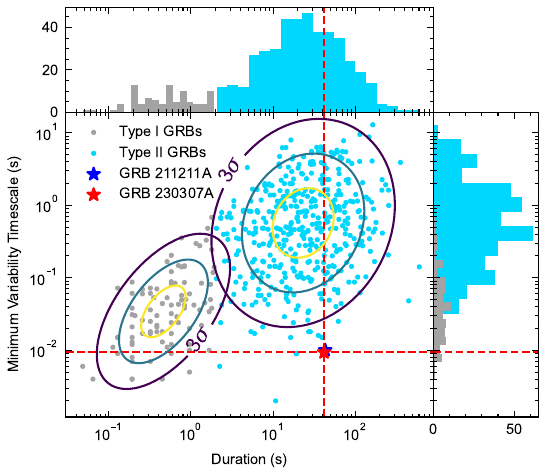}\put(2, 80){\bf a}\end{overpic} &
\begin{overpic}[width=0.45\textwidth]{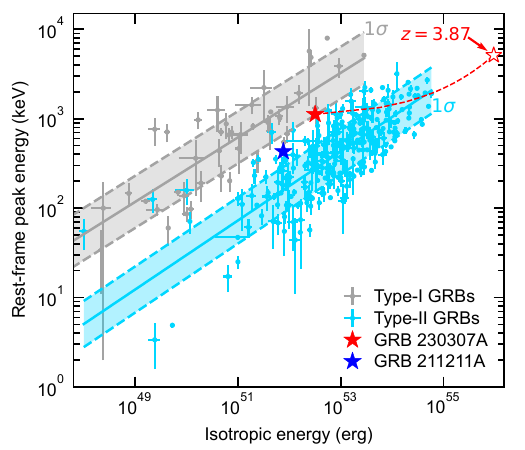}\put(2, 80){\bf b}\end{overpic} \\
\begin{overpic}[width=0.45\textwidth]{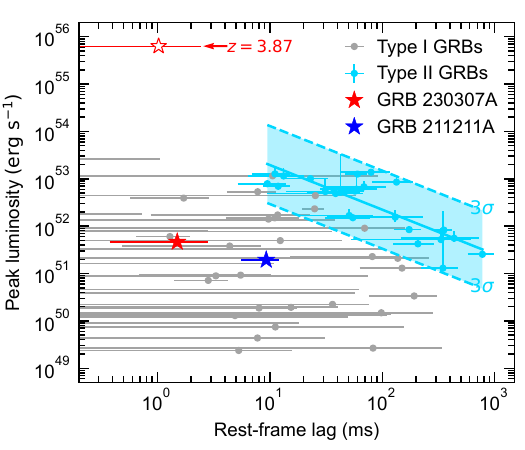}\put(2, 80){\bf c}\end{overpic} &
\begin{overpic}[width=0.45\textwidth]{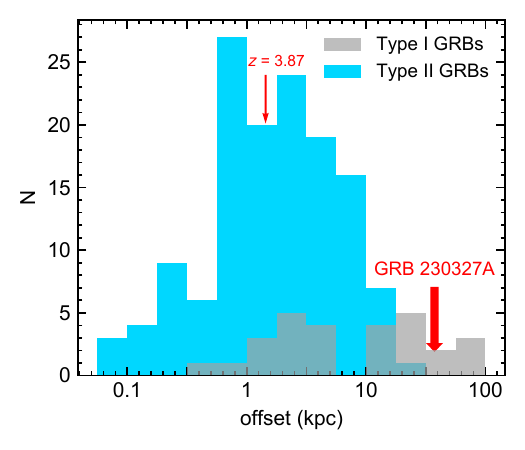}\put(2, 80){\bf d}\end{overpic} \\
\end{tabular}
\caption{\noindent\textbf{Position of GRB 230307A in various GRB classification schemes.} \textbf{a}, The minimum variability timescale versus duration diagram. \textbf{b}, The rest-frame peak energy versus isotropic energy correlation diagram. The best-fit correlations and 1$\sigma$ regions of intrinsic scatter are presented by solid lines and shaded areas, respectively. \textbf{c}, The peak luminosity versus rest-frame lag diagram. The best-fit correlation and 3$\sigma$ region of intrinsic scatter for type II GRBs are presented by cyan line and shaded area, respectively. In \textbf{a}--\textbf{c}, type I and type II GRBs are represented by grey and cyan solid circles, respectively. GRB 230307A ($z=0.065$) and GRB 211211A are highlighted by red and blue filled stars, respectively. GRB 230307A ($z=3.87$) is marked by a red unfilled star. \textbf{d}, The distribution of the physical offset of type I and type II GRBs from the center of the host galaxy\cite{Li2016ApJS}. GRB 230307A is overplotted in red, showing its consistency with type I GRBs. In all the figures, the cyan and grey colors represent type II and type I GRBs, respectively, with the assumption that all the bursts under each color belong to their physical categories, namely, with massive star core collapse and compact star merger origins, respectively. All error bars on data points represent their 1$\sigma$ confidence level.}
\label{fig:classification}
\end{figure}

\clearpage

\begin{figure}
\centering
\begin{tabular}{ccc}
\begin{overpic}[width=0.32\textwidth]{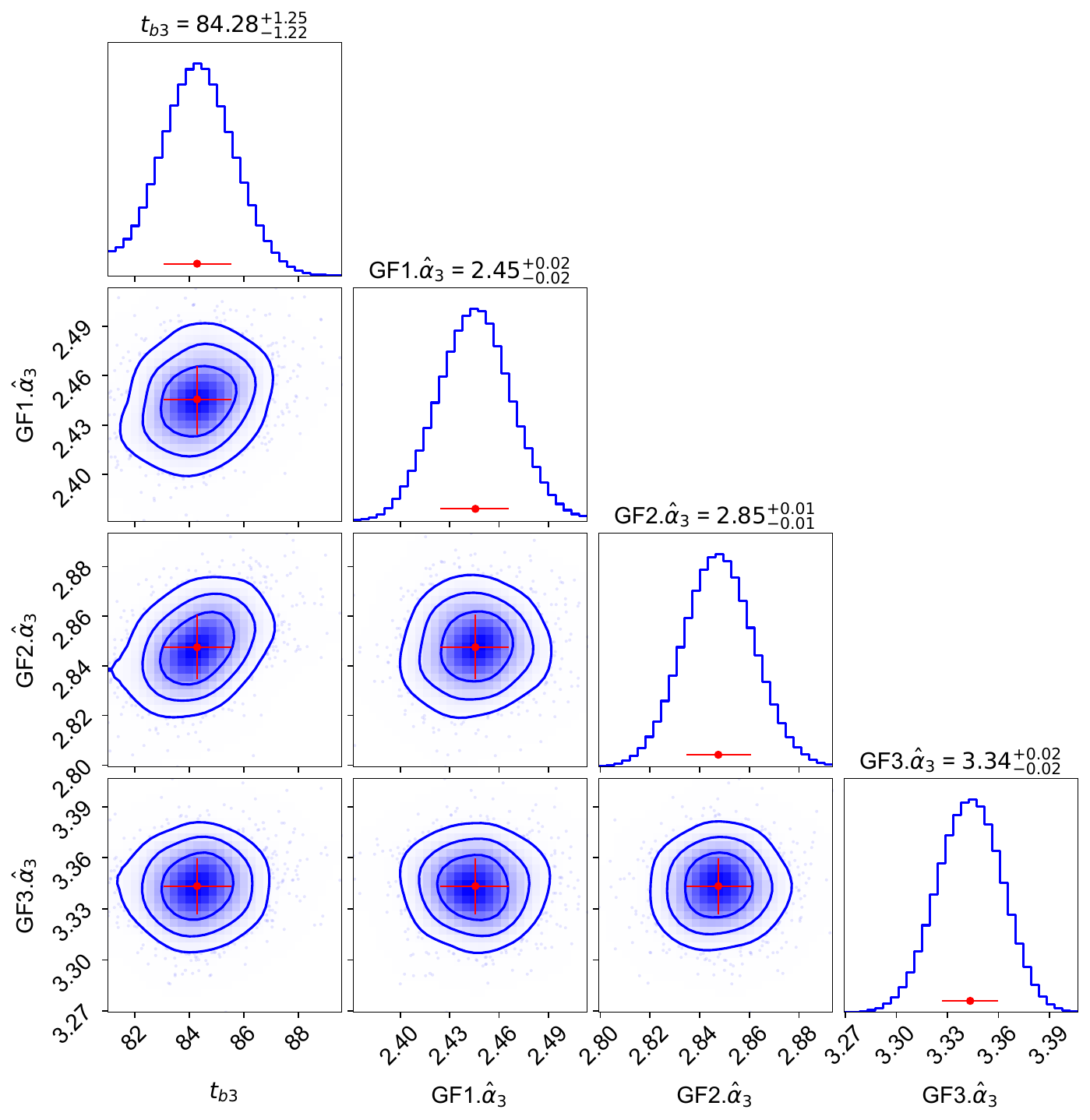}\put(2, 95){\bf a}\end{overpic}
\begin{overpic}[width=0.32\textwidth]{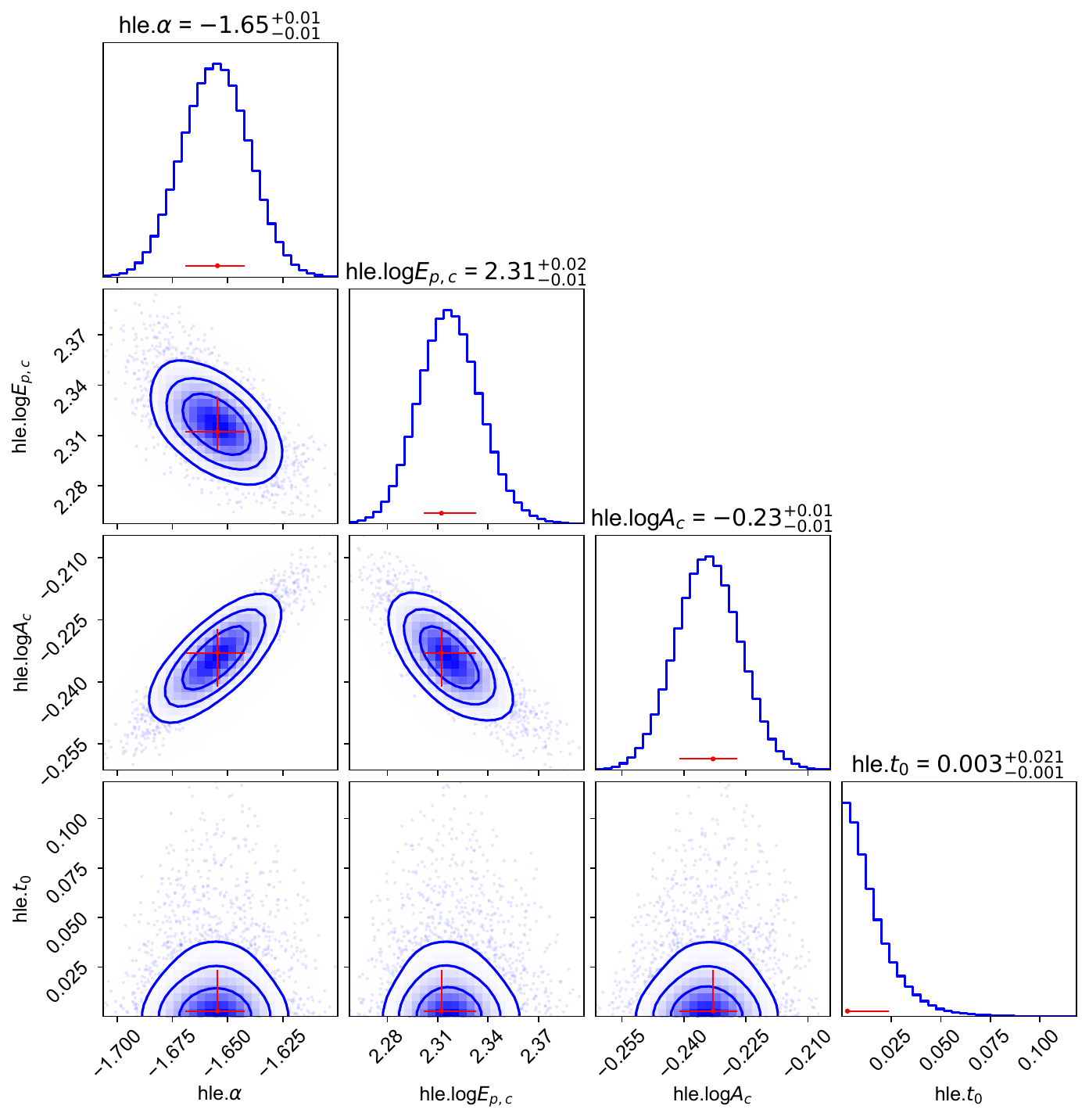}\put(2, 95){\bf b}\end{overpic}
\begin{overpic}[width=0.3\textwidth]{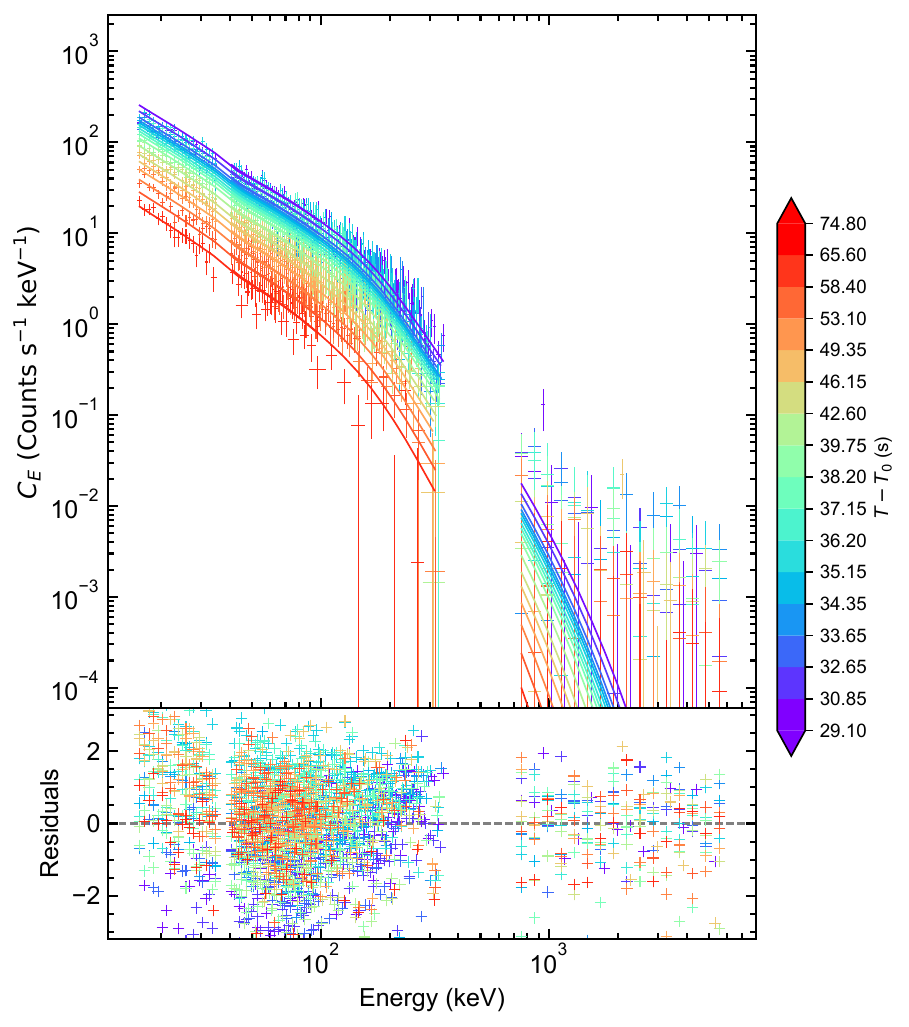}\put(1, 95){\bf c}\end{overpic}
\end{tabular}
\caption{\noindent\textbf{The SBPL model and curvature effect model fits to observed GECAM data.} \textbf{a}, The corner plot of the posterior probability distributions of the parameters for the global SBPL fit to GECAM multi-wavelength (including 15--30, 30--100 and 100--350 keV, denoted by GF1, GF2 and GF3, respectively) flux light curves. It should be noted that for visual clarity, only the achromatic break time ($t_{\rm b3}$) and its previous decay indices ($\hat{\alpha}_3$) are displayed. The red error bars represent 1$\sigma$ uncertainties. \textbf{b}, The corner plot of the posterior probability distributions of the parameters of curvature effect model (denoted by ``hle'') for the fit to the GECAM S-II spectra. The red error bars represent 1$\sigma$ uncertainties. \textbf{c}, The comparison between observed (points) and model-predicted (lines) count spectra as well as the residuals (pluses) based on the fit of curvature effect model to the GECAM S-II spectra.}
\label{fig:lc_sbpl}
\end{figure}

\clearpage

\begin{figure}
\centering
\begin{tabular}{cc}
\begin{overpic}[width=0.40\textwidth]{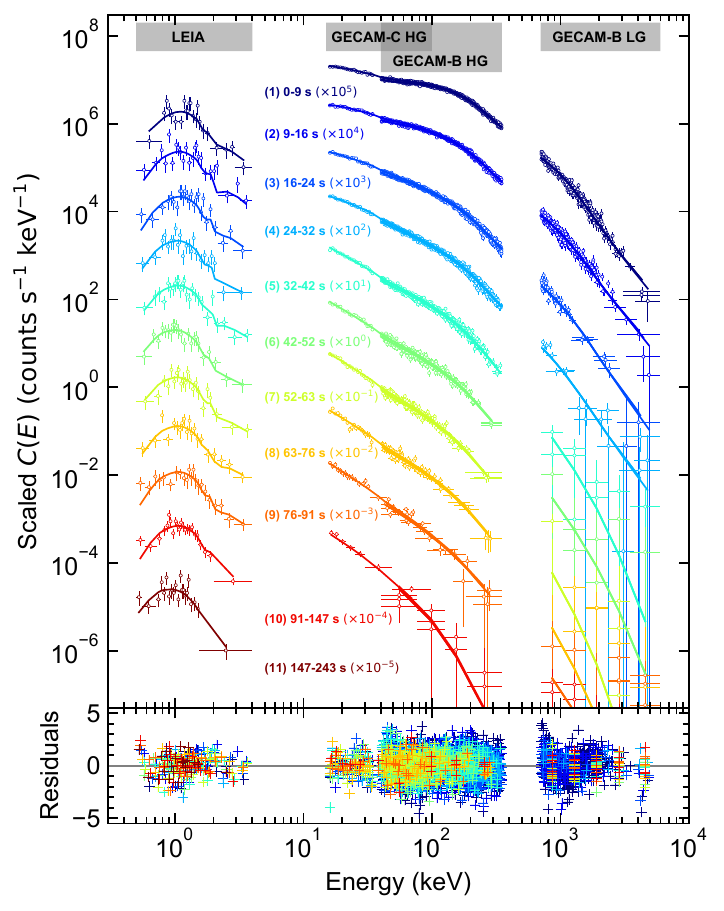}\put(2, 95){\bf a}\end{overpic} & 
\begin{overpic}[width=0.40\textwidth]{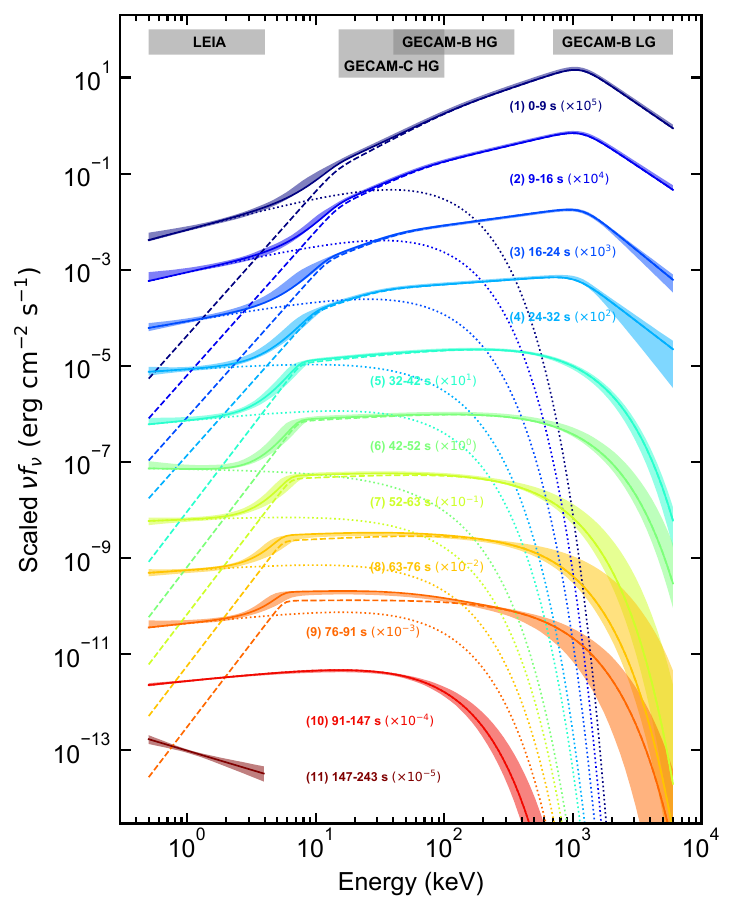}\put(2, 95){\bf b}\end{overpic} \\
\multicolumn{2}{c}{\begin{overpic}[width=0.5\textwidth]{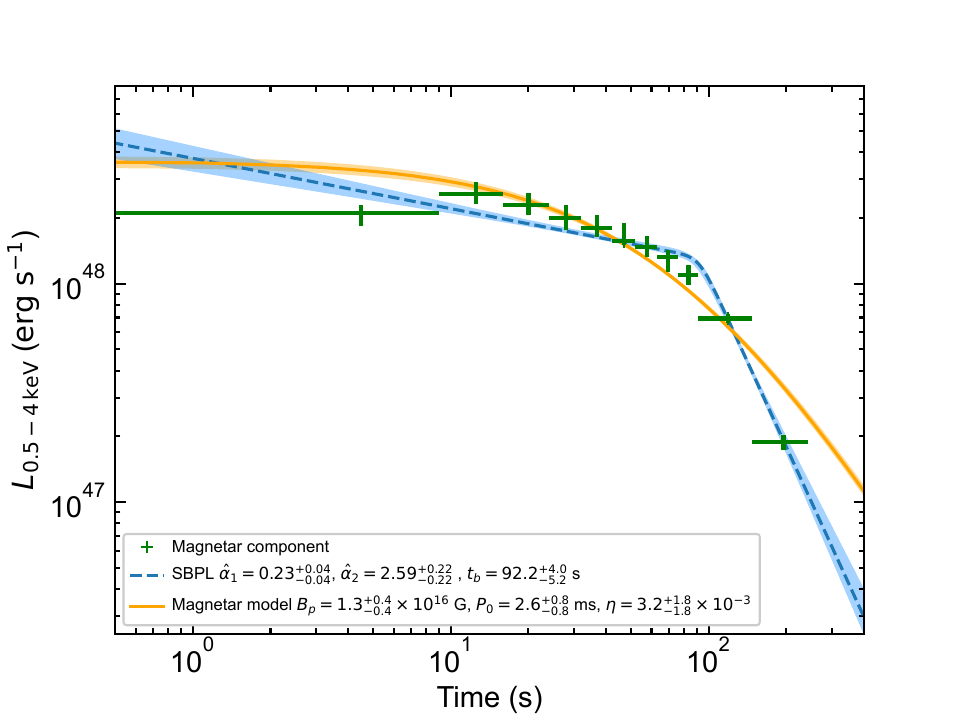}\put(2, 65){\bf c}\end{overpic}} \\
\end{tabular}
\caption{\noindent\textbf{Fit of spectral energy distribution.} \textbf{a}, The comparison between observed (points) and model-predicted (solid lines) count spectra as well as the residuals (pluses) based on the joint spectral fittings of LEIA S-III and GECAM S-III. \textbf{b}, The model-predicted SEDs derived from the joint spectral fittings of LEIA S-III and GECAM S-III. The high-energy (jet) and low-energy (magnetar) spectral components are represented with the dashed and dotted lines, respectively. The superimposed models are shown with the solid lines. \textbf{c}, The X-ray light curve of the magnetar component after the subtraction of the prompt emission. The green data are calculated from the low-energy CPL modelled magnetar-only component. The yellow and blue dashed lines represent the  magnetar dipole radiation model and smoothly broken power law model fitted to those calculated magnetar-only data. All error bars and shaded areas represent 1$\sigma$ uncertainties.}
\label{fig:SED}
\end{figure}

\clearpage

\begin{figure}
\centering
\begin{tabular}{c}
\includegraphics[width=\textwidth]{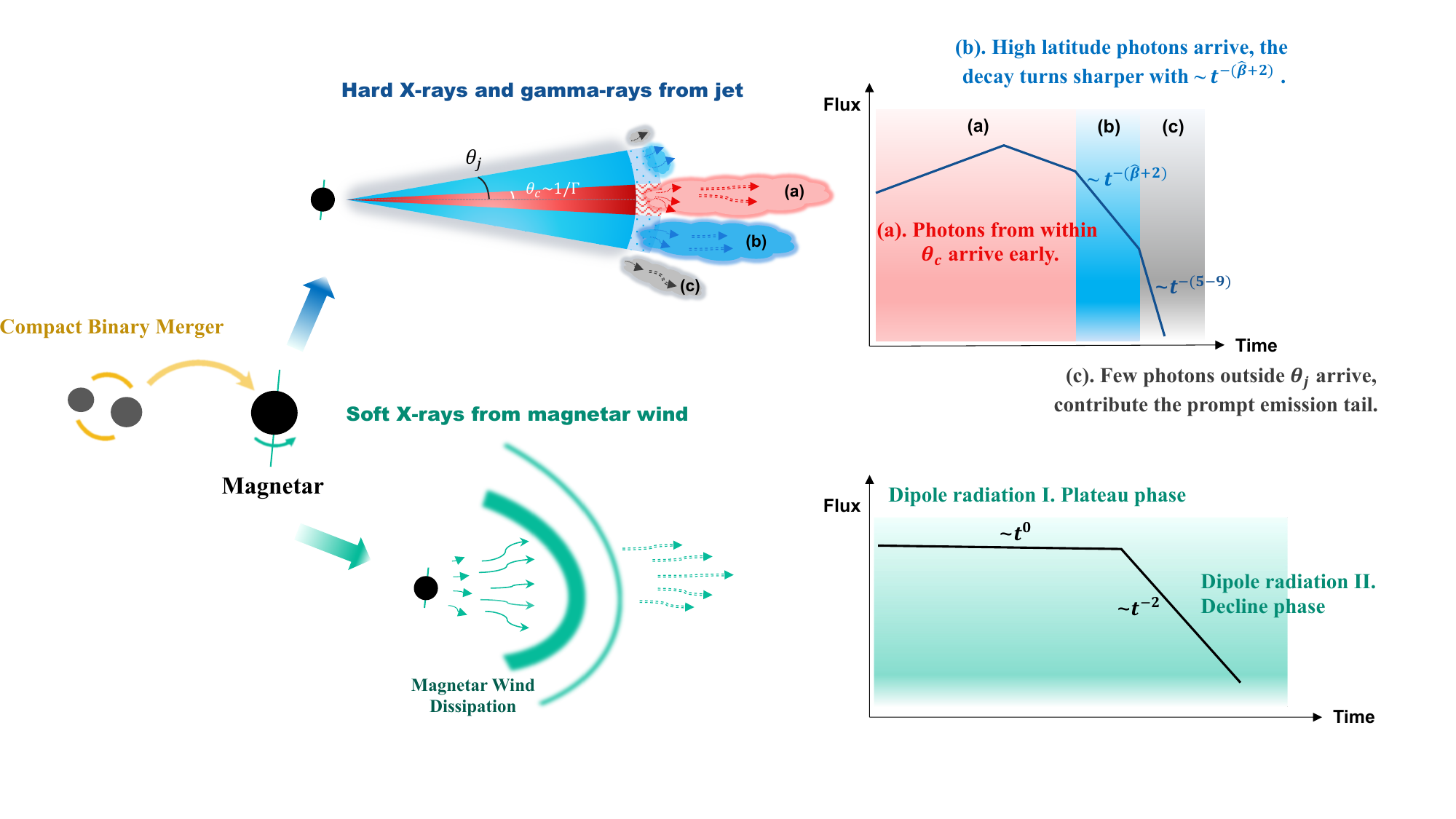} \\ 
\end{tabular}
\caption{\noindent\textbf{Schematic diagram of the two-component prompt emission of GRB 230307A.} GRB 230307A is suggested to be powered by a millisecond magnetar resulting from a compact binary star merger. In such a scenario, the hard X-rays and gamma-ray are powered by a narrow jet. The early bright emission of the light curve is mainly contributed by photons from the jet within a cone of $\theta_c \sim 1/ \Gamma$ at the emission radius $R_{\rm GRB}$ (Phase a). When the jet ceases, photons from higher-latitude parts arrive at later times, resulting in a decay in the light curve with a slope consistent with the curvature effect prediction (Phase b). After an achromatic break signaling the end of the curvature effect, the light curve decays with an even steeper slope. This enables us to estimate the jet opening angle, for the first time, during the prompt emission phase of a GRB. The tail of the prompt emission is possibly contributed by some weak emission from regions beyond the jet opening angle (Phase c). In contrast, the soft X-ray emission is powered by a presumably more isotropic magnetar wind. The light curve is consistent with prediction of the spin-down law with a plateau followed by a shallow decline. All the light curves are displayed in logarithmic coordinate space.}
\label{fig:schematic}
\end{figure}

\clearpage

\begin{figure}
\centering
\begin{tabular}{c}
\includegraphics[width=0.8\textwidth]{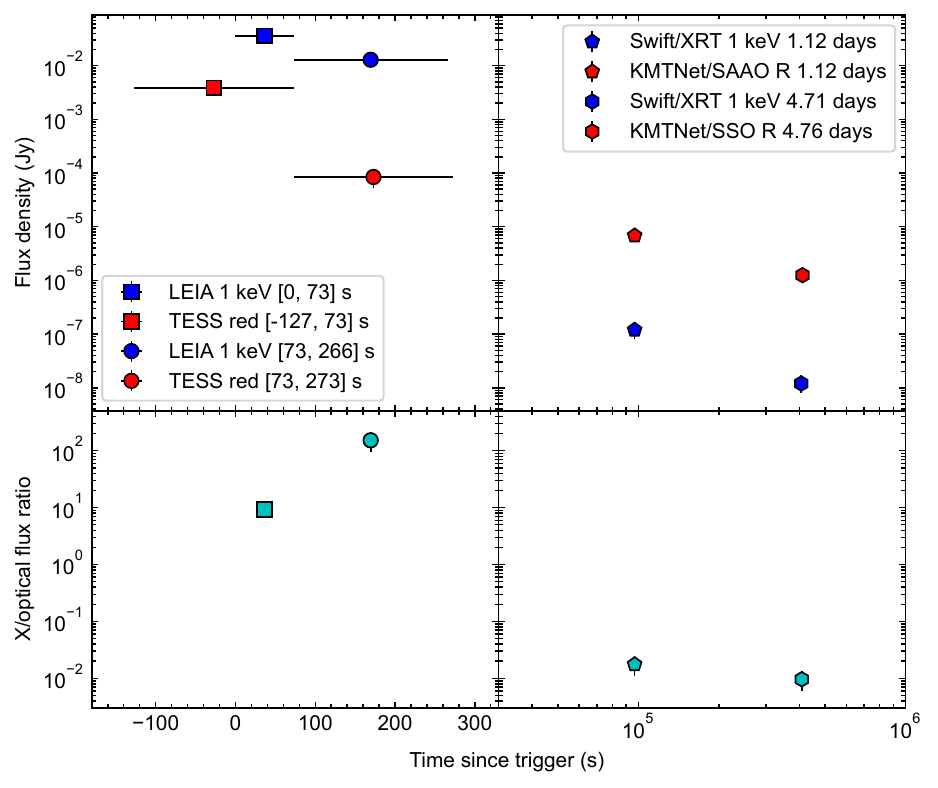} \\ 
\end{tabular}
\caption{\noindent\textbf{The flux density ratios between X-ray and optical bands for GRB 230307A}. GRB 230307A was simultaneously detected in the X-ray (blue markers) and optical (red markers) bands during the prompt emission (LEIA and TESS\cite{Fausnaugh2023RNAAS}) and the following afterglow phases (e.g., Swift/XRT and KMTNet\cite{YHYang2023prep}). The flux density ratios during the early prompt emission phase are more than three orders of magnitude higher than those during later times. The abnormally large ratios during the first two epochs imply that there is an additional component contributing to the early X-ray data. This provides strong evidence that the soft X-ray observations from LEIA do not originate from the afterglow emission.}
\label{fig:xoptratio}
\end{figure}

\clearpage

\begin{figure}
\centering
\begin{tabular}{cc}
\begin{overpic}[width=0.5\textwidth]{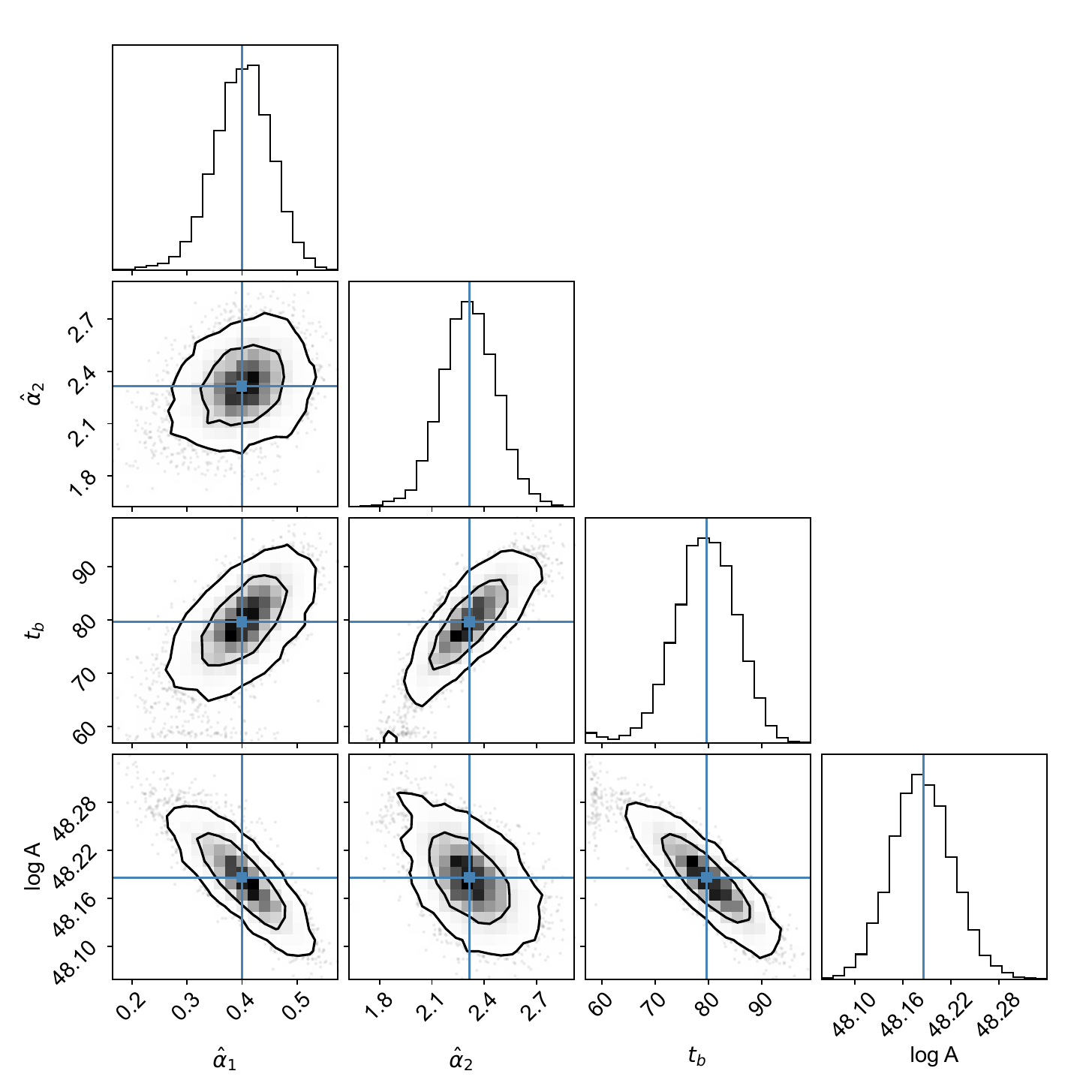}\put(2, 95){\bf a}\end{overpic} & 
\begin{overpic}[width=0.5\textwidth]{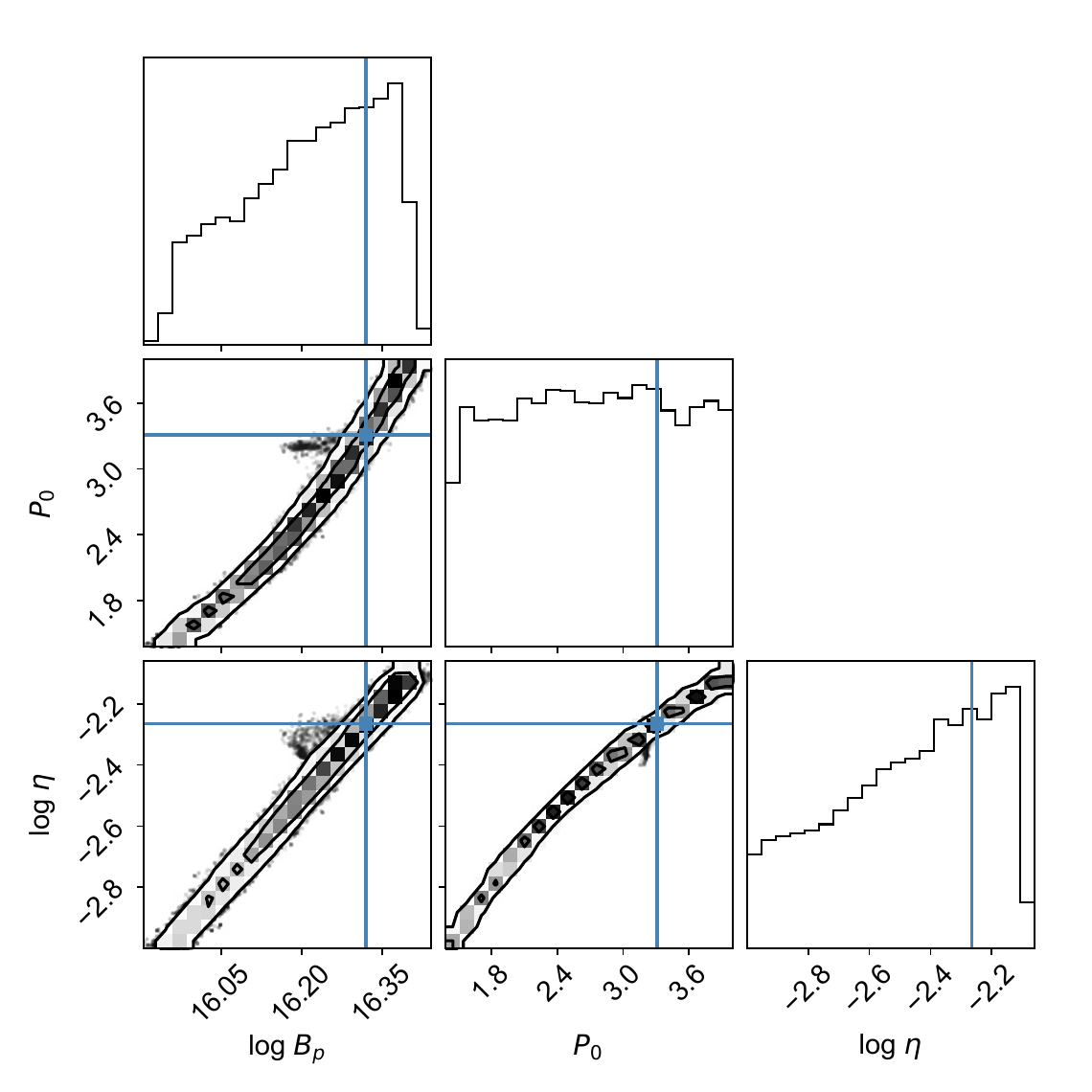}\put(2, 95){\bf b}\end{overpic} 
\end{tabular}
\caption{\noindent\textbf{Corner plots of the posterior probability distributions of the parameters for the fit of the X-ray light curve (0.5--4 keV).} \textbf{a}, Fit with the smoothly broken power law model. \textbf{b}, Fit with the magnetar dipole radiation model.}
\label{fig:magnetar2}
\end{figure}

\clearpage

\begin{table*}
\centering
\scriptsize
\caption{\textbf{Spectral fitting results and corresponding fitting statistics for LEIA S-II spectra.} All errors represent the 1$\sigma$ uncertainties.}
\label{tab:leia_spec_fit}
\begin{tabular}{cccccc}
\hline
$t_1$ & $t_2$ & $\Gamma_{\rm ph}$ & Normalization & Unabsorbed flux & CSTAT/dof \\
(s) & (s) & & (${\rm photons~cm^{-2}~s^{-1}~keV^{-1}}$) & (${\rm erg~cm^{-2}~s^{-1}}$) &\\
\hline
$0$ & $6$ & $0.99_{-0.35}^{+0.36}$ & $29.65_{-4.54}^{+5.01}$ & $1.68_{-0.27}^{+0.33}\times 10^{-7}$ & $7.64/6$ \\
$6$ & $9$ & $1.15_{-0.35}^{+0.36}$ & $70.88_{-12.07}^{+13.87}$ &$3.60_{-0.46}^{+0.55}\times 10^{-7}$ & $15.87/7$ \\
$9$ & $13$ & $1.43_{-0.28}^{+0.29}$ & $66.73_{-8.29}^{+9.08}$ & $2.87_{-0.33}^{+0.38}\times 10^{-7}$& $9.67/9$ \\
$13$ & $17$ & $1.03_{-0.24}^{+0.24}$ & $52.53_{-6.91}^{+7.45}$ & $2.89_{-0.34}^{+0.38}\times 10^{-7}$& $3.11/9$ \\
$17$ & $21$ & $0.96_{-0.29}^{+0.30}$ & $44.93_{-7.29}^{+8.22}$ & $2.58_{-0.35}^{+0.39}\times 10^{-7}$& $9.95/7$ \\
$21$ & $25$ & $1.38_{-0.33}^{+0.33}$ & $55.87_{-7.79}^{+8.51}$ & $2.43_{-0.31}^{+0.37}\times 10^{-7}$& $10.01/7$ \\
$25$ & $29$ & $1.80_{-0.32}^{+0.33}$ & $58.68_{-7.40}^{+8.06}$ & $2.11_{-0.24}^{+0.28}\times 10^{-7}$& $7.77/6$ \\
$29$ & $34$ & $1.59_{-0.29}^{+0.30}$ & $50.42_{-6.02}^{+6.52}$ & $2.00_{-0.23}^{+0.27}\times 10^{-7}$& $6.36/8$ \\
$34$ & $39$ & $1.30_{-0.26}^{+0.27}$ & $48.26_{-6.24}^{+6.84}$ & $2.23_{-0.25}^{+0.28}\times 10^{-7}$& $1.75/9$ \\
$39$ & $45$ & $1.54_{-0.32}^{+0.34}$ & $40.59_{-5.11}^{+5.57}$ & $1.64_{-0.21}^{+0.24}\times 10^{-7}$& $6.32/8$ \\
$45$ & $50$ & $2.10_{-0.39}^{+0.41}$ & $52.03_{-6.68}^{+7.14}$ & $1.68_{-0.20}^{+0.23}\times 10^{-7}$& $7.58/7$ \\
$50$ & $55$ & $1.30_{-0.26}^{+0.26}$ & $47.19_{-6.31}^{+6.43}$ & $2.20_{-0.25}^{+0.28}\times 10^{-7}$& $18.18/8$ \\
$55$ & $61$ & $1.76_{-0.38}^{+0.39}$ & $41.07_{-5.77}^{+6.38}$ & $1.50_{-0.18}^{+0.22}\times 10^{-7}$& $2.88/6$ \\
$61$ & $68$ & $1.39_{-0.27}^{+0.28}$ & $35.74_{-4.57}^{+4.97}$ & $1.57_{-0.18}^{+0.20}\times 10^{-7}$& $9.96/9$ \\
$68$ & $76$ & $1.23_{-0.43}^{+0.46}$ & $25.73_{-4.73}^{+5.33}$ &$1.24_{-0.19}^{+0.23}\times 10^{-7}$ & $12.58/7$ \\
$76$ & $84$ & $1.30_{-0.30}^{+0.31}$ & $25.48_{-3.49}^{+3.78}$ & $1.18_{-0.15}^{+0.18}\times 10^{-7}$& $5.08/7$ \\
$84$ & $92$ & $1.49_{-0.31}^{+0.31}$ & $29.16_{-3.67}^{+4.00}$ & $1.21_{-0.15}^{+0.18}\times 10^{-7}$& $4.26/7$ \\
$92$ & $103$ & $1.40_{-0.32}^{+0.34}$ & $19.38_{-2.48}^{+2.71}$ &$8.47_{-1.17}^{+1.40}\times 10^{-8}$ & $3.55/7$ \\
$103$ & $116$ & $2.02_{-0.36}^{+0.38}$ & $18.84_{-2.36}^{+2.54}$ & $6.23_{-0.77}^{+0.88}\times 10^{-8}$& $3.84/7$ \\
$116$ & $131$ & $2.14_{-0.33}^{+0.35}$ & $16.48_{-2.06}^{+2.21}$ & $5.25_{-0.60}^{+0.65}\times 10^{-8}$& $5.31/7$ \\
$131$ & $153$ & $2.10_{-0.41}^{+0.43}$ & $13.27_{-1.63}^{+1.77}$ &$4.28_{-0.48}^{+0.58}\times 10^{-8}$ & $1.95/7$ \\
$153$ & $187$ & $2.41_{-0.38}^{+0.39}$ & $7.96_{-0.89}^{+0.95}$ & $2.37_{-0.26}^{+0.29}\times 10^{-8}$& $7.90/8$ \\
$187$ & $266$ & $3.16_{-0.36}^{+0.37}$ & $4.33_{-0.45}^{+0.47}$ &$1.22_{-0.13}^{+0.14}\times 10^{-8}$ & $7.04/10$ \\
\hline
\end{tabular}
\end{table*}

\clearpage

\begingroup\tiny
\begin{ThreePartTable}
\begin{TableNotes} 
\item[*] In cases where the 1$\sigma$ lower limit of $\alpha_2(\alpha)$ nears or falls below $-2$, we use cutoff energy as a substitute for peak energy.
\end{TableNotes} 
\begin{longtable}{cccccccccc}
\caption{\textbf{Spectral fitting results and corresponding fitting statistics for GECAM S-I and S-II spectra.} All errors represent the 1$\sigma$ uncertainties.}
\label{tab:gecam_spec_fit} \\
\hline
$t_1$ & $t_2$ & Best model & $\alpha_1$ & $\alpha_2(\alpha)$ & $\beta$ & ${\rm log}E_{\rm b}$ & ${\rm log}E_{\rm p}({\rm log}E_{\rm c})$ & ${\rm log}A$ & PGSTAT/dof \\
(s) & (s) & & & & & (${\rm keV}$) & (${\rm keV}$) & (${\rm photons~cm^{-2}~s^{-1}~keV^{-1}}$) & \\
\hline
\endhead 
\multicolumn{6}{r}{\textit{continued}}
\endfoot
\insertTableNotes \\ 
\endlastfoot 
\multicolumn{2}{c}{\textbf{GECAM S-I:}} & & & & & & & & \\
$1.6$ & $43.2$ & 2SBPL & $-0.92_{-0.03}^{+0.05}$ & $-1.274_{-0.008}^{+0.005}$ & $-3.85_{-0.09}^{+0.03}$ & $1.39_{-0.03}^{+0.05}$ & $3.022_{-0.004}^{+0.007}$ & $1.83_{-0.06}^{+0.04}$ & $2209/960$ \\
\hline
\multicolumn{2}{c}{\textbf{GECAM S-II:}} & & & & & & & & \\
$0.0$ & $0.7$ & CPL & -- & $-0.96_{-0.05}^{+0.04}$ & -- & -- & $2.16_{-0.01}^{+0.02}$ & $-0.05_{-0.03}^{+0.03}$ & $854/963$ \\
$0.7$ & $1.0$ & CPL & -- & $-0.57_{-0.03}^{+0.03}$ & -- & -- & $2.76_{-0.02}^{+0.02}$ & $0.02_{-0.01}^{+0.01}$ & $754/963$ \\
$1.0$ & $1.4$ & CPL & -- & $-0.66_{-0.03}^{+0.03}$ & -- & -- & $2.72_{-0.02}^{+0.02}$ & $0.01_{-0.01}^{+0.01}$ & $808/963$ \\
$1.4$ & $1.6$ & CPL & -- & $-0.48_{-0.03}^{+0.02}$ & -- & -- & $3.01_{-0.01}^{+0.02}$ & $0.12_{-0.01}^{+0.01}$ & $886/963$ \\
$1.6$ & $1.8$ & SBPL & -- & $-0.66_{-0.03}^{+0.01}$ & $-3.82_{-0.33}^{+0.14}$ & -- & $3.01_{-0.01}^{+0.02}$ & $1.35_{-0.03}^{+0.07}$ & $835/962$ \\
$1.8$ & $2.0$ & CPL & -- & $-0.25_{-0.03}^{+0.03}$ & -- & -- & $3.10_{-0.01}^{+0.01}$ & $0.12_{-0.01}^{+0.01}$ & $954/963$ \\
$2.0$ & $2.3$ & SBPL & -- & $-0.64_{-0.02}^{+0.02}$ & $-4.07_{-0.33}^{+0.20}$ & -- & $2.86_{-0.01}^{+0.02}$ & $1.28_{-0.04}^{+0.04}$ & $881/962$ \\
$2.3$ & $2.7$ & SBPL & -- & $-0.69_{-0.02}^{+0.01}$ & $-4.22_{-0.25}^{+0.31}$ & -- & $2.83_{-0.02}^{+0.02}$ & $1.33_{-0.02}^{+0.05}$ & $935/962$ \\
$2.7$ & $3.2$ & CPL & -- & $-0.73_{-0.03}^{+0.02}$ & -- & -- & $2.82_{-0.02}^{+0.02}$ & $-0.04_{-0.01}^{+0.01}$ & $929/963$ \\
$3.2$ & $3.4$ & CPL & -- & $-0.42_{-0.02}^{+0.02}$ & -- & -- & $3.08_{-0.01}^{+0.01}$ & $0.304_{-0.006}^{+0.005}$ & $905/963$ \\
$3.4$ & $3.6$ & CPL & -- & $-0.43_{-0.02}^{+0.03}$ & -- & -- & $3.09_{-0.02}^{+0.01}$ & $0.17_{-0.01}^{+0.01}$ & $955/963$ \\
$3.6$ & $4.0$ & CPL & -- & $-0.58_{-0.02}^{+0.03}$ & -- & -- & $2.82_{-0.02}^{+0.01}$ & $0.06_{-0.01}^{+0.01}$ & $817/963$ \\
$4.0$ & $4.3$ & SBPL & -- & $-0.81_{-0.02}^{+0.02}$ & $-3.36_{-0.23}^{+0.18}$ & -- & $2.93_{-0.01}^{+0.02}$ & $1.54_{-0.04}^{+0.04}$ & $929/962$ \\
$4.3$ & $4.7$ & CPL & -- & $-0.72_{-0.02}^{+0.02}$ & -- & -- & $2.88_{-0.02}^{+0.02}$ & $0.04_{-0.01}^{+0.01}$ & $873/963$ \\
$4.7$ & $4.9$ & CPL & -- & $-0.54_{-0.03}^{+0.02}$ & -- & -- & $2.98_{-0.01}^{+0.02}$ & $0.22_{-0.01}^{+0.01}$ & $896/963$ \\
$4.9$ & $5.0$ & CPL & -- & $-0.64_{-0.04}^{+0.03}$ & -- & -- & $2.84_{-0.02}^{+0.02}$ & $0.30_{-0.01}^{+0.01}$ & $773/963$ \\
$5.0$ & $5.2$ & CPL & -- & $-0.72_{-0.02}^{+0.02}$ & -- & -- & $3.00_{-0.01}^{+0.02}$ & $0.30_{-0.01}^{+0.01}$ & $896/963$ \\
$5.2$ & $5.4$ & CPL & -- & $-0.65_{-0.02}^{+0.02}$ & -- & -- & $3.01_{-0.02}^{+0.01}$ & $0.32_{-0.01}^{+0.01}$ & $884/963$ \\
$5.4$ & $5.7$ & CPL & -- & $-0.74_{-0.02}^{+0.02}$ & -- & -- & $2.97_{-0.02}^{+0.01}$ & $0.20_{-0.01}^{+0.01}$ & $869/963$ \\
$5.7$ & $5.8$ & CPL & -- & $-0.62_{-0.03}^{+0.03}$ & -- & -- & $3.05_{-0.02}^{+0.02}$ & $0.25_{-0.01}^{+0.01}$ & $821/963$ \\
$5.8$ & $5.9$ & CPL & -- & $-0.52_{-0.03}^{+0.03}$ & -- & -- & $3.13_{-0.02}^{+0.02}$ & $0.27_{-0.01}^{+0.01}$ & $866/963$ \\
$5.9$ & $6.0$ & CPL & -- & $-0.50_{-0.02}^{+0.03}$ & -- & -- & $3.10_{-0.02}^{+0.02}$ & $0.33_{-0.01}^{+0.01}$ & $956/963$ \\
$6.0$ & $6.2$ & CPL & -- & $-0.57_{-0.02}^{+0.02}$ & -- & -- & $3.01_{-0.01}^{+0.01}$ & $0.36_{-0.01}^{+0.01}$ & $859/963$ \\
$6.2$ & $6.3$ & CPL & -- & $-0.56_{-0.02}^{+0.03}$ & -- & -- & $3.09_{-0.02}^{+0.02}$ & $0.42_{-0.01}^{+0.01}$ & $965/963$ \\
$6.3$ & $6.4$ & CPL & -- & $-0.55_{-0.03}^{+0.03}$ & -- & -- & $3.06_{-0.02}^{+0.02}$ & $0.36_{-0.01}^{+0.01}$ & $916/963$ \\
$6.4$ & $6.6$ & CPL & -- & $-0.62_{-0.02}^{+0.02}$ & -- & -- & $3.11_{-0.01}^{+0.01}$ & $0.293_{-0.006}^{+0.005}$ & $994/963$ \\
$6.6$ & $6.7$ & CPL & -- & $-0.53_{-0.03}^{+0.03}$ & -- & -- & $3.12_{-0.02}^{+0.02}$ & $0.36_{-0.01}^{+0.01}$ & $886/963$ \\
$6.7$ & $6.9$ & CPL & -- & $-0.59_{-0.02}^{+0.02}$ & -- & -- & $3.01_{-0.01}^{+0.01}$ & $0.35_{-0.01}^{+0.01}$ & $925/963$ \\
$6.9$ & $7.2$ & CPL & -- & $-0.81_{-0.02}^{+0.02}$ & -- & -- & $2.89_{-0.02}^{+0.02}$ & $0.09_{-0.01}^{+0.01}$ & $812/963$ \\
$7.2$ & $7.4$ & CPL & -- & $-0.75_{-0.02}^{+0.03}$ & -- & -- & $2.91_{-0.02}^{+0.02}$ & $0.20_{-0.01}^{+0.01}$ & $808/963$ \\
$7.4$ & $7.7$ & CBAND & $-0.68_{-0.07}^{+0.05}$ & $-1.27_{-0.17}^{+0.13}$ & -- & $2.29_{-0.13}^{+0.12}$ & $2.82_{-0.03}^{+0.04}$ & $1.61_{-0.08}^{+0.10}$ & $880/961$ \\
$7.7$ & $7.9$ & CSBPL & $-0.43_{-0.14}^{+0.67}$ & $-1.00_{-0.05}^{+0.05}$ & -- & $1.52_{-0.16}^{+0.11}$ & $2.89_{-0.02}^{+0.03}$ & $1.35_{-0.86}^{+0.18}$ & $863/961$ \\
$7.9$ & $8.1$ & CPL & -- & $-0.77_{-0.02}^{+0.02}$ & -- & -- & $3.00_{-0.02}^{+0.02}$ & $0.24_{-0.01}^{+0.01}$ & $882/963$ \\
$8.1$ & $8.3$ & CPL & -- & $-0.73_{-0.02}^{+0.02}$ & -- & -- & $3.07_{-0.02}^{+0.02}$ & $0.27_{-0.01}^{+0.01}$ & $920/963$ \\
$8.3$ & $8.4$ & CPL & -- & $-0.75_{-0.03}^{+0.03}$ & -- & -- & $3.07_{-0.03}^{+0.02}$ & $0.25_{-0.01}^{+0.01}$ & $870/963$ \\
$8.4$ & $8.6$ & CPL & -- & $-0.70_{-0.02}^{+0.02}$ & -- & -- & $3.15_{-0.01}^{+0.02}$ & $0.26_{-0.01}^{+0.01}$ & $976/963$ \\
$8.6$ & $8.9$ & CPL & -- & $-0.74_{-0.02}^{+0.02}$ & -- & -- & $2.96_{-0.02}^{+0.02}$ & $0.19_{-0.01}^{+0.01}$ & $923/963$ \\
$8.9$ & $9.1$ & CPL & -- & $-0.77_{-0.03}^{+0.02}$ & -- & -- & $2.95_{-0.02}^{+0.02}$ & $0.13_{-0.01}^{+0.01}$ & $754/963$ \\
$9.1$ & $9.3$ & CSBPL & $-0.39_{-0.26}^{+0.13}$ & $-1.00_{-0.11}^{+0.04}$ & -- & $1.54_{-0.06}^{+0.31}$ & $2.92_{-0.04}^{+0.03}$ & $1.18_{-0.19}^{+0.35}$ & $782/961$ \\
$9.3$ & $9.5$ & CPL & -- & $-0.75_{-0.02}^{+0.03}$ & -- & -- & $2.88_{-0.02}^{+0.02}$ & $0.25_{-0.01}^{+0.01}$ & $870/963$ \\
$9.5$ & $9.7$ & CPL & -- & $-0.87_{-0.02}^{+0.02}$ & -- & -- & $3.04_{-0.02}^{+0.02}$ & $0.14_{-0.01}^{+0.01}$ & $933/963$ \\
$9.7$ & $10.0$ & CPL & -- & $-0.87_{-0.02}^{+0.02}$ & -- & -- & $3.10_{-0.02}^{+0.02}$ & $0.04_{-0.01}^{+0.01}$ & $863/963$ \\
$10.0$ & $10.3$ & 2SBPL & $-0.61_{-0.07}^{+0.02}$ & $-1.26_{-0.06}^{+0.05}$ & $-4.12_{-0.26}^{+0.47}$ & $2.03_{-0.05}^{+0.08}$ & $3.14_{-0.04}^{+0.03}$ & $1.45_{-0.04}^{+0.12}$ & $973/960$ \\
$10.3$ & $10.5$ & CPL & -- & $-0.76_{-0.02}^{+0.03}$ & -- & -- & $2.95_{-0.03}^{+0.02}$ & $0.15_{-0.01}^{+0.01}$ & $893/963$ \\
$10.5$ & $10.7$ & CSBPL & $-0.14_{-0.08}^{+0.52}$ & $-0.96_{-0.01}^{+0.06}$ & -- & $1.57_{-0.12}^{+0.03}$ & $3.00_{-0.04}^{+0.01}$ & $0.85_{-0.70}^{+0.11}$ & $880/961$ \\
$10.7$ & $10.9$ & CPL & -- & $-0.85_{-0.02}^{+0.03}$ & -- & -- & $2.93_{-0.02}^{+0.03}$ & $0.12_{-0.01}^{+0.01}$ & $796/963$ \\
$10.9$ & $11.2$ & CSBPL & $-0.80_{-0.05}^{+0.06}$ & $-1.39_{-0.09}^{+0.13}$ & -- & $2.08_{-0.13}^{+0.07}$ & $2.85_{-0.04}^{+0.03}$ & $1.73_{-0.10}^{+0.08}$ & $885/961$ \\
$11.2$ & $11.4$ & CBAND & $-0.36_{-0.17}^{+0.07}$ & $-1.11_{-0.08}^{+0.04}$ & -- & $1.99_{-0.04}^{+0.13}$ & $2.96_{-0.03}^{+0.03}$ & $1.25_{-0.09}^{+0.22}$ & $849/961$ \\
$11.4$ & $11.7$ & CBAND & $-0.57_{-0.12}^{+0.17}$ & $-1.21_{-0.06}^{+0.04}$ & -- & $1.93_{-0.09}^{+0.12}$ & $2.86_{-0.03}^{+0.04}$ & $1.54_{-0.20}^{+0.15}$ & $862/961$ \\
$11.7$ & $12.0$ & CPL & -- & $-1.05_{-0.01}^{+0.02}$ & -- & -- & $2.95_{-0.03}^{+0.02}$ & $0.09_{-0.01}^{+0.01}$ & $935/963$ \\
$12.0$ & $12.2$ & CPL & -- & $-1.04_{-0.02}^{+0.03}$ & -- & -- & $2.96_{-0.04}^{+0.03}$ & $0.06_{-0.01}^{+0.01}$ & $842/963$ \\
$12.2$ & $12.6$ & 2SBPL & $-0.84_{-0.03}^{+0.08}$ & $-1.47_{-0.03}^{+0.07}$ & $-3.89_{-0.94}^{+0.19}$ & $1.94_{-0.10}^{+0.04}$ & $3.06_{-0.03}^{+0.07}$ & $1.81_{-0.13}^{+0.05}$ & $910/960$ \\
$12.6$ & $12.9$ & 2SBPL & $-0.74_{-0.04}^{+0.14}$ & $-1.36_{-0.05}^{+0.04}$ & $-4.11_{-0.83}^{+0.33}$ & $1.79_{-0.09}^{+0.08}$ & $3.00_{-0.04}^{+0.05}$ & $1.70_{-0.21}^{+0.06}$ & $896/960$ \\
$12.9$ & $13.1$ & CSBPL & $-0.52_{-0.15}^{+0.10}$ & $-1.15_{-0.08}^{+0.04}$ & -- & $1.64_{-0.09}^{+0.14}$ & $2.85_{-0.02}^{+0.03}$ & $1.56_{-0.13}^{+0.21}$ & $841/961$ \\
$13.1$ & $13.3$ & 2SBPL & $-0.56_{-0.11}^{+0.09}$ & $-1.39_{-0.04}^{+0.03}$ & $-4.89_{-0.76}^{+0.48}$ & $1.71_{-0.06}^{+0.06}$ & $3.05_{-0.04}^{+0.04}$ & $1.56_{-0.13}^{+0.17}$ & $860/960$ \\
$13.3$ & $13.5$ & 2SBPL & $-0.87_{-0.09}^{+0.04}$ & $-1.55_{-0.06}^{+0.06}$ & $-5.07_{-0.50}^{+1.10}$ & $1.93_{-0.07}^{+0.10}$ & $3.17_{-0.09}^{+0.04}$ & $1.96_{-0.06}^{+0.14}$ & $847/960$ \\
$13.5$ & $14.0$ & CBAND & $-0.31_{-0.32}^{+0.06}$ & $-1.29_{-0.04}^{+0.03}$ & -- & $1.77_{-0.03}^{+0.11}$ & $2.88_{-0.03}^{+0.04}$ & $1.26_{-0.07}^{+0.35}$ & $885/961$ \\
$14.0$ & $14.4$ & 2SBPL & $-0.63_{-0.05}^{+0.22}$ & $-1.39_{-0.02}^{+0.03}$ & $-4.17_{-0.48}^{+0.58}$ & $1.61_{-0.08}^{+0.03}$ & $3.02_{-0.03}^{+0.04}$ & $1.59_{-0.30}^{+0.08}$ & $948/960$ \\
$14.4$ & $14.9$ & CBAND & $-0.55_{-0.17}^{+0.06}$ & $-1.53_{-0.06}^{+0.03}$ & -- & $1.89_{-0.03}^{+0.09}$ & $2.80_{-0.04}^{+0.04}$ & $1.61_{-0.08}^{+0.20}$ & $884/961$ \\
$14.9$ & $15.3$ & CSBPL & $-0.57_{-0.20}^{+0.22}$ & $-1.28_{-0.07}^{+0.03}$ & -- & $1.51_{-0.08}^{+0.12}$ & $2.80_{-0.03}^{+0.04}$ & $1.57_{-0.29}^{+0.27}$ & $817/961$ \\
$15.3$ & $15.9$ & CBAND & $-0.63_{-0.10}^{+0.26}$ & $-1.46_{-0.02}^{+0.05}$ & -- & $1.84_{-0.10}^{+0.04}$ & $2.94_{-0.06}^{+0.02}$ & $1.71_{-0.29}^{+0.12}$ & $883/961$ \\
$15.9$ & $16.6$ & 2SBPL & $-0.82_{-0.08}^{+0.13}$ & $-1.46_{-0.03}^{+0.02}$ & $-4.29_{-1.20}^{+0.06}$ & $1.61_{-0.07}^{+0.07}$ & $3.02_{-0.01}^{+0.07}$ & $1.74_{-0.18}^{+0.11}$ & $869/960$ \\
$16.6$ & $17.4$ & CBAND & $-0.31_{-0.23}^{+0.61}$ & $-1.47_{-0.01}^{+0.04}$ & -- & $1.67_{-0.10}^{+0.03}$ & $2.83_{-0.05}^{+0.03}$ & $1.32_{-0.59}^{+0.25}$ & $841/961$ \\
$17.4$ & $18.6$ & CPL & -- & $-1.51_{-0.03}^{+0.03}$ & -- & -- & $2.57_{-0.06}^{+0.07}$ & $-0.55_{-0.02}^{+0.02}$ & $939/963$ \\
$18.6$ & $19.3$ & CBAND & $-0.06_{-0.37}^{+0.42}$ & $-1.68_{-0.04}^{+0.03}$ & -- & $1.63_{-0.05}^{+0.05}$ & $2.71_{-0.06}^{+0.13}$ & $1.02_{-0.41}^{+0.41}$ & $871/961$ \\
$19.3$ & $19.9$ & CSBPL & $-0.90_{-0.11}^{+0.10}$ & $-1.60_{-0.05}^{+0.05}$ & -- & $1.64_{-0.07}^{+0.08}$ & $2.79_{-0.04}^{+0.08}$ & $1.89_{-0.14}^{+0.16}$ & $900/961$ \\
$19.9$ & $20.4$ & CBAND & $-0.63_{-0.13}^{+0.16}$ & $-1.52_{-0.04}^{+0.03}$ & -- & $1.83_{-0.05}^{+0.07}$ & $3.07_{-0.08}^{+0.04}$ & $1.69_{-0.18}^{+0.15}$ & $923/961$ \\
$20.4$ & $21.1$ & CBAND & $-0.81_{-0.12}^{+0.54}$ & $-1.39_{-0.02}^{+0.05}$ & -- & $1.78_{-0.19}^{+0.06}$ & $2.84_{-0.05}^{+0.03}$ & $1.77_{-0.56}^{+0.14}$ & $930/961$ \\
$21.1$ & $21.5$ & CBAND & $-0.56_{-0.20}^{+0.40}$ & $-1.38_{-0.04}^{+0.03}$ & -- & $1.76_{-0.11}^{+0.08}$ & $2.87_{-0.02}^{+0.07}$ & $1.58_{-0.42}^{+0.24}$ & $831/961$ \\
$21.5$ & $22.1$ & CBAND & $-0.87_{-0.14}^{+0.16}$ & $-1.41_{-0.04}^{+0.04}$ & -- & $1.80_{-0.08}^{+0.12}$ & $2.82_{-0.04}^{+0.04}$ & $1.90_{-0.19}^{+0.16}$ & $893/961$ \\
$22.1$ & $22.9$ & 2SBPL & $-0.83_{-0.15}^{+0.11}$ & $-1.63_{-0.04}^{+0.01}$ & $-5.60_{-0.13}^{+1.31}$ & $1.54_{-0.04}^{+0.07}$ & $2.97_{-0.06}^{+0.03}$ & $1.70_{-0.15}^{+0.20}$ & $957/960$ \\
$22.9$ & $23.5$ & CBAND & $-0.88_{-0.13}^{+0.12}$ & $-1.54_{-0.12}^{+0.09}$ & -- & $1.95_{-0.10}^{+0.09}$ & $2.40_{-0.03}^{+0.08}$ & $1.82_{-0.13}^{+0.16}$ & $846/961$ \\
$23.5$ & $24.3$ & CBAND & $-0.56_{-0.12}^{+0.27}$ & $-1.68_{-0.02}^{+0.04}$ & -- & $1.75_{-0.06}^{+0.03}$ & $2.73_{-0.07}^{+0.07}$ & $1.60_{-0.29}^{+0.14}$ & $962/961$ \\
$24.3$ & $25.0$ & CBAND & $-1.08_{-0.08}^{+0.08}$ & $-1.63_{-0.08}^{+0.04}$ & -- & $2.00_{-0.07}^{+0.13}$ & $2.90_{-0.08}^{+0.06}$ & $2.14_{-0.11}^{+0.10}$ & $779/961$ \\
$25.0$ & $25.6$ & CBAND & $-0.55_{-0.23}^{+0.28}$ & $-1.61_{-0.06}^{+0.07}$ & -- & $1.74_{-0.08}^{+0.09}$ & $2.42_{-0.05}^{+0.05}$ & $1.62_{-0.30}^{+0.25}$ & $792/961$ \\
$25.6$ & $26.3$ & CPL & -- & $-1.40_{-0.03}^{+0.04}$ & -- & -- & $2.24_{-0.03}^{+0.03}$ & $-0.20_{-0.02}^{+0.03}$ & $824/963$ \\
$26.3$ & $27.8$ & CBAND & $-0.70_{-0.27}^{+0.21}$ & $-1.93_{-0.04}^{+0.02}$ & -- & $1.62_{-0.03}^{+0.06}$ & $2.38_{-0.11}^{+0.28}$ & $1.88_{-0.21}^{+0.27}$ & $945/961$ \\
$27.8$ & $28.5$ & CBAND & $-0.71_{-0.08}^{+0.58}$ & $-1.77_{-0.03}^{+0.03}$ & -- & $1.70_{-0.10}^{+0.02}$ & $2.96_{-0.07}^{+0.22}$ & $1.86_{-0.59}^{+0.09}$ & $949/961$ \\
$28.5$ & $29.7$ & CPL & -- & $-1.60_{-0.02}^{+0.02}$ & -- & -- & $2.70_{-0.07}^{+0.08}$ & $-0.58_{-0.01}^{+0.01}$ & $992/963$ \\
$29.7$ & $32.0$ & CPL & -- & $-1.68_{-0.02}^{+0.02}$ & -- & -- & $2.51_{-0.06}^{+0.07}$ & $-0.71_{-0.01}^{+0.01}$ & $1027/963$ \\
$32.0$ & $33.3$ & CPL & -- & $-1.83_{-0.06}^{+0.03}$ & -- & -- & $2.03_{-0.04}^{+0.11}$ & $-0.77_{-0.04}^{+0.02}$ & $911/963$ \\
$33.3$ & $34.0$ & CPL & -- & $-1.78_{-0.07}^{+0.02}$ & -- & -- & $2.45_{-0.05}^{+0.53}$ & $-0.75_{-0.04}^{+0.01}$ & $926/963$ \\
$34.0$ & $34.7$ & CPL & -- & $-1.65_{-0.04}^{+0.04}$ & -- & -- & $2.29_{-0.05}^{+0.09}$ & $-0.64_{-0.03}^{+0.02}$ & $864/963$ \\
$34.7$ & $35.6$ & CPL & -- & $-1.68_{-0.04}^{+0.03}$ & -- & -- & $2.41_{-0.06}^{+0.12}$ & $-0.60_{-0.02}^{+0.02}$ & $897/963$ \\
$35.6$ & $36.8$ & CPL & -- & $-1.61_{-0.05}^{+0.05}$ & -- & -- & $2.11_{-0.04}^{+0.05}$ & $-0.68_{-0.03}^{+0.03}$ & $935/963$ \\
$36.8$ & $37.5$ & CPL & -- & $-1.65_{-0.06}^{+0.04}$ & -- & -- & $2.19_{-0.04}^{+0.09}$ & $-0.64_{-0.04}^{+0.02}$ & $884/963$ \\
$37.5$ & $38.9$ & CPL & -- & $-1.64_{-0.04}^{+0.04}$ & -- & -- & $2.23_{-0.05}^{+0.06}$ & $-0.80_{-0.03}^{+0.03}$ & $937/963$ \\
$38.9$ & $40.6$ & CPL & -- & $-1.83_{-0.05}^{+0.02}$ & -- & -- & $2.32_{-0.08}^{+0.29}$ & $-1.03_{-0.03}^{+0.02}$ & $895/963$ \\
$40.6$ & $44.6$ & CPL & -- & $-1.95_{-0.03}^{+0.03}$ & -- & -- & $1.69_{-0.24}^{+0.11}$ & $-1.13_{-0.02}^{+0.02}$ & $1007/963$ \\
$44.6$ & $47.7$ & CPL & -- & $-1.88_{-0.04}^{+0.03}$ & -- & -- & $2.00_{-0.07}^{+0.12}$ & $-1.14_{-0.03}^{+0.02}$ & $954/963$ \\
$47.7$ & $51.0$ & CPL & -- & $-1.73_{-0.05}^{+0.04}$ & -- & -- & $2.16_{-0.05}^{+0.09}$ & $-1.08_{-0.03}^{+0.02}$ & $996/963$ \\
$51.0$ & $55.2$ & CPL & -- & $-1.86_{-0.06}^{+0.04}$ & -- & -- & $1.70_{-0.11}^{+0.06}$ & $-1.24_{-0.04}^{+0.03}$ & $1016/963$ \\
$55.2$ & $61.6$ & CPL & -- & $-1.94_{-0.03}^{+0.06}$ & -- & -- & $1.35_{-0.20}^{+0.19}$ & $-1.39_{-0.02}^{+0.04}$ & $979/963$ \\
$61.6$ & $69.6$ & CPL & -- & $-1.94_{-0.09}^{+0.04}$ & -- & -- & ($2.66_{-0.10}^{+0.39}$)\tnote{*} & $-1.55_{-0.06}^{+0.03}$ & $988/963$ \\
$69.6$ & $80.0$ & CPL & -- & $-1.96_{-0.15}^{+0.07}$ & -- & -- & ($2.42_{-0.12}^{+0.52}$) & $-1.81_{-0.12}^{+0.05}$ & $944/963$ \\
$80.0$ & $90.0$ & PL & -- & $-2.18_{-0.05}^{+0.04}$ & -- & -- & -- & $-2.05_{-0.02}^{+0.02}$ & $910/964$ \\
$90.0$ & $100.0$ & PL & -- & $-2.33_{-0.07}^{+0.08}$ & -- & -- & -- & $-2.36_{-0.04}^{+0.04}$ & $908/964$ \\
$100.0$ & $115.0$ & PL & -- & $-2.53_{-0.19}^{+0.13}$ & -- & -- & -- & $-2.86_{-0.13}^{+0.08}$ & $789/964$ \\
$115.0$ & $147.0$ & PL & -- & $-3.30_{-0.56}^{+0.27}$ & -- & -- & -- & $-3.66_{-0.42}^{+0.16}$ & $542/964$ \\
\hline
\end{longtable} 
\end{ThreePartTable}
\endgroup

\clearpage

\begin{table*}
\tiny
\centering
\scalebox{0.85}{
\begin{threeparttable}
\caption{\textbf{Spectral fitting results and corresponding fitting statistics for LEIA S-III and GECAM S-III spectra.} All errors represent the 1$\sigma$ uncertainties.}
\label{tab:sed}
\begin{tabular}{lccccccccccc}
\hline
& \multicolumn{3}{c}{\textbf{LEIA S-III:\tnote{*}}} & & \multicolumn{7}{c}{\textbf{GECAM S-III:}} \\
\cline{2-4} \cline{6-12}
Time & Best model & $\alpha$ & CSTAT/dof & & Best model & $\alpha_{1}$ & $\alpha_2(\alpha)$ & $\beta$ & ${\rm log}E_{\rm b}$ & ${\rm log}E_{\rm p}({\rm log}E_{\rm c})$ & PGSTAT/dof \\
(s) & & & & & & & & & (${\rm keV}$) & (${\rm keV}$) & \\
\hline
(1) 0--9 & PL & $-1.34_{-0.18}^{+0.19}$  & $27/20$ & & 2SBPL & $-0.71_{-0.01}^{+0.01}$ & $-1.15_{-0.01}^{+0.03}$ & $-3.90_{-0.03}^{+0.10}$ & $2.17_{-0.05}^{+0.02}$ & $3.023_{-0.012}^{+0.004}$ & $1718/960$ \\
(2) 9--16 & PL & $-1.24_{-0.20}^{+0.18}$ & $36/22$ & & 2SBPL & $-0.78_{-0.03}^{+0.01}$ & $-1.37_{-0.02}^{+0.01}$ & $-3.79_{-0.09}^{+0.11}$ & $1.80_{-0.02}^{+0.03}$ & $3.00_{-0.01}^{+0.01}$ & $1491/960$ \\
(3) 16--24 & PL & $-1.37_{-0.19}^{+0.19}$ & $24/23$ & & 2SBPL & $-0.98_{-0.02}^{+0.04}$ & $-1.65_{-0.01}^{+0.01}$ & $-4.31_{-0.42}^{+0.27}$ & $1.63_{-0.03}^{+0.01}$ & $2.98_{-0.02}^{+0.03}$ & $1176/960$ \\
(4) 24--32 & PL & $-1.65_{-0.21}^{+0.18}$ & $25/21$ & & 2SBPL & $-1.11_{-0.06}^{+0.04}$ & $-1.81_{-0.01}^{+0.01}$ & $-4.32_{-1.04}^{+0.47}$ & $1.48_{-0.02}^{+0.03}$ & $2.89_{-0.05}^{+0.06}$ & $1084/960$ \\
(5) 32--42 & PL & $-1.53_{-0.20}^{+0.17}$ & $19/23$ & & CPL & -- & $-1.73_{-0.02}^{+0.01}$ & -- & -- & $2.22_{-0.02}^{+0.03}$ & $1056/963$ \\
(6) 42--52 & PL & $-1.91_{-0.22}^{+0.19}$ & $13/21$ & & CPL & -- & $-1.86_{-0.03}^{+0.02}$ & -- & -- & $2.02_{-0.04}^{+0.05}$ & $1034/963$ \\
(7) 52--63 & PL & $-1.51_{-0.20}^{+0.17}$ & $16/20$ & & CPL & -- & $-1.92_{-0.04}^{+0.04}$ & -- & -- & $1.52_{-0.17}^{+0.10}$ & $1014/963$ \\
(8) 63--76 & PL & $-1.46_{-0.18}^{+0.18}$ & $17/18$ & & CPL & -- & $-1.95_{-0.05}^{+0.05}$ & -- & -- & ($2.66_{-0.12}^{+0.20}$)\tnote{\dag} & $889/963$ \\
(9) 76--91 & PL & $-1.56_{-0.22}^{+0.18}$ & $18/20$ & & CPL & -- & $-2.03_{-0.09}^{+0.09}$ & -- & -- & ($2.48_{-0.18}^{+0.31}$) & $886/963$ \\
(10) 91--147 & PL & $-2.00_{-0.16}^{+0.14}$ & $23/21$ & & CPL & -- & $-2.51_{-0.05}^{+0.30}$ & -- & -- & ($2.83_{-0.76}^{+0.02}$) & $399/963$ \\
(11) 147--243 & PL & $-2.80_{-0.27}^{+0.26}$ & $26/19$ & & -- & -- & -- & -- & -- & -- & -- \\
\hline
& & & & & & & & & & \\
\hline
& \multicolumn{11}{c}{\textbf{LEIA S-III \& GECAM S-III:}} \\
\cline{2-12}
Time & Best model & ${\rm CPL}.\alpha$ & ${\rm PL_{sa}}.\alpha_{\rm sa}$ & & ${\rm PL_{sa}}.{\rm log}E_{\rm sa}$ & $\alpha_{1}$ & $\alpha_2(\alpha)$ & $\beta$ & ${\rm log}E_{\rm b}$ & ${\rm log}E_{\rm p}({\rm log}E_{\rm c})$ & STAT/dof \\
(s) & & & & & (${\rm keV}$) & & & & (${\rm keV}$) & (${\rm keV}$) & \\
\hline
(1) 0--9 & CPL+PL$_{\rm sa}$-2SBPL\tnote{\ddag} & $-1.37_{-0.22}^{+0.14}$ & (1) & & $1.13_{-0.19}^{+0.03}$ & $-0.63_{-0.04}^{+0.04}$ & $-1.12_{-0.02}^{+0.03}$ & $-3.86_{-0.03}^{+0.09}$ & $2.09_{-0.04}^{+0.03}$ & $3.018_{-0.011}^{+0.004}$ & $(27+1715)/977$\tnote{**} \\
(2) 9--16 & CPL+PL$_{\rm sa}$-2SBPL & $-1.50_{-0.26}^{+0.11}$ & (1) & & $1.16_{-0.11}^{+0.01}$ & $-0.76_{-0.02}^{+0.03}$ & $-1.37_{-0.02}^{+0.01}$ & $-3.75_{-0.13}^{+0.07}$ & $1.82_{-0.02}^{+0.03}$ & $3.00_{-0.01}^{+0.01}$ & $(38+1487)/979$ \\
(3) 16--24 & CPL+PL$_{\rm sa}$-2SBPL & $-1.53_{-0.20}^{+0.15}$ & (1) & & $1.09_{-0.21}^{+0.03}$ & $-0.93_{-0.03}^{+0.05}$ & $-1.63_{-0.01}^{+0.02}$ & $-4.18_{-0.45}^{+0.20}$ & $1.64_{-0.03}^{+0.02}$ & $2.97_{-0.02}^{+0.03}$ & $(26+1176)/980$ \\
(4) 24--32 & CPL+PL$_{\rm sa}$-2SBPL & $-1.83_{-0.27}^{+0.20}$ & (1) & & $1.00_{-0.13}^{+0.08}$ & $-1.08_{-0.07}^{+0.04}$ & $-1.81_{-0.01}^{+0.02}$ & $-4.05_{-1.25}^{+0.23}$ & $1.50_{-0.03}^{+0.02}$ & $2.86_{-0.02}^{+0.08}$ & $(25+1084)/978$ \\
(5) 32--42 & CPL+PL$_{\rm sa}$-CPL & $-1.77_{-0.24}^{+0.11}$ & (1.5) & & $0.87_{-0.05}^{+0.09}$ & -- & $-1.70_{-0.02}^{+0.02}$ & -- & -- & ($2.78_{-0.04}^{+0.05}$)\tnote{\dag} & $(18+1055)/983$ \\
(6) 42--52 & CPL+PL$_{\rm sa}$-CPL & $-2.19_{-0.16}^{+0.26}$ & (1.5) & & $0.84_{-0.05}^{+0.10}$ & -- & $-1.83_{-0.03}^{+0.02}$ & -- & -- & ($2.83_{-0.06}^{+0.11}$) & $(12+1034)/981$ \\
(7) 52--63 & CPL+PL$_{\rm sa}$-CPL & $-1.89_{-0.16}^{+0.21}$ & (1.5) & & $0.80_{-0.06}^{+0.10}$ & -- & $-1.86_{-0.05}^{+0.04}$ & -- & -- & ($2.59_{-0.07}^{+0.17}$) & $(17+1015)/980$ \\
(8) 63--76 & CPL+PL$_{\rm sa}$-CPL & $-1.83_{-0.21}^{+0.13}$ & (1.5) & & $0.69_{-0.03}^{+0.16}$ & -- & $-1.82_{-0.11}^{+0.02}$ & -- & -- & ($2.60_{-0.05}^{+0.35}$) & $(18+890)/978$ \\
(9) 76--91 & CPL+PL$_{\rm sa}$-CPL & $-1.71_{-0.33}^{+0.03}$ & (1.5) & & $0.72_{-0.07}^{+0.10}$ & -- & $-1.93_{-0.09}^{+0.06}$ & -- & -- & ($2.80_{-0.32}^{+0.06}$) & $(18+887)/980$ \\
(10) 91--147 & CPL & $-1.71_{-0.07}^{+0.04}$ & \multicolumn{8}{c}{${\rm log}E_{\rm p}=1.19_{-0.05}^{+0.05}$} & $(26+403)/986$ \\
\hline
\end{tabular}
\begin{tablenotes}
\item [*] To align with the temporal divisions of the GECAM S-III spectra, we merge two bins, namely 91–113 s and 113–147 s, into a single bin. Consequently, there are 11 LEIA S-III spectra available for fitting the spectral energy distributions.
\item [\dag] In cases where the 1$\sigma$ lower limit of $\alpha_2(\alpha)$ nears or falls below $-2$, we use cutoff energy as a substitute for peak energy.
\item [\ddag] For the CPL components in the joint fittings, their cutoff energies are fixed at 53 keV to be same with that of the tenth SED's CPL model.
\item [**] In cases where LEIA and GECAM are jointly fitted, the reduced statistic is expressed as $({\rm CSTAT~of~LEIA} + {\rm PGSTAT~of~GECAM}) / {\rm dof}$. 
\end{tablenotes}
\end{threeparttable}}
\end{table*}

\clearpage

\begin{table*}
\centering
\tiny
\caption{\textbf{Smoothly broken power law fitting results for the multi-wavelength flux light curves.} All errors represent the 1$\sigma$ uncertainties.}
\label{tab:lc_sbpl}
\begin{tabular}{lccccccccc}
\hline
Energy & \multicolumn{2}{c}{$\hat{\alpha}_{1}$} & \multicolumn{2}{c}{$\hat{\alpha}_{2}$} & \multicolumn{3}{c}{$t_{\rm b}$} & ${\rm log}A$ & $\chi^2$/dof \\
(keV) & & & & & & (s) & & (${\rm erg~cm^{-2}~s^{-1}}$) & \\
\hline
LF: 0.5--4 & \multicolumn{2}{c}{$0.40_{-0.05}^{+0.05}$} & \multicolumn{2}{c}{$2.32_{-0.15}^{+0.16}$} & \multicolumn{3}{c}{$79.6_{-5.8}^{+5.5}$} & $-6.85_{-0.04}^{+0.04}$ & $12/17$ \\
\hline
\\
\hline
Energy & $\hat{\alpha}_{1}$ & $\hat{\alpha}_{2}$ & $\hat{\alpha}_{3}$ & $\hat{\alpha}_{4}$ & $t_{\rm b1}$ & $t_{\rm b2}$ & $t_{\rm b3}$ & ${\rm log}A$ & $\chi^2$/dof \\
(keV) & & & & & (s) & (s) & (s) & (${\rm erg~cm^{-2}~s^{-1}}$) & \\
\hline
GF1: 15--30 & $-0.70_{-0.04}^{+0.04}$ & $0.11_{-0.02}^{+0.02}$ & $2.45_{-0.02}^{+0.02}$ & $5.2_{-0.3}^{+0.4}$ & $7.5_{-0.3}^{+0.4}$ & $26.3_{-0.2}^{+0.2}$ & & $-5.537_{-0.005}^{+0.005}$ & \\
GF2: 30--100 & $-0.76_{-0.01}^{+0.01}$ & $0.47_{-0.01}^{+0.01}$ & $2.85_{-0.01}^{+0.01}$ & $6.9_{-0.4}^{+0.5}$ & $6.47_{-0.06}^{+0.06}$ & $22.0_{-0.1}^{+0.1}$ & $84.3_{-1.2}^{+1.2}$ & $-4.755_{-0.002}^{+0.002}$ & $40864/249$ \\
GF3: 100--350 & $-0.46_{-0.01}^{+0.01}$ & $1.30_{-0.01}^{+0.01}$ & $3.34_{-0.02}^{+0.02}$ & $8.8_{-0.8}^{+0.9}$ & $6.75_{-0.03}^{+0.03}$ & $21.0_{-0.1}^{+0.1}$ & & $-4.104_{-0.002}^{+0.002}$ & \\
\hline
GF4: 350--700 & $-0.37_{-0.01}^{+0.01}$ & $1.89_{-0.01}^{+0.01}$ & $4.12_{-0.06}^{+0.06}$ & $>6.5$ & $6.73_{-0.03}^{+0.03}$ & $22.0_{-0.2}^{+0.2}$ & ($84.3$) & $-3.976_{-0.002}^{+0.002}$ & $17452/86$ \\
GF5: 700--2000 & $-0.24_{-0.02}^{+0.02}$ & $2.24_{-0.04}^{+0.04}$ & $5.3_{-0.3}^{+0.3}$ & -- & $6.69_{-0.04}^{+0.05}$ & $22.0_{-1.4}^{+1.3}$ & -- & $-3.636_{-0.004}^{+0.004}$ & $8638/87$ \\
\hline
\end{tabular}
\end{table*}

\end{document}